\documentclass[twocolumn]{aastex631}
\usepackage{natbib}
\usepackage{multirow}
\usepackage{CJKutf8}
\usepackage{tablefootnote}
\usepackage{textcomp}
\usepackage{graphicx}% Use pdf, png, jpg, or eps§ with pdflatex; use eps in DVI mode
\usepackage{amsmath}
\usepackage{amssymb}

\newcommand{\msunyr}{{M$_{\sun}$ yr$^{-1}$}}

\newcommand{\kms}{{km s$^{-1}$}}

\def\lya{\mbox {Ly$\alpha$}}
\def\lyb{\mbox {Ly$\beta$}}

\def\ha{\mbox {H$\alpha$}}
\def\hb{\mbox {H$\beta$}}
\def\cm3{~cm$^{-3}$}

\newcommand{\mum}{\ifmmode{\rm \mu m}\else{$\mu$m}\fi}
\newcommand{\wbaha}{$W_{80,H\alpha}$}
\newcommand{\vbaha}{$v_{80,H\alpha}$}
\newcommand{\vwuha}{{$v_{50,H\alpha}$}}
\newcommand{\wbao}{$W_{80,[O~\sc{III}]}$}
\newcommand{\vbao}{$v_{80,[O~\sc{III}]}$}
\newcommand{\vwuo}{{$v_{50,[O~\sc{III}]}$}}
\newcommand{\wba}{$W_{80}$}
\newcommand{\vba}{$v_{80}$}
\newcommand{\vwu}{{$v_{50}$}}
\newcommand{\vjiu}{{$v_{90}$}}
\newcommand{\vyi}{{$v_{10}$}}
\newcommand{\ajiuyi}{{$A_{91}$}}
\newcommand{\ajiuyiha}{{$A_{91,H\alpha}$}}
\newcommand{\ajiuyio}{{$A_{91,[O~\sc{III}]}$}}

\newcommand{\flux}{erg cm$^{-2}$ s$^{-1}$}
\newcommand{\lum}{erg s$^{-1}$}

\newcommand{\oiii}{[O~{\sc iii}] $\lambda$5007}
\newcommand{\oiiia}{[O~{\sc iii}] $\lambda$4959}

\newcommand{\niiab}{[N~{\sc ii}] $\lambda\lambda$6548, 6583}

\newcommand{\lagn}{L$_{AGN}$}
\newcommand{\lbol}{L$_{bol}$}

\newcommand{\fagn}{f$_{AGN}$}
\newcommand{\ewlya}{EW$_{Ly\alpha}$}

\newcommand{\logluv}{log($\lambda L_{1125})$}
\newcommand{\luv}{$\lambda L_{1125}$}

\newcommand{\cf}{C$_{f}$}

\newcommand{\lir}{L$_{IR}$}

\newcommand{\NHxray}{N$_{H}$({\rm X-ray})}

\newcommand{\aox}{$\alpha_{OX}$}
\newcommand{\aux}{$\alpha_{UV,X}$}
\newcommand{\ebv}{$E(B-V)$}
\newcommand{\weq}{$W_{eq}$}
\newcommand{\vavg}{$v_{wtavg}$}
\newcommand{\savg}{$\sigma_{rms}$}

\newcommand{\hahb}{H$\alpha$/H$\beta$}

\newcommand{\niuv}{N~I 1200}
\newcommand{\niiuv}{N~II 1084}

\newcommand{\siiic}{Si~II 1260}
\newcommand{\siiiabc}{Si~II 1190, 1193, 1260}

\newcommand{\siiii}{Si~III 1206}

\newcommand{\nad}{Na~I~D $\lambda\lambda$5890, 5896}
\newcommand{\nadb}{Na~I~D}

\newcommand{\ovi}{O~VI\ 1032, 1038}

\newcommand{\nv}{N~V 1238, 1243}

\urldef{\kcwiurl}\url{https://github.com/Keck-DataReductionPipelines/KcwiDRP/blob/master/AAAREADME}
\urldef{\dimmurl}\url{http://mkwc.ifa.hawaii.edu/current/seeing/index.cgi}
\urldef{\irsaurl}\url{https://irsa.ipac.caltech.edu/cgi-bin/Gator/nph-scan?mission=irsa&submit=Select&projshort=2MASS}
\urldef{\nsaurl}\url{http://www.nsatlas.org/data}

\def \fesc {\mbox{$f_{esc}({\rm Ly}\alpha)$}}

\begin{document}

\title{Galactic Winds across the Gas-Rich Merger Sequence \\
II. Lyman Alpha Emission and Highly Ionized O~VI and N~V Outflows in Ultraluminous Infrared Galaxies}

\correspondingauthor{Weizhe Liu}
\email{oscarlwz@gmail.com}

\author[0000-0003-3762-7344]{Weizhe Liu \begin{CJK}{UTF8}{gbsn}(刘伟哲)\end{CJK}}
\affiliation{Department of Astronomy, University of Maryland, College Park, MD 20742, USA}

\author[0000-0002-3158-6820]{Sylvain Veilleux} 
\affiliation{Department of Astronomy, University of Maryland, College Park, MD 20742, USA}
\affiliation{Joint Space-Science
  Institute, University of Maryland, College Park, MD 20742, USA}

\author[0000-0002-1608-7564]{David S. N. Rupke}
\affiliation{Department of Physics, Rhodes College, Memphis, TN 38112, USA}

\author[0000-0002-1218-640X]{Todd M. Tripp}
\affiliation{Department of Astronomy, University of Massachusetts, Amherst, MA 01003, USA}

\author{Frederick Hamann}
\affiliation{Department of Physics and Astronomy, University of California, Riverside, CA 92507, USA} 

\author[0000-0001-9189-7818]{Crystal Martin}
\affiliation{
Department of Physics, University of California, Santa Barbara, CA, 93106, USA}

\begin{abstract}
This paper is the second in a series aimed at examining the gaseous environments of z$\le$0.3 quasars and ultraluminous infrared galaxies (ULIRGs) as a function of AGN/host galaxy properties across the merger sequence. This second paper focuses on the \lya\ emission and \ovi\ and \nv\ absorption features, tracers of highly ionized gas outflows, in ULIRGs observed with HST/COS. \lya\ emission is detected in 15 out of 19 ULIRGs, and 12 of the 14 clear \lya\ detections show emission with blueshifted velocity centroids and/or wings. The equivalent widths of the \lya\ emission increase with increasing AGN luminosities and AGN bolometric fractions. The blueshifts of the \lya\ emission correlate positively with those of the \oiii\ emission, where the latter traces the ionized gas outflows. The \lya\ escape fractions tend to be slightly larger in objects with stronger AGN and larger outflow velocities, but they do not correlate with nebular line reddening. Among the 12 ULIRGs with good continuum signal-to-noise ratios, O~VI and/or N~V absorption features are robustly detected in 6 of them, all of which are blueshifted, indicative of outflows. In the combined ULIRG $+$ quasar sample, the outflows are more frequently detected in the X-ray weak or absorbed sources. The absorption equivalent widths, velocities and velocity dispersions of the outflows are also higher in the X-ray weak sources. No other strong correlations are visible between the properties of the outflows and those of the AGN or host galaxies.
\end{abstract}


\section{Introduction} \label{1}

Major mergers of gas-rich galaxies, both near and far, are the paradise for magnificent starbursts and rapid growth of supermassive black holes. In the local universe, the majority of the ultraluminous infrared galaxies (ULIRGs) are mergers of gas-rich galaxies. The merger process drives the gas and dust to the central region of the system, fueling the (circum)nuclear starbursts and the rapid accretion of the supermassive black holes. As described by a popular evolution scenario, the merger system advances from the ULIRG phase to the dusty quasar phase, and then to a fully-exposed quasar phase, with the gas and dust either transformed into stars or expelled and/or heated by the galactic winds triggered by the quasar and starburst activities \citep[e.g.,][]{Sanders1988,Veilleux2009,Hickox2018}. The ubiquity of galactic winds in local ULIRGs, dusty quasars, and luminous post-starburst galaxies supports this scenario. The observed winds extend over a large physical scale, from fast, nuclear winds on  $\lesssim$pc scales all the way to galactic winds reaching $\gtrsim$10 kpc, blending smoothly with the circumgalactic medium \citep[e.g.,][for a review]{Martin2005,Rupke2005c,Martin2006,Tremonti2007,Martin2009,Sturm2011,RupkeVeilleux2013b,Veilleux2013a,Veilleuxherschel,Cicone2014,Veilleux2014,Tombesi2015,Rupke2017,Liu2019,Fluetsch2019,Fluetsch2020,Lutz2020,Veilleux2020}.

While the cooler, neutral and/or molecular phases on larger scale ($\gtrsim$kpc) often dominate the outflow energetics, it is the hotter, ionized phase of the wind that serves as the best probe for the driving mechanism of these winds. ULIRG F11119$+$3257, arguably the best example so far, possesses a massive, galactic scale (1--10 kpc) molecular and neutral-gas outflow apparently driven by the fast ($>$0.1 c), highly ionized (Fe {\sc XXV} and Fe {\sc XXVI} at $\sim$7 keV) nuclear wind \citep{Tombesi2015,Tombesi2017,Veilleux2017}. While this result is intriguing, the faintness of the majority of ULIRGs at $\sim$ 7 keV, unlike many quasars, has impeded a statistically meaningful study of this phenomenon in most ULIRGs with current X-ray facilities.       

The superb far-ultraviolet (FUV) sensitivity of the Cosmic Origins Spectrograph (COS) on the Hubble Space Telescope (HST) provides a powerful alternative tool for such study in the low-z universe. Rest-frame FUV spectroscopy has enabled a comprehensive study of the multi-phase nature of outflows, built upon the abundant spectral features arising from the high-ionization, low-ionization, and neutral phases of the outflowing gas \citep[e.g.,][]{Chisholm2015,Tripp2011,Heckman2015,Hamann2019b,Arav2020}. Up to now, only about a dozen ULIRGs have been studied with HST/COS data, but the results are fascinating. In Mrk 231, highly blueshifted \lya\ emission (with respect to systemic velocity) is observed to coincide in velocity with the highly blueshifted absorption features tracing the fast outflow in this galaxy, suggesting an outflow-related origin for the \lya\ emission \citep[][]{Veilleux2013a,Veilleux2016}. With a larger sample of 11 ULIRGs, \citet[][hereafter M15]{Martin2015} have shown that prominent, blueshifted \lya\ emission down to $-$1000 \kms\ exists in about half of the objects, and they argued that the blueshifted \lya\ emission originates from the clumps of gas condensing out of hot winds driven by the central starbursts \citep{Thompson2016}. In addition, blueshifted absorption features from high-ionization species like O$^{5+}$ and/or N$^{4+}$ (114 and 77 eV are needed to produce these ions, respectively) and low-ionization species like Si$^+$ and Fe$^+$ are also detected in a few objects, providing unambiguous evidence of outflowing gas.

Despite the tantalizing evidence of FUV-detected outflows in the ULIRGs described above, the sample examined so far is small and incomplete, where AGN-dominated ULIRGs and matched quasars are largely missing. To address this issue, we have selected a more complete sample of ULIRGs and quasars to systematically study the gaseous environments along the merger sequence, from ULIRGs to quasars. In \citet{Veilleux2021} (hereafter Paper I), we presented the results from the first part of our study, focused on the highly-ionized gas outflows, traced by \ovi\ and \nv\ absorption features, in a sample of 33 local quasars. We found that the O~VI and N~V outflows are present in $\sim$61\% of the sample, and the incidence rate and equivalent widths (EWs) of these highly ionized outflows are higher among X-ray weak or absorbed sources. Similarly, the flux-weighted outflow velocity dispersions are also the highest among the X-ray weakest sources. However, no significant correlation is visible between the flux-weighted outflow velocities/velocity dispersions and the other properties of the quasars and host galaxies. 

In this paper, we report the results from an analysis of the \lya\ emission and O~VI and N~V absorption features of the 21 ULIRGs in the sample \footnote{While FUV studies analyze Si~IV and C~IV transitions along with the lines of O~VI and N~V, our spectra do not cover the Si~IV and C~IV doublets at the target redshifts.}, expanding on the results from Paper I by considering the combined ULIRG $+$ quasar sample. In Sec. \ref{2}, we describe the HST/COS observations of the ULIRG sample, the reduction of the data sets, and the ancillary data from the literature. In Sec. \ref{3}, we present the analysis of the FUV spectra of the ULIRGs, focusing on \lya\ emission in the first part and the O~VI and N~V absorption features in the second. In Sec. \ref{4}, we discuss the potential key factors that control the observed \lya\ properties, and in Sec. \ref{5}, we examine the incidence rates and properties of the O~VI and N~V outflows. In Sec. \ref{6}, we search for trends between the O~VI and/or N~V outflow properties and the AGN/galaxy properties in the ULIRG$+$quasar sample. In Sec. \ref{7}, we summarize the main results of this paper. Throughout the paper, we adopt a $\Lambda$CDM cosmology with $H_0$ = 75 km s$^{-1}$ Mpc$^{-1}$, $\Omega_{\rm m}$ = 0.3, and $\Omega_{\rm \Lambda} = 0.7$.

\section{HST and Ancillary Data on the ULIRGs} \label{2}

\begin{deluxetable*}{ccccc cccccccc}
\tablecolumns{13}
\tabletypesize{\scriptsize}
\tablewidth{\textwidth}
\tablecaption{Basic Properties of the ULIRGs in the Sample\label{tab:targets}}
\tablehead{\colhead{Name} & \colhead{Short} & \colhead{z} & \colhead{log($\frac{L_{bol}}{L_\odot})$} & \colhead{Spectral} & \colhead{Merger} &  \colhead{AGN} & \colhead{m$_{FUV}$} & \colhead{log(SX)} & \colhead{log(HX)} & \NHxray & \logluv  & \aux \\
& Name & &  & Type & Class &  Fraction & (AB)& [\lum] & [\lum] & [10$^{22}$ cm$^{-2}$] & & }
\colnumbers
\startdata
F01004$-$2237  &  F01004      & 0.117701\tablenotemark{a}   & 12.36 & HII &  V    & 55 & 18.5 &  41.0           & 42.1      & 0.16  & 44.2       &    $-$3.0 \\
QSO-B0157$+$001   &  Mrk~1014 & 0.16311\tablenotemark{b}   & 12.70 & S1  &  IVb  & 73$_{-27}^{+24}$ & 16.6 &  43.92$\pm{0.04}$           & 43.83$^{+0.11}_{-0.25}$   & $<$0.009   & 45.3                  & $-$1.2 \\
... & ... & ... & ... & ... & ... & ... & ... & 44.00$^{+0.04}_{-0.08}$ & 43.86$^{+0.04}_{-0.04}$ & $<$0.009 & ... & ...  \\
F04103$-$2838     &  F04103   & 0.117464\tablenotemark{b}   & 12.30 & L   &  IVb  & 49 & 20.2 &  41.8           & 42.1     & 0.190     & 42.6          & $-$1.4 \\
F05189$-$2524     &  F05189   & 0.04288\tablenotemark{c} & 12.22 & S2  &  IVb  & 71 & 19.1 &  43.4           & 43.4         & 7.83     & 42.8             & $-$1.1 \\ 
F07599$+$6508     &  F07599   & 0.1483\tablenotemark{c} & 12.58 & S1  &  IVb  & 75 & 17.6 &  43.5           & 42.9         & 52.2     & ...              & $-$2.5 \\ 
F08572$+$3915:NW &  F08572    & 0.0584\tablenotemark{d} & 12.22 & L   &  IIIb & 72 & 19.1 &  40.0           & 41.4          & 2.40    & 42.0             & $-$3.4 \\
F11119$+$3257     &  F11119   & 0.189\tablenotemark{e} & 12.64 & S1  &  IVb  & 80 & 21.4 &  44.2           & 44.2         & 1.71     & ...              & 0.1 \\  
Z11598$-$0112  &  Z11598      & 0.150694\tablenotemark{a} & 12.49 & S1  &  IVb  & 74 & 17.9 &  43.6           & 43.1           & ...     & 44.6                  & $-$0.9 \\ 
F12072$-$0444     &  F12072       & 0.128360\tablenotemark{a}   & 12.45 & S2  &  IIIb & 65 & 20.5 &  41.4           & 41.2     & $<$0.01      & 42.9             & $-$1.8 \\ 
3C~273           &  3C~273    & 0.158339\tablenotemark{b}   & 13.03 & S1  &  IVb  & 95$_{-13}^{+5}$ & 13.3 &  45.491$_{-0.002}^{+0.002}$  &  45.742$_{-0.008}^{+0.008}$  & $<$0.01 & 46.5        & $-$0.9 \\
... & ... & ... & ... & ... & ... & ... & ... &                            45.461$_{-0.004}^{+0.004}$  &  45.722$_{-0.006}^{+0.005}$  & $<$0.01 & ... & ... \\
... & ... & ... & ... & ... & ... & ... & ... &                            45.591$_{-0.002}^{+0.002}$  &  45.820$_{-0.005}^{+0.004}$  & $<$0.01 & ... & ... \\
... & ... & ... & ... & ... & ... & ... & ... &                            45.663$_{-0.001}^{+0.002}$  &  45.825$_{-0.006}^{+0.006}$  & $<$0.01 & ... & ... \\
... & ... & ... & ... & ... & ... & ... & ... &                            45.461$_{-0.004}^{+0.003}$  &     45.67$_{-0.02}^{+0.01}$  & $<$0.01 & ... & ... \\
... & ... & ... & ... & ... & ... & ... & ... &                            45.544$_{-0.003}^{+0.003}$  &  45.941$_{-0.007}^{+0.010}$  & $<$0.01 & ... & ... \\
Mrk~231         &  Mrk~231    & 0.0422\tablenotemark{c} & 12.61 & S1  &  IVb  & 71$_{-7}^{+7}$ & 19.0 &  42.13$_{-0.04}^{+0.01}$           & 42.58$_{-0.11}^{+0.01}$   & 9.5$_{-1.9}^{+2.3}$ & 42.7        & $-$1.3 \\
... & ... & ... & ... & ... & ... & ... & ... &  ... & ...  & 19.4$_{-4.4}^{+5.7}$ & ... & ... \\
F13218$+$0552     &  F13218   & 0.2047\tablenotemark{c} & 12.68 & S1  &  V    & 83 & 19.4 &  42.2           & 42.7   & 0.17     & 43.8              & $-$2.0 \\
F13305$-$1739     &  F13305   & 0.148365\tablenotemark{b} & 12.27 & S2  &  V    & 88 & 19.5 &  ...            & ...    & ...      & ...                   & ... \\ 
Mrk~273         &  Mrk~273    & 0.03778\tablenotemark{b} & 12.24 & S2  &  IVb  & 46 & 18.6 &  42.7     & 43.0   & 41.3     & 42.0               & $-$1.6 \\
F14070$+$0525     &  F14070   & 0.26438\tablenotemark{b} & 12.84 & S2  &  V    & 41 & 21.0 &  ...            & ...        & ...      & ...                    & ... \\ 
F15001$+$1433:E  &  F15001    & 0.162746\tablenotemark{b} & 12.51 & S2  &  Tpl  & 43 & 20.3 &  ...            & ...         & ...     & ...                    & ... \\ 
F15250+3608     &  F15250     & 0.05515\tablenotemark{b} & 12.12 & L   &  IVa  & 45 & 18.5 &  43.1           & 42.8       & 132      & 42.7              & $-$2.3 \\ 
F16156$+$0146:NW &  F16156    & 0.132\tablenotemark{e} & 12.19 & S2  &  IIIb & 77 & 20.9 &  ...            & ...         & ...     & ...                    & ... \\ 
F21219$-$1757     &  F21219   & 0.1127\tablenotemark{c} & 12.17 & S1  &  V    & 78 & 17.3 &  ...            & ...        & ...      & 44.7                   & ... \\ 
F23060$+$0505     &  F23060   & 0.173\tablenotemark{e} & 12.50 & S2  &  IVb  & 75 & 20.2 &  ...            & ...        & ...      & ...                    & ... \\ 
F23233$+$2817     &  F23233   & 0.114009\tablenotemark{b} & 12.06 & S2  &  Iso  & 72 & 20.9 &  ...            & ...        & ...      & 41.0                   & ...  
\enddata
\label{targets}
\tablecomments{Column (1): Name of the object; Column (2): Short names of the objects adopted in this paper; Column (3): Redshift. The label on the upper-right corner of each entry indicates the reference of the redshift. $a$: Based on the narrow optical emission lines from \citet{Martin2015}. $b$: The best redshift adopted by NASA/IPAC Extragalactic Database\footnote{The NASA/IPAC Extragalactic Database (NED)
is operated by the Jet Propulsion Laboratory, California Institute of Technology,
under contract with the National Aeronautics and Space Administration.}. $c$: Based on the narrow optical emission lines from \citet{Rupke2017}.
$d$: Based on the narrow optical emission lines From \citet{RupkeVeilleux2013a}.
$e$: Based on the optical emission lines from \citet{kim+sanders98}; Column (4): Log of bolometric luminosity in solar units based on. For ULIRGs, we assumed \lbol $=$ 1.15 \lir\ where \lir\ is the 8-1000 \mum\ infrared luminosity from \citet{kim+sanders98}. For the QUEST quasars, we assumed \lbol $=$ 7 L(5100\AA) $+ $ \lir\ based on \citet{Netzer2007}; Column (5): Optical spectral type from \citet{Veilleux1999}: S1 means Seyfert 1, S2 means Seyfert 2, L means LINER, HII means star-forming. Column (6): Interaction class from Veilleux+02: I–First approach, II–First contact, III(a/b)–Pre-merger (Wide binary/Close binary), IV(a/b)–Merger (Diffuse/Compact), V–Old Merger, Iso–Isolated, Tpl–triplet. Column (7): Fraction (in percentage) of
the bolometric luminosity produced by the AGN, based on the mean values derived in \citet{Veilleux2009a}. The typical uncertainties are 10--15\%.  Column (8):
FUV AB magnitudes from GALEX. Column (9): Soft (0.5--2 keV) X-ray luminosity. For X-ray related quantities (Column (9), (10), (11) and (13)), different rows $=$ different observations dates. The references for these quantities are described in Sec. \ref{232}. The errors are omitted whenever they are not available from the literature or public data; Column (10) Hard (2--10 keV) X-ray luminosity; Column (11):  The X-ray absorbing column density in units of 10$^{22}$ cm$^{-2}$; Column (12): Monochromatic continuum luminosity at the rest-frame 1125 \AA; Column (13): X-ray-to-UV index, \aux, as defined in Sec. \ref{232}.}
\end{deluxetable*}

\subsection{HST/COS G130M Observations of ULIRGs} \label{21}

Our sample is selected based on three criteria: (1) They are part of the 1-Jy sample of 118 local ULIRGs with z $<$ 0.3 and $|$b$|$ $>$ 30$^\circ$ \citep[hence modest Galactic extinctions;][]{Kim1998}; (2) In order to address the role of AGN feedback in these systems, the ULIRGs have AGN signatures in the optical (AGN Type 1 or 2) or in the mid-infrared \citep[Spitzer-derived AGN bolometric fraction $\gtrsim$40\%;][]{Veilleux2009}. (3) They are the FUV-brightest ULIRGs of the 1-Jy sample with FUV magnitudes (AB) m$_{FUV}$ $\lesssim$ 21. These criteria result in a sample of 21 objects (Table \ref{tab:targets}): 15 of which were observed through the HST cycle 26 program (PID:15662; PI: Sylvain Veilleux), and the remaining 6 objects have archival COS G130M spectra of sufficient quality from three programs (PID: 12533, PI: C.\ Martin; PID: 12569, PI: S.\ Veilleux; PID: 12038: PI: J.\ Green). Among these 6 objects, QSO-B0157$+$001, 3C~273, and Mrk~231 were also studied in Paper I as they meet the criteria for QUEST quasars, and F01004$-$2237, Z11598$-$0112, and F12072$-$0444 were also studied in M15. In the following sections, we adopt the short names listed in Table \ref{tab:targets} when referring to the objects in our sample. 

The Cycle 26 HST/COS spectra presented in this paper were obtained in TIME-TAG mode through the PSA using the medium resolution FUV grating, G130M. Four focal plane offset positions were adopted to reduce the impact of fixed-pattern noise associated with the micro-channel plate.  We got all four FP-POS settings for all targets, except for targets F04103, F14070, F21219, and F23233 with a central wavelength of 1291 \AA. For these objects, we followed the COS2025 recommendations and used FP-POS $=$ 3 and 4 to get equal exposures for segments A and B. The wavelength setting was adjusted according to the redshift of the target and was selected to optimize the number of strong lines that can be observed with G130M. At all but the highest redshifts (z $<$ 0.2), \lya, the high-ionization lines from N~V, and low-ionization lines from \siiic\ and \siiii\ fit within the wavelength coverage of the data. At the highest redshifts (z $\simeq$ 0.20-0.27), we lose N~V and \lya\ but gain \lyb\ and \ovi. The wavelength coverage for individual observations is summarized in Table \ref{tab:obs}.

\begin{deluxetable*}{ccccc ccc}
\tablecolumns{8}
\tabletypesize{\scriptsize}
\tablecaption{Summary of the HST/COS  G130M Observations\label{tab:obs}}
\tablehead{\colhead{Name} & \colhead{PID} & \colhead{PI} & \colhead{Date} & \colhead{R.A.} &  \colhead{Dec.} & \colhead{Wavelength Coverage} & \colhead{t$_{exp}$} }
\colnumbers
\startdata
   F01004  & 12533 & C. Martin   & 2011-12-03 & 01 02 49.9631 & $-$22 21 57.02 & 1137-1274/1292-1432 & 1716  \\
   Mrk~1014   & 12569 & S. Veilleux & 2012-01-25  &  01 59 50.250  & $+$00 23 41.30     &  1154-1468  & 1961 \\
   F04103    & 15662  & S. Veilleux &  2020-01-23  &  12 19.413 & -28 30 24.64 & 1137-1274/1292-1432 & 10798  \\
   F05189     & 15662 & S. Veilleux &  2019-07-18  &  05 21 01.388 & -25 21 45.10 & 1069-1207/1223-1363  & 7802  \\ 
   F07599     & 15662 & S. Veilleux &  2019-02-21  &  04 30.487 & +64 59 52.75 & 1173-1312/1328-1468  &    2148 \\
  F08572     & 15662  & S. Veilleux &  2019-10-17  &  00 25.281 & +39 03 54.83 & 1069-1207/1223-1363  &    7906 \\
   F11119     & 15662 & S. Veilleux &  2020-01-20 &  14 38.908 & +32 41 33.04 & 1173-1312/1328-1468  &    13537 \\ 
  Z11598   & 12533 & C. Martin   & 2011-11-18  & 12 02 26.7505  & $-$01 29 15.49  &  1154-1468 & 1304 \\  
   F12072     & 12533 & C. Martin   & 2013-01-24 & 12 09 45.1000  & $-$05 01 13.20  & 1137-1448  & 1176   \\
   3C~273      & 12038 & J. Green    &  2012-04-22 &  12 29 06.695  & $+$02 03 08.66   & 1137-1408 &  4515   \\
   Mrk~231    & 12569 & S. Veilleux &  2011-10-15 & 12 56 14.111 & $+$56 52 24.70  & 1154-1468  &  12851  \\
   F13218     & 15662 & S. Veilleux &  2020-05-21  &  24 19.897 & +05 37 05.06 & 1069-1207/1223-1363  &    7742 \\
   F13305     & 15662 & S. Veilleux &  2019-12-17  &  33 16.529 & $-$17 55 10.52 &  1173-1312/1328-1468 &    7776 \\
   Mrk~273     & 15662 & S. Veilleux &  2019-07-23 & 13 44 42.080 & $+$55 53 12.99 &  1069-1207/1223-1363 &    4086 \\ 
   F14070     & 15662 & S. Veilleux  &  2020-06-30  &  09 31.249 & +05 11 31.45 &  1137-1274/1292-1432 &    13377\\ 
   F15001     & 15662 & S. Veilleux  &  2020-03-01  &  02 31.936 & +14 21 35.15 &  1173-1312/1328-1468 &    10719 \\
   F15250     & 15662 & S. Veilleux  &  2019-07-11  &  26 59.463 & +35 58 37.47 & 1069-1207/1223-1363 &    4951 \\
   F16156     & 15662 & S. Veilleux  &  2020-07-08  &  18 09.426 & +01 39 21.66 &  1173-1312/1328-1468 &    13346 \\
   F21219     & 15662 & S. Veilleux  &  2019-04-13  &  24 41.606 & $-$17 44 45.52 & 1137-1274/1292-1432 &    2210 \\
   F23060     & 15662 & S. Veilleux  &  2019-12-05  &  08 33.947 & +05 21 29.95 & 1173-1312/1328-1468 &    10703 \\
   F23233     & 15662 & S. Veilleux  &  2019-10-14,2020-01-2223  &  25 49.406 & +28 34 20.84 &  1137-1274/1292-1432 & 13477
\enddata
\tablecomments{Column (1): Name of the object; Column (2): HST Program ID; Column (3): Principal Investigator; Column (4): Date of Observation; Column (5) \& (6): J2000 coordinates; Column (7): Wavelength coverage of the observations in \AA. The values are formatted as A/B segments or entire wavelength ranges; Column (8) Total exposure time in seconds.} 
\end{deluxetable*}

\subsection{HST/COS Data Reduction} \label{22}

Among the 6 objects with archival data, we retrieved the fully reduced spectra for 5 of them from the Hubble Legacy Spectrum Archive \citep{hsla}, and obtained the fully reduced spectrum of F12072$-$0444 published in M15 from C.\ L.\ Martin. For the other 15 newly-observed objects presented in this paper, the raw data were processed and combined by the CALCOS pipeline v3.3.10. CALCOS corrects the data for instrumental effects, assigns a vacuum wavelength scale, and extracts flux-calibrated spectra. It applies a heliocentric correction to the final x1d files for each exposure, and combines the individual exposures to a single spectra when possible.  

The COS aperture is filled with emission from geocoronal airglow, so the observed wavelengths of these lines are independent of target position in the PSA. By inspecting the velocity offsets of theoretical and observed wavelength of strong geocoronal lines, we can examine the potential systematic errors in the wavelength calibration. The measured velocity offsets are randomly distributed with absolute values $<$ 30 \kms, and we take these measurements as typical errors of the wavelength calibration from the pipeline.

Finally, all spectra are corrected for foreground Galactic extinctions from \citet{GalacticExtinction} and the reddening curve with $R_V = $3.1 of \citet{Fitzpatrick1999}.

\subsection{Ancillary Data and Measurements} \label{23}

\subsubsection{General Physical Properties} \label{231}
For the bolometric luminosities of our sources, we adopt \lbol\ $=$ 1.15 \lir\ where \lir\ is the 8-1000 \mum\ infrared luminosity retrieved  from \citet{Kim1998}, except for the three sources that are also QUEST quasars, which are quoted from Paper I where we assume \lbol $=$ 7 L(5100\AA) $+ $ \lir\ based on \citet{Netzer2007}.  The L(5100\AA) is the continuum luminosity $\lambda$L$_\lambda$ at 5100 \AA\ rest wavelength and \lir\ is the 1 -- 1000 \mum\ infrared luminosity (the details can be found in the notes of Table 1 in Paper I).

The optical spectral classifications are quoted from \citet{Veilleux1999}: S1 means Seyfert 1, S2 means Seyfert 2, L means LINER, HII means star-forming galaxies.  The interaction classes (or merger classes) are from \citet{Veilleux2002}: I–First approach, II–First contact, III(a/b)–Pre-merger (Wide binary/Close binary), IV(a/b)–Merger (Diffuse/Compact), V–Old Merger, Iso–Isolated, Tpl–triplet. The fraction of the bolometric luminosity produced by the AGN, or simply AGN fractions, are the average values derived in \citet{Veilleux2009}. The AGN luminosities are defined as the bolometric luminosities multiplied by the AGN fractions.

\subsubsection{X-ray Data} \label{232}

Published X-ray data and measurements exist for 14 out of the 21 objects: Those for the 3 quasars (QSO-B0157$+$001, 3C~273, Mrk~231) also studied in Paper I are from \citet{Teng2010,Teng2014, Veilleux2014,Ricci2017} Those for the remaining 11 sources are from a series of X-ray studies of ULIRGs and quasars \citep{Teng2005,Teng2010}.

Following Paper I, the X-ray weakness of AGN/quasars can be described with the X-ray to optical spectral index, \aox\ $\equiv 0.372 log[(F(2\ {\rm keV})/F(3000 {\rm \AA})]$ \citep[e.g.,][]{Brandt2000}. While \aox\ is measured for most nearby quasars, there are virtually no published measurements for the ULIRGs in our sample. Instead, we have defined an alternative X-ray to FUV spectral index, \aux, based on the ratio of the soft X-ray (0.5--2 keV) flux to the FUV flux from GALEX \citep{GALEX}, where \aux\ $\equiv log[F(0.5-2\ {\rm keV})/F(FUV)]$. These results are listed in Table \ref{tab:targets}.

\begin{figure}[!htb]   
\plotone{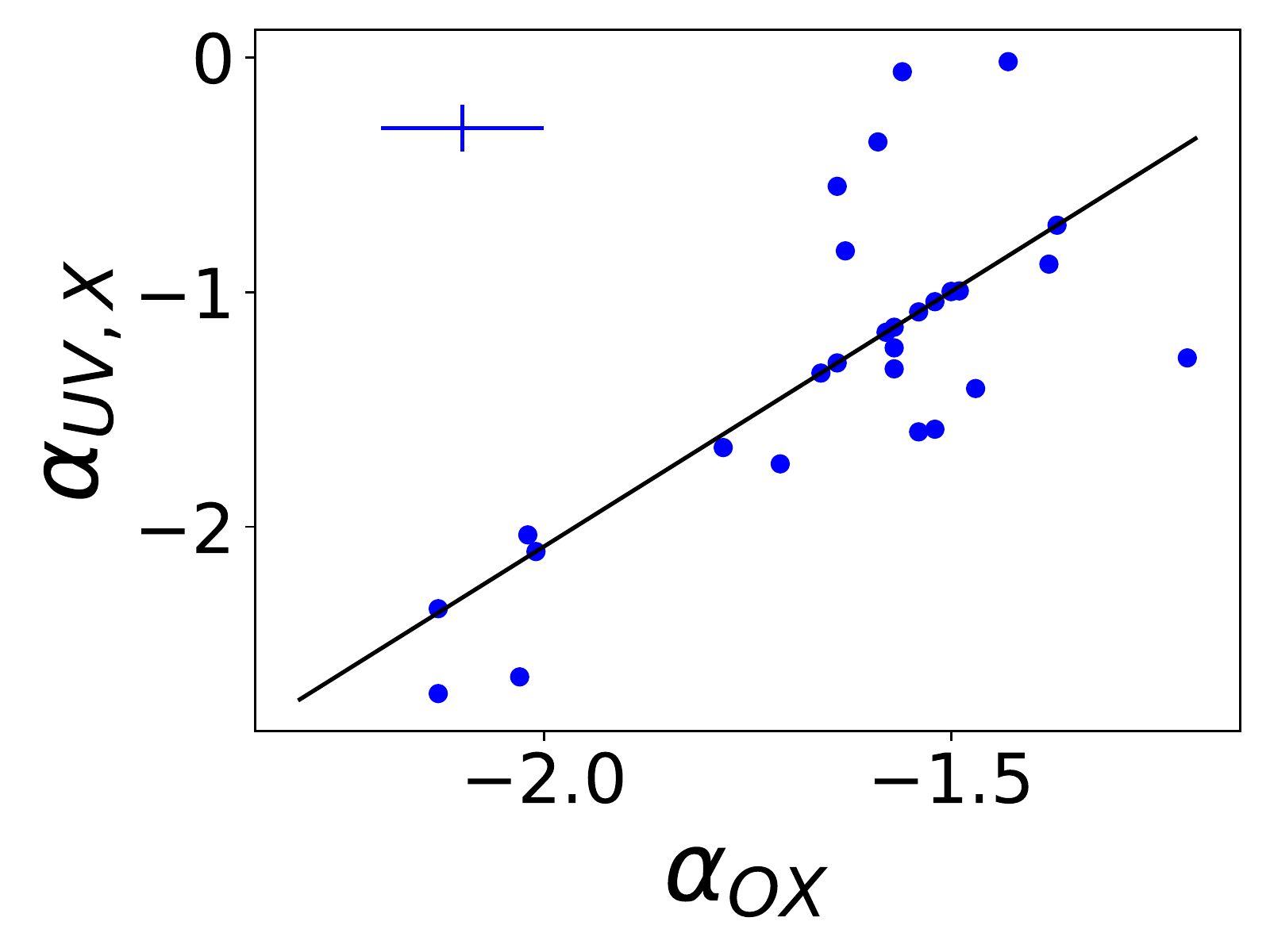}
\caption{The X-ray to optical spectral index \aox\ versus the X-ray to FUV spectral index \aux\ for the quasars in Paper I. The two indices are defined as \aox\ $\equiv 0.372 log[(F(2\ {\rm keV})/F(3000 {\rm \AA})]$ and \aux\ $\equiv log[F(0.5-2\ {\rm keV})/F(FUV)]$, respectively. The solid line is a linear fit to the data points. The errors on \aox\ and \aux\ are uncertain, and largely associated with the uncertainties in the analyses of the X-ray spectra, as described in, e.g., \citet[][]{Teng2010}. The cross in the upper left corner of the figure indicates $\pm$ 0.1 dex errors adopted in the Kendall tau test and the fit.}
\label{fig:aox_aux}
\end{figure}

For the QUEST quasars studied in Paper I, there is a clear positive correlation between \aux\ and \aox\ (see Fig. \ref{fig:aox_aux}), with a $p$-value $\simeq$ 2$\times$10$^{-5}$ from the Kendall tau test (the null hypothesis is no correlation), which demonstrates that \aux\ is indeed a good surrogate for \aox. For the quasars without published 0.5--2 keV flux from Chandra or XMM-Newton and/or FUV flux from GALEX, we convert their \aox\ listed in Paper I to \aux\ by adopting the relation \aux$=2.17$\aox$+2.26$ , which is obtained from a linear fit to the quasars with both \aux\ and \aox\ measurements. 

\subsubsection{Optical Spectra} \label{233}

The Gemini/GMOS IFU spectra from \citet{RupkeVeilleux2013a}, \citet{Rupke2017}, or the SDSS spectra \citep{sdss3} are adopted as the optical spectra for our objects by default unless explicitly stated otherwise (the GMOS data are adopted by default whenever both GMOS and SDSS data are available for the same object). The long-slit, optical spectra of all objects but 3C~273 are retrieved from \citet{Veilleux1999}, and the optical spectrum of 3C~273 is retrieved from \citet{3C273}. They are adopted as the default optical spectra whenever the Gemini/GMOS and SDSS spectra are not available.

For these spectra, the continua are modeled with either stellar population synthesis (SPS) models \citep{Delgado2005} adopting pPXF \citep{ppxf} or 4th-order polynomials and/or power-law function with customized Python software utilizing LMFIT \citep{lmfit}, on a case-by-case basis. The properties of the \oiii\ and \ha\ emission lines are then measured from these continuum-subtracted optical spectra.

\subsubsection{AGN Fractions of Starburst-dominated ULIRGs} \label{234}

Most of the key AGN and host galaxy properties of the QUEST quasars from Paper I and the starburst-dominated ULIRGs from M15 are tabulated in these papers. One exception is the AGN fractions of the ULIRGs in M15, which are estimated based on the IRAS-based 25 \mum\ to 60 \mum\ flux ratios (F25/F60) listed in Table 1 of M15. Specifically, the AGN fraction is calculated adopting the best-fit to the trend presented in Fig. 36(c) in \citet{Veilleux2009}:

\begin{eqnarray}
f_{AGN} (\%) = 
\begin{cases} 
67.1x+100.2,\ if\ x \geq -1.1   \\
\leq 27.3,\ if\ x < -1.1 
\end{cases} 
\end{eqnarray}
where $x = {\rm log}(F25/F60)$.

\section{Results from the HST Data Analysis} \label{3}

In this section, we present the major results from our analysis of the \textit{HST/COS} spectra. First, the properties regarding \lya\ emission are examined in Sec. \ref{31}; Second, the measurement of FUV continuum luminosity is briefly described in Sec. \ref{32}; Finally, the properties of \ovi\ and \nv\ absorbers are discussed in Sec. \ref{33}.

\subsection{\lya\ Emission} \label{31}
\subsubsection{Detection Rates} \label{311}

The \lya\ transition falls within the wavelength range of the observations for 19 of the 21 objects. Among the 19 objects, the \lya\ emission is detected (S/N $>$ 3) in 15 of them, including F07599 where the \lya\ emission is heavily affected by deep, broad and narrow \nv\ absorption features  (and perhaps also broad \lya\ absorption). After \lya, \nv\ and \ovi\ are the most frequently detected emission lines in our sample (notice that our G130M spectra do not cover Si~IV and C~IV). Descriptions about the presence of emission/absorption features other than \lya\ in our objects may be found in Appendix \ref{A3}.

\subsubsection{Line Profiles} \label{312}

\begin{figure*}[!htb]   
\epsscale{1.2}
\plotone{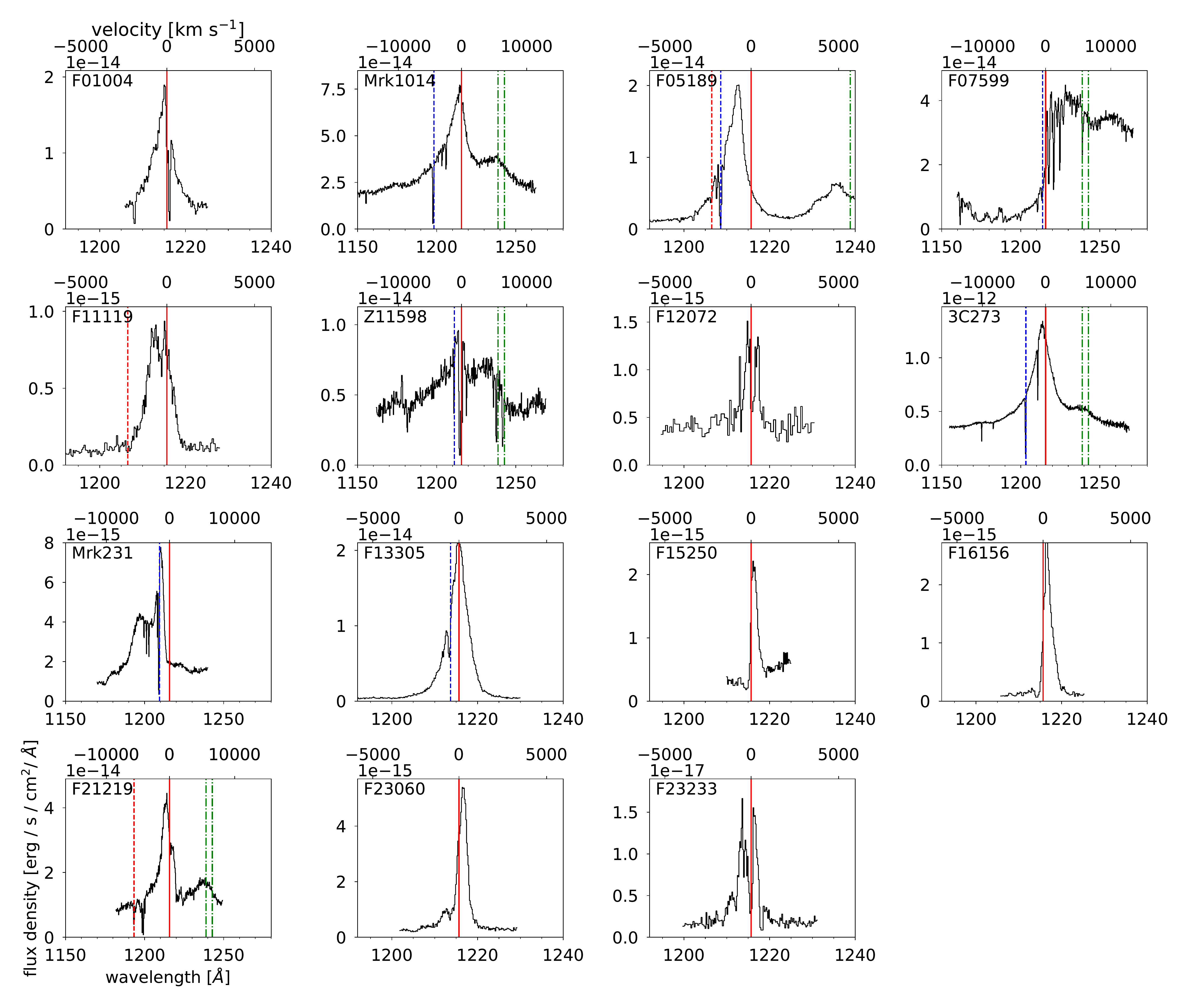}
\caption{\lya\ profiles of the 14 objects with clear \lya\ emission and the one with a potential detection F07599. The spectra are shown in each object's rest-frame. The velocities relative to the \lya\ transition are shown on the top of each panel, and the zero velocities are indicated by the vertical solid lines. The zero-velocities of the \nv\ doublet are indicated by the vertical, green dash-dotted lines when they are blended with the \lya\ emission. Other strong absorption features at the systemic velocities are indicated by vertical dashed lines in red. Foreground absorption features at z$\simeq$0 due to the Milky Way are indicated by dashed lines in blue. The flux scale for the vertical axis in units of erg s$^{-1}$ cm$^{-2}$ \AA$^{-1}$ is listed in the upper left corner above each panel. To help with the visual comparison among different objects, the x-axes are set to two fixed wavelength ranges (1192--1240 \AA\ or 1150--1280 \AA) depending on the widths of the lines.}
\label{fig:lyaloopAGN}
\end{figure*}

The \lya\ profiles of the 15 \lya\ detections are presented in Fig.\ \ref{fig:lyaloopAGN}. For these objects, the \lya\ emission is the most prominent feature in the observed spectral range, except for F07599, where the \lya\ line is severely suppressed by a highly blueshifted broad absorption line (BAL) and several less blueshifted narrow N V 1238, 1243 absorption features. As a result, a relatively robust measurement of the \lya\ profile of F07599 is impossible, and this source is left out from the analyses to characterize the \lya\ profiles in the following sections.

The flux and EWs of the entire \lya\ profiles are measured in wavelength windows customized for each object based on their line widths. The local continuum of each object is determined by fitting the line-free windows adjacent to the \lya\ features with power-law or low-order polynomials (order $\le$ 2). The contamination from nearby N~V emission is subtracted for sources Mrk~1014, F05189, 3C~273, and F21219, where the N~V doublet are modeled as Gaussian profiles. The foreground absorption features and absorption lines from other species at the systemic velocity are interpolated over with cubic splines. For the non-detections, the 3-$\sigma$ upper limits on the flux of \lya\ are estimated in a velocity window of $-$1000 to $+$1000 \kms. 

\begin{figure}[!htb]   
\plotone{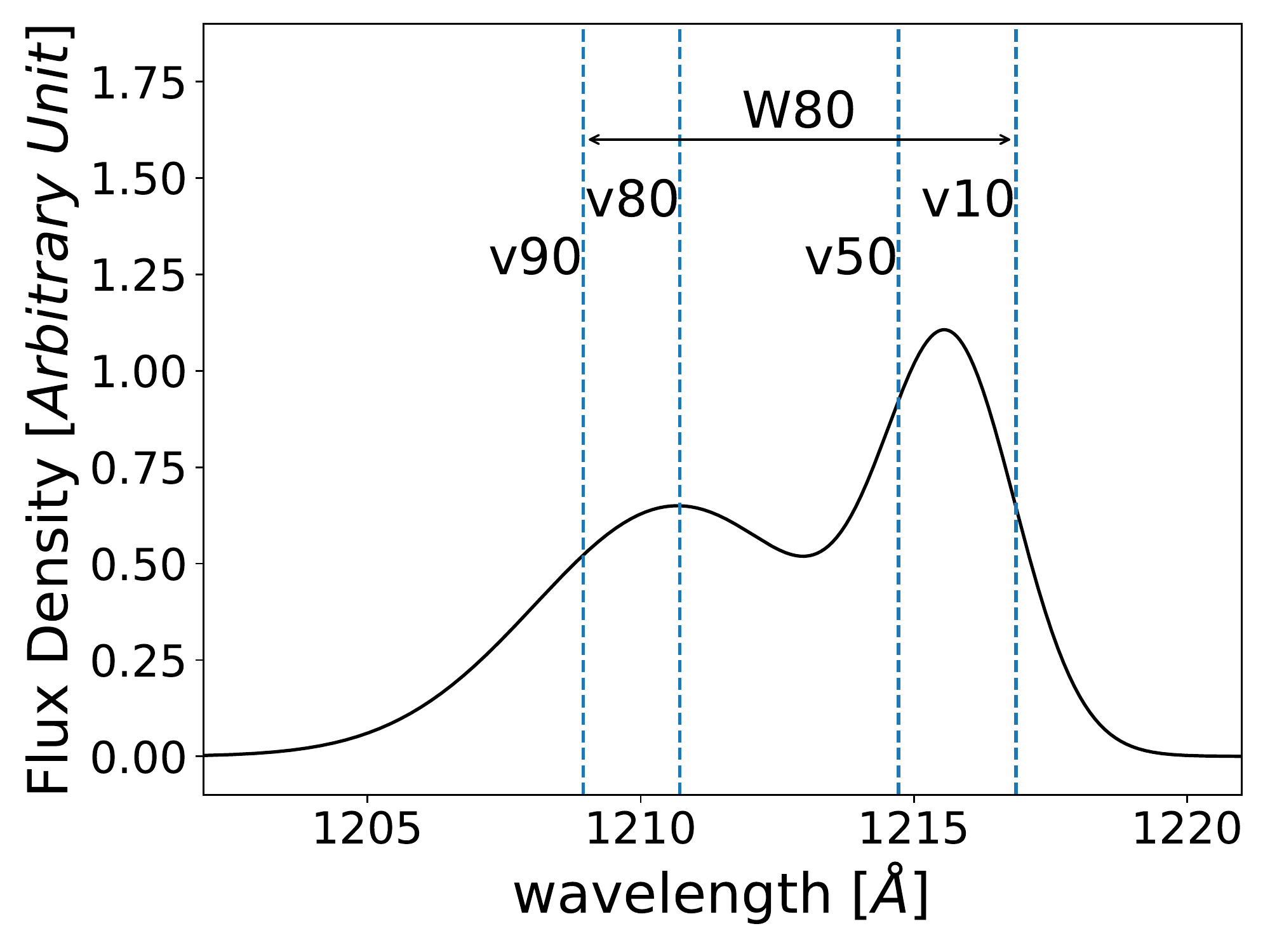}
\caption{Example of a line profile illustrating the various non-parametric kinematic parameters used in this paper. The vertical dashed lines mark the locations of \vjiu, \vba, \vwu, and \vyi\ for the mock emission-line profile shown in the figure. \wba\ is defined as the line width between \vjiu\ and \vyi, and the line asymmetry \ajiuyi\ is [\vjiu$+$\vyi]/\wba\ (not shown in the figure).}
\label{fig:nonpar}
\end{figure}

We adopt a non-parametric approach to characterize the \lya\ profiles quantitatively: the velocities \vwu\ and \vba\ (velocities at 50 and 80 percentiles of the total flux calculated from the red side of the line), the line width \wba\ (which encloses the central 80 percent of the total flux), and the line asymmetry \ajiuyi\ ([\vjiu$+$\vyi]/\wba), where \vjiu\ and \vyi\ are the velocities at 90 and 10 percentiles of the total flux calculated from the red side of the line) are the 
primary measurements adopted for our analyses in the following sections. All velocities are calculated with respect to the systemic redshifts listed in Table \ref{tab:targets}. Table \ref{tab:lya} summarizes the results. A visualization of these non-parametric measurements is shown in Fig. \ref{fig:nonpar}.

Blueshifted wings of the \lya\ emission are often seen across our sample (12 out of the 14 objects with robustly measured \lya\ profiles show \vba\ $<$ 0 and line asymmetry \ajiuyi\ $<$ 0), while the P-Cygni-like profile (blueshifted absorption accompanied by redshifted emission) is only seen in F15250 (in F16156, a weak blueshifted wing is seen in the \lya\ emission blueward of the strong blueshifted absorption feature, and the overall \lya\ profile is thus not P-Cygni-like). This is in clear contrast with the prevalence of P-Cygni-like profiles seen in low-redshift star-forming galaxies \citep[e.g.,][]{Wofford2013}. The blueshift of the velocity centroid of the \lya\ line is also prevalent in our sample, where 11 out of the 14 objects with \lya\ detections show \vwu\ $<$ 0. Moreover, in two Type 1 sources, 3C~273 and Mrk~1014, the broad \lya\ emission (accompanied by broad \nv\ emission, when the spectrum covers this region) shows line width typical for the broad emission line region (BELR) of AGN, which is often the case for low-redshift Type 1 AGN and quasars \citep[e.g.,][]{Shull2012}. Finally, \lya\ absorption features near systemic velocities are clearly seen in 9 objects.

\begin{deluxetable*}{ccccc cccc}
\tablecolumns{10}
\tabletypesize{\scriptsize}
\tablecaption{\lya\ Measurements \label{tab:lya}}
\tablehead{\colhead{Name} & \colhead{log(Flux)} & \colhead{log(Lum.)} & \colhead{EW} & \colhead{\vwu} &  \colhead{\vba} & \colhead{\wba}  & \colhead{\ajiuyi} & \colhead{log[\fesc]} \\
& [\flux] & [\lum] & [\AA] & [\kms] &  [\kms] & [\kms] &  }
\colnumbers
\startdata
F01004         & -13.09$\pm{0.02}$ & 42.48$\pm{0.02}$ & 27.8$\pm{4.7}$ & -228$\pm{42}$  & -788$\pm{78}$  & 1792$\pm{51}$  & -0.25$\pm{0.14}$ & 0.12$\pm{0.49}$  \\
Mrk~1014        & -11.74$\pm{0.006}$ & 44.13$\pm{0.006}$ & 102.7$\pm{0.3}$ & -1274$\pm{153}$ & -5453$\pm{654}$ & 10876$\pm{1305}$ & -0.62$\pm{0.12}$ & -0.22$\pm{0.32}$  \\
F04103         & $<$-14.07      & $<$41.49       & ...            & ...            & ...            & ...            & ...            & $<$-2.36  \\ 
F05189         & -12.79$\pm{0.02}$ & 41.86$\pm{0.02}$ & 164.0$\pm{56.8}$ & -913$\pm{71}$  & -1782$\pm{159}$ & 2748$\pm{50}$  & -0.87$\pm{0.22}$ & -1.29$\pm{0.30}$  \\
F07599         & ...            & ...            & ...            & ...            & ...            & ...            & ...            & ...\\ 
F08572         & $<$-12.86      & $<$42.05       & ...            & ...            & ...            & ...            & ...            & $<$-0.28  \\ 
F11119         & -14.17$\pm{0.05}$ & 41.85$\pm{0.05}$ & 80.8$\pm{39.8}$ & -467$\pm{211}$ & -1104$\pm{245}$ & 2170$\pm{548}$ & -0.53$\pm{0.97}$ & -3.89$\pm{0.32}$  \\
Z11598         & -12.65$\pm{0.003}$ & 43.15$\pm{0.003}$ & 70.5$\pm{2.3}$ & -1739$\pm{591}$ & -6517$\pm{2216}$ & 12766$\pm{4340}$ & -0.51$\pm{0.31}$ & 0.34$\pm{0.44}$  \\
F12072         & -14.23$\pm{0.07}$ & 41.41$\pm{0.07}$ & 17.2$\pm{10.2}$ & -75$\pm{100}$  & -743$\pm{150}$ & 1601$\pm{122}$ & -0.25$\pm{0.65}$ & -0.62$\pm{0.07}$  \\
3C~273          & -10.67$\pm{0.04}$ & 45.18$\pm{0.04}$ & 53.0$\pm{8.3}$ & -624$\pm{340}$ & -2988$\pm{792}$ & 7875$\pm{10}$  & -0.32$\pm{0.53}$ & -0.08$\pm{0.36}$  \\
Mrk~231         & -13.08$\pm{0.03}$ & 41.54$\pm{0.03}$ & 53.2$\pm{10.2}$ & -2771$\pm{371}$ & -4731$\pm{309}$ & 4698$\pm{20}$  & -1.34$\pm{0.31}$ & -2.64$\pm{0.80}$  \\
F13218         & ...            & ...            & ...            & ...            & ...            & ...            & ...            & ...\\ 
F13305         & -12.72$\pm{0.02}$ & 43.06$\pm{0.02}$ & 307.1$\pm{257.9}$ & -20$\pm{58}$   & -599$\pm{97}$  & 1775$\pm{18}$  & -0.15$\pm{0.15}$ & -0.19$\pm{0.48}$  \\
Mrk~273         & $<$-12.85      & $<$41.68       & ...            & ...            & ...            & ...            & ...            & $<$-0.91  \\ 
F14070         & ...            & ...            & ...            & ...            & ...            & ...            & ...            & ...\\ 
F15001         & $<$-14.09      & $<$41.78       & ...            & ...            & ...            & ...            & ...            & $<$-0.80  \\ 
F15250         & -15.32$\pm{0.11}$ & 39.55$\pm{0.11}$ & 0.7$\pm{18.0}$ & 394$\pm{7}$    & 359$\pm{7}$    & 112$\pm{78}$   & 7.28$\pm{5.09}$ & -1.41$\pm{0.37}$  \\
F16156         & -13.94$\pm{0.06}$ & 41.73$\pm{0.06}$ & 62.8$\pm{66.8}$ & 331$\pm{78}$   & 142$\pm{85}$   & 749$\pm{62}$   & 1.12$\pm{2.33}$ & -1.33$\pm{0.34}$  \\
F21219         & -12.21$\pm{0.02}$ & 43.31$\pm{0.02}$ & 46.3$\pm{4.0}$ & -384$\pm{86}$  & -1488$\pm{180}$ & 4036$\pm{48}$  & -0.20$\pm{0.12}$ & -0.81$\pm{0.44}$  \\
F23060         & -13.57$\pm{0.05}$ & 42.36$\pm{0.05}$ & 90.4$\pm{80.2}$ & 219$\pm{69}$   & -228$\pm{201}$ & 1535$\pm{89}$  & -0.10$\pm{1.06}$ & -2.47$\pm{0.36}$  \\
F23233         & -15.88$\pm{0.06}$ & 39.65$\pm{0.06}$ & 36.8$\pm{18.2}$ & -278$\pm{188}$ & -709$\pm{166}$ & 1600$\pm{582}$ & -0.45$\pm{1.20}$ & -2.10$\pm{1.10}$
\enddata
\label{lya}
\tablecomments{Columnn (2): Observed \lya\ flux (cgs units) in logarithm. Column (3): \lya\ luminosity (cgs units) in logarithm. Column (4): \lya\ EW in units of \AA. Column (5): \vwu\ of \lya\ profile in units of \kms. Column (6): \vba\ of \lya\ profile in units of \kms. Column (7): \wba\ of \lya\ profile in units of \kms. Column (8): \ajiuyi\ of \lya. Column (9): Logarithm of \lya\ escape fraction as defined in Sec. \ref{314}. Note that for F13218 and F14070, the \lya\ transition is not covered by the observations, while for F07599, the \lya\ feature cannot be measured robustly mainly due to contamination from the N V BAL.} 
\end{deluxetable*}

\subsubsection{Comparison of the \lya\ and Optical Emission Line Profiles} \label{313}

\begin{figure*}[!htb]   
\epsscale{1.1}
\plotone{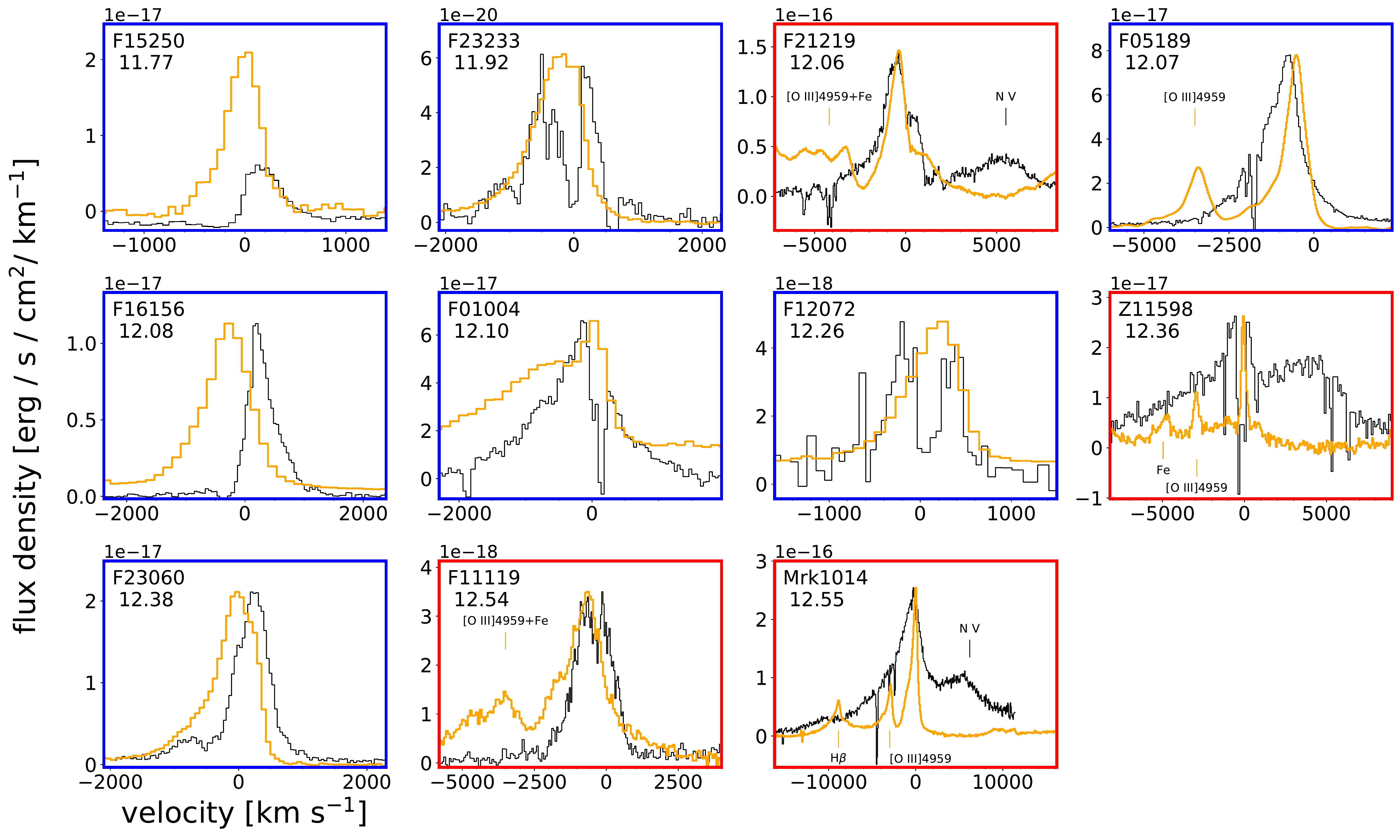}
\caption{Comparison of \lya\ profiles (black) with \oiii\ emission lines (orange) for the 11 objects in our sample with both \lya\ detections and \oiii\ observations. The flux scale on the vertical axis refers to the \lya\ line and the scale factor in units of erg s$^{-1}$ cm$^{-2}$ km$^{-1}$ s is listed in the upper left corner above each panel. The [O III] emission line profiles are re-scaled for better visualization. Nearby emission features from N\ V in the FUV and those from \hb, \oiiia, and Fe in the optical are marked with black and orange vertical bars when those features are present, respectively. The panels are ordered in increasing AGN luminosities, which are indicated in the top left corner of each panel under the object name (in log units of solar luminosities). The panel frames of Type 1 AGN are marked in red and those of type 2 AGN are marked in blue.}
\label{fig:O3loopAGN}
\end{figure*}

\begin{figure*}[!htb]   
\centering
\epsscale{1.3}
\plotone{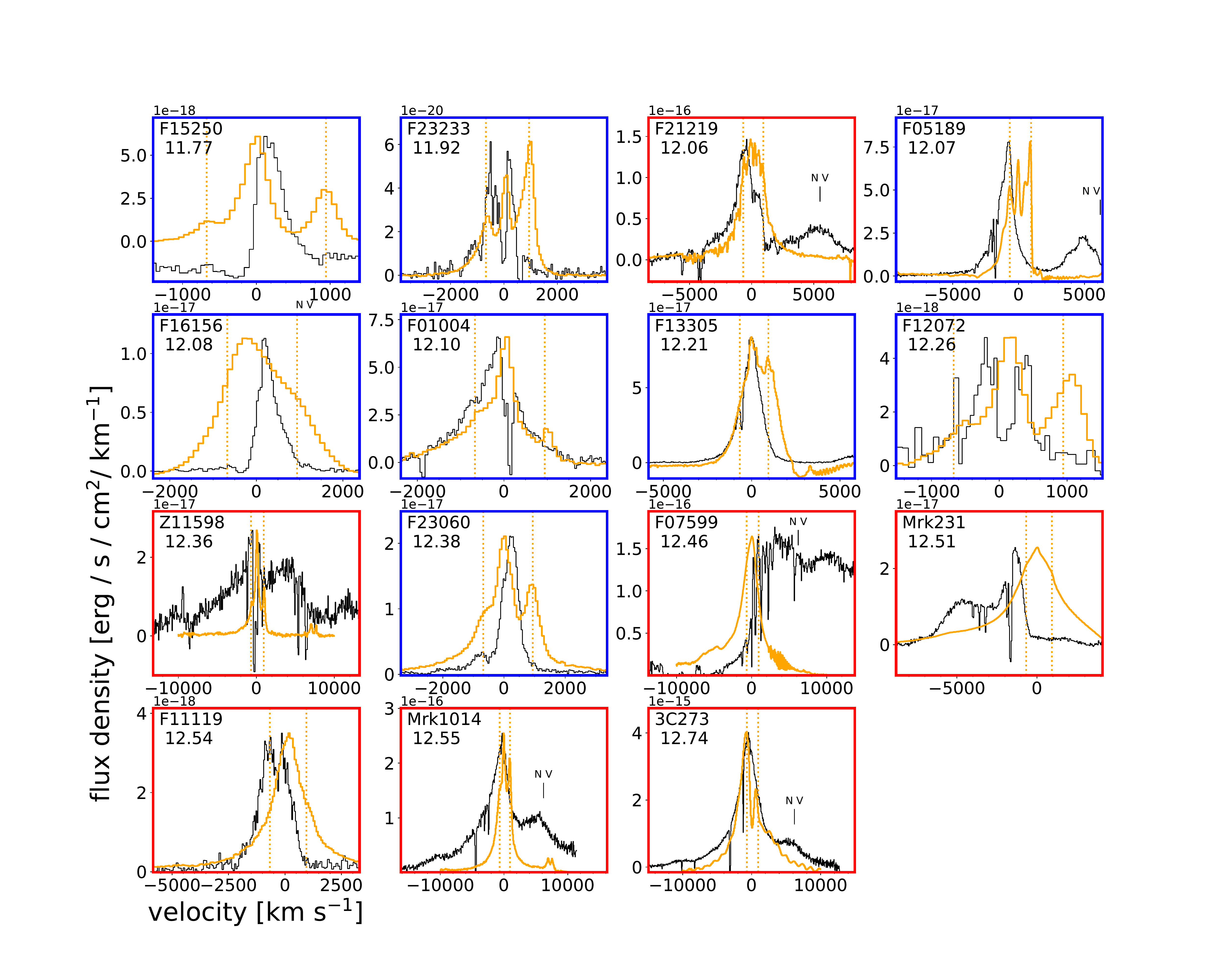}
\caption{Same as Fig. \ref{fig:O3loopAGN} but for the comparison of the \lya\ (black) and \ha\ (orange) emission lines for the 15 objects with \lya\ detections. The nearby N~V emission are marked with black bars when those features are present. Note that, in some cases,  the strong \niiab\ emission lines in the optical near $-$674 \kms\ and $+$944 \kms\ (marked by orange dotted lines) make this comparison difficult. }
\label{fig:HaloopAGN}
\end{figure*}

The observed \lya\ profiles in these dusty ULIRGs are likely affected by complex radiative transfer effects due to the resonant nature of the transition. While it is impossible to derive the intrinsic profile of the resonant \lya\ line with our data, the non-resonant optical emission lines (e.g. \oiii\ forbidden line, \ha\ recombination line) provide a point of reference since they are less affected by absorption and scattering. Qualitatively, we may therefore infer the extent to which \lya\ photons are absorbed and/or scattered, by comparing the \lya\ and optical line profiles.

In Fig.\ \ref{fig:O3loopAGN} and \ref{fig:HaloopAGN}, we plot the \lya\ profiles in comparison with the \oiii\ profiles and the \ha\ profiles. All of these line profiles are continuum-subtracted and rescaled for better visualization. In total, 11 objects with \oiii\ observations and 15 objects with \ha\ observations show \lya\ detections. The measurements of the \oiii\ and \ha\ profiles are summarized in Table \ref{tab:optical}. At first glance, one similarity between the \lya\ profiles and the non-resonant line profiles is the occurrence of blueshifted emission line wings in many objects.  Blueshifted \oiii\ (and, to a lesser extent, \ha) emission is a tell-tale signature of ionized gas outflows, so the blueshifted \lya\ emission may also arise from the outflowing gas. In addition, \lya\ emission is generally broader than or comparable in width to the \oiii\ and \ha\ emission. Otherwise, the \lya\ profiles are diverse among the objects in our sample and no apparent trend is seen between the profiles of \lya\ and the non-resonant optical lines. We will discuss these results further in Sec. \ref{42}.

\subsubsection{\lya\ Escape Fraction} \label{314}

At first, it may be surprising that significant \lya\ emission is observed from these dusty ULIRGs given the implied huge optical depth to \lya\ photons. In this section, we quantify the escape of \lya\ photons by calculating the \lya\ escape fraction.

Under Case~B recombination, the ionized region is optically thick to the Lyman series, and the intrinsic F(\lya)/F(\ha) flux ratio depends only on the electron density and temperature. Adopting the low density limit where $n_e << n_{e,crit} = 1.55 \times 10^4$\cm3, the collisions can be safely neglected. The 
intrinsic \lya\ flux is then predicted to be 8.1 times the intrinsic \ha\ flux \citep{Hummer1987,Draine2011}. 

Galaxies usually show \lya-to-\ha\ flux ratios much less 
than the predicted Case~B ratio. We can therefore describe the suppression of \lya\ photons 
by comparing the observed \lya\ flux to the intrinsic values indicated by the \ha\ emission.
The escape fraction of \lya\ photons can thus be defined as
\begin{eqnarray}
\fesc\ = \frac{F(\lya)}{8.1 \times F(\ha)_{cor}}.
\label{eqn:fesclya} 
\end{eqnarray}

The intrinsic \ha\ flux is calculated from the observed \ha\ flux, corrected for the nebular reddening using the \citet{Cardelli1989} reddening curve, namely
\begin{eqnarray}
F_{cor}(\ha) = F_{obs}(\ha) \times 10^{1.012 E(B-V)}.
\end{eqnarray}
and  the \ebv\ is calculated from the Balmer decrement:
\begin{eqnarray}
E(B-V) = \frac{1}{0.9692} \log \left [ \frac{(F(\ha)/F(\hb))_{\rm obs}}{(F(\ha)/F(\hb))_{\rm int}} \right ]
\end{eqnarray}
We set the intrinsic \hahb\ ratio to 3.1 for all but one objects as they show optical spectral features consistent with AGN activity. The only exception is F01004: it is located in the star-forming region in the BPT and VO87 diagnostic line ratio diagrams \citep{bpt,Veilleux1987,Osterbrock2006}, and we set the intrinsic \hahb\ ratio of this object to 2.87.

\begin{deluxetable*}{ccccc ccccccc}
\tablecolumns{12}
\tabletypesize{\scriptsize}
\tablecaption{Optical Spectroscopic Measurements \label{tab:optical}}
\tablehead{\colhead{Name} & \colhead{\ebv} & \colhead{log[Flux(\ha)]} & \colhead{cor,\ha} &  \colhead{\vwuha} &  \colhead{\vbaha} & \colhead{\wbaha} & \colhead{\ajiuyiha} & \colhead{\vwuo} & \colhead{\vbao} & \colhead{\wbao} & \colhead{\ajiuyio} \\
& & [\flux] & & [\kms] &  [\kms] & [\kms] & & [\kms]  & [\kms] & [\kms] & }
\colnumbers
\startdata
F01004         & 0.05$\pm{0.004}$ & -14.05$\pm{0.03}$ & 0.86           & 111$\pm{100}$  & -688$\pm{100}$ & 1398$\pm{141}$ & -0.54$\pm{0.11}$ & -563$\pm{199}$ & -1560$\pm{200}$ & 2591$\pm{282}$ & -0.58$\pm{0.13}$ \\
Mrk~1014        & 0.24$\pm{0.02}$ & -12.42$\pm{0.04}$ & 0.98           & 113$\pm{69}$   & -1061$\pm{69}$ & 4901$\pm{98}$  & 0.19$\pm{0.02}$ & -147$\pm{69}$  & -768$\pm{69}$  & 1519$\pm{98}$  & -0.42$\pm{0.07}$ \\
F04103         & 1.29$\pm{0.13}$ & -12.54$\pm{0.03}$ & 0.82           & ...            & ...            & ...            & ...            & ...            & ...            & ...            & ... \\ 
F05189         & 1.02$\pm{0.10}$ & -12.34$\pm{0.04}$ & 0.86           & -354$\pm{20}$  & -904$\pm{20}$  & 1588$\pm{29}$  & -0.90$\pm{0.02}$ & -661$\pm{27}$  & -1195$\pm{27}$ & 1415$\pm{38}$  & -1.38$\pm{0.05}$ \\
F07599         & ...            & -10.03$\pm{0.03}$ & 0.68           & 642$\pm{3760}$ & 323$\pm{4079}$ & 957$\pm{5393}$ & 1.48$\pm{10.06}$ & ...            & ...            & ...            & ... \\ 
F08572         & ...            & -13.36$\pm{0.03}$ & 0.75           & ...            & ...            & ...            & ...            & ...            & ...            & ...            & ... \\ 
F11119         & ...            & -11.19$\pm{0.03}$ & 1.00           & 144$\pm{69}$   & -1029$\pm{69}$ & 4970$\pm{98}$  & 0.00$\pm{0.02}$ & -667$\pm{69}$  & -1427$\pm{69}$ & 2278$\pm{98}$  & -0.62$\pm{0.05}$ \\
Z11598         & ...            & -13.69$\pm{0.03}$ & 0.62           & 172$\pm{69}$   & -1071$\pm{69}$ & 4901$\pm{98}$  & -0.38$\pm{0.02}$ & -64$\pm{60}$   & -1734$\pm{60}$ & 4057$\pm{84}$  & -0.31$\pm{0.02}$ \\
F12072         & 0.10$\pm{0.001}$ & -14.80$\pm{0.001}$ & 1.33           & 49$\pm{87}$    & -213$\pm{88}$  & 1051$\pm{124}$ & -0.03$\pm{0.12}$ & -121$\pm{37}$  & -831$\pm{37}$  & 1642$\pm{53}$  & -0.38$\pm{0.03}$ \\
3C~273          & ...            & -11.50$\pm{0.05}$ & 1.00           & 637$\pm{89}$   & -2031$\pm{89}$ & 7277$\pm{125}$ & 0.00$\pm{0.02}$ & ...            & ...            & ...            & ... \\ 
Mrk~231         & ...            & -11.14$\pm{0.04}$ & 0.62           & -321$\pm{20}$  & -2421$\pm{20}$ & 6396$\pm{29}$  & -0.29$\pm{0.00}$ & ...            & ...            & ...            & ... \\ 
F13218         & 0.73$\pm{0.07}$ & -10.09$\pm{0.04}$ & 0.80           & -400$\pm{18}$  & -1730$\pm{18}$ & 2924$\pm{25}$  & -0.76$\pm{0.01}$ & -1064$\pm{69}$ & -1755$\pm{69}$ & 2140$\pm{98}$  & -0.98$\pm{0.06}$ \\
F13305         & 1.18$\pm{0.12}$ & -13.45$\pm{0.04}$ & 1.01           & 17$\pm{97}$    & -371$\pm{97}$  & 1066$\pm{137}$ & 0.15$\pm{0.13}$ & 55$\pm{127}$   & -453$\pm{127}$ & 1398$\pm{180}$ & 0.14$\pm{0.13}$ \\
Mrk~273         & 1.94$\pm{0.19}$ & -12.70$\pm{0.03}$ & 0.72           & ...            & ...            & ...            & ...            & ...            & ...            & ...            & ... \\ 
F14070         & 2.11$\pm{0.21}$ & -12.82$\pm{0.04}$ & 1.22           & ...            & ...            & ...            & ...            & ...            & ...            & ...            & ... \\ 
F15001         & 1.04$\pm{0.10}$ & -14.00$\pm{0.05}$ & 0.62           & -54$\pm{69}$   & -330$\pm{69}$  & 897$\pm{98}$   & -0.37$\pm{0.12}$ & -107$\pm{69}$  & -590$\pm{69}$  & 1036$\pm{98}$  & -0.69$\pm{0.11}$ \\
F15250         & 0.62$\pm{0.06}$ & -14.60$\pm{0.04}$ & 0.60           & 51$\pm{69}$    & -87$\pm{69}$   & 552$\pm{98}$   & 0.35$\pm{0.19}$ & 67$\pm{69}$    & -71$\pm{69}$   & 483$\pm{98}$   & 0.32$\pm{0.21}$ \\
F16156         & 0.71$\pm{0.07}$ & -13.39$\pm{0.03}$ & 0.75           & 123$\pm{112}$  & -439$\pm{113}$ & 1909$\pm{159}$ & 0.19$\pm{0.08}$ & -54$\pm{147}$  & -644$\pm{148}$ & 1474$\pm{208}$ & -0.27$\pm{0.15}$ \\
F21219         & ...            & -12.17$\pm{0.04}$ & 0.73           & 11$\pm{4285}$  & -788$\pm{5083}$ & 2627$\pm{6278}$ & 0.20$\pm{2.44}$ & -490$\pm{131}$ & -1147$\pm{132}$ & 3015$\pm{185}$ & 0.20$\pm{0.06}$ \\
F23060         & 1.15$\pm{0.11}$ & -12.03$\pm{0.03}$ & 1.04           & 133$\pm{69}$   & -419$\pm{69}$  & 2071$\pm{98}$  & 0.13$\pm{0.05}$ & 12$\pm{69}$    & -403$\pm{69}$  & 1104$\pm{98}$  & -0.35$\pm{0.09}$ \\
F23233         & 0.59$\pm{0.06}$ & -14.70$\pm{0.05}$ & 1.02           & 70$\pm{69}$    & -344$\pm{69}$  & 828$\pm{98}$   & -0.32$\pm{0.12}$ & -189$\pm{69}$  & -673$\pm{69}$  & 1381$\pm{98}$  & -0.57$\pm{0.08}$ 
\enddata
\tablecomments{Columnn (2): Color excess based on the Balmer decrement measured from the optical spectra described in Sec. \ref{233}. Columnn (3): Extinction-corrected \ha\ flux in logarithm and cgs units. Columnn (4): Appeture correction factor applied to the \ha\ flux in the calculation of \fesc. Column (5)--(8): Kinematic properties \vwu, \vba, \wba, and \ajiuyi\ of \ha\ emission. Column (9)--(12): Kinematic properties \vwu, \vba, \wba, and \ajiuyi\ of \oiii\ emission.}
\end{deluxetable*}

The \ha\ fluxes are measured from spectra gathered from literature, as described in Sec. \ref{233}, and can be divided into three groups: (1) 
Gemini/GMOS IFU observations; (2) SDSS spectra; (3) long-slit spectra from \citet{Veilleux1999}. Some of our objects are point sources or show very compact morphology in the narrow-band \ha\ images (based on the GMOS data) or R-band images \citep[from SDSS or][]{Veilleux1999}, so no aperture corrections are needed for their \ha\ flux in the calculation of \lya\ escape fraction. However, for the more extended objects, the aperture difference between the COS FUV spectroscopy and the optical observations need to be taken into account, as the throughput of the 2.5\arcsec\ COS aperture drops sharply beyond the central 0.4\arcsec.  

In practice, the aperture correction is negligible if at least one of the following three criteria is met: (1) it is a Type 1 AGN; (2) the PSF contribution to its overall flux in the R-band image is more than 50\% based on the measurements in \citet{Veilleux2002}; (3) the R-band effective radius is less than 1\arcsec\ based on the measurements in \citet{Veilleux2002}. For the other objects, aperture corrections are needed to account for the rapid decrease of COS throughput at large radius as mentioned above. Specifically, we calculate the aperture correction factors for each group of optical observations separately: (i) for the GMOS IFU observations, we generate the \ha\ flux maps based on the data cube, and use the COS throughput function to vignette the \ha\ flux maps within the region with radius r $<$ 1.25\arcsec. The aperture correction factor is then the vignetted \ha\ flux within r $<$ 1.25\arcsec (corresponding to the COS aperture) divided by the original \ha\ flux within the same aperture; (ii) for the SDSS spectra, we adopt the r-band images as surrogates for the \ha\ flux maps. We then vignette the r-band images within the same COS throughput function, and the aperture correction factor is the vignetted r-band flux within r $<$ 1.25\arcsec\ divided by the original r-band flux within SDSS aperture (D$=$3\arcsec\ or D$=$2\arcsec); (iii) for the long-slit spectra from \citet{Veilleux1999}, we follow the same logic as adopted for the SDSS spectra but use the R-band images in \citet{Kim2002}. The aperture correction factor is thus the vignetted R-band flux within r $<$ 1.25\arcsec\ divided by the original R-band flux within the extraction region of the long-slit spectra (2\arcsec$\times$5 kpc). 

The aperture correction factors adopted in the calculations and resulting \lya\ escape fractions are listed in Tables \ref{tab:optical} and \ref{tab:lya}, respectively.

\subsection{Continuum Luminosity at 1125 \AA} \label{32}

As a surrogate for the FUV continuum luminosity adopted in Paper I, the monochromatic luminosities at rest-frame 1125 \AA, \logluv, are measured whenever the continuum is detected, with a bandpass of 20 \AA. These results are recorded in Table \ref{tab:targets}.

\subsection{FUV Absorption Features} \label{33}

The focus of this section is the strongest metal absorption features detected in our objects, \ovi\ and \nv, tracers of the highly ionized gas in these systems. Only 12 out of the 21 objects have continuum S/N in the vicinity of \ovi\ and/or \nv\ that are high enough (S/N $\gtrsim$ 10 in a 500 \kms\ window) to allow for the detection of corresponding absorption lines. Out of these 12 objects, 6 objects show O~VI and/or N~V absorption features associated with the galaxy (velocity centroid $<$ 13000 \kms\ and not from intervening systems), and the velocity centroids of these absorption features are all blueshifted. One more object, F15250, may display a N~V absorption feature, but the doublet is so close to a group of geo-coronal emission lines that the N~V 1239 transition is heavily contaminated and no robust measurements of the N~V feature can be made. Our estimates for the EW and centroid velocity of the N~V 1242 absorber alone are $\sim$0.3 \AA\ and $-$500 \kms, respectively, which has not taken into account the infilling from the \nv\ emission. This source is excluded from the discussions related to the absorption features in the following sections.

The properties of these detected O~VI and/or N~V absorption features vary wildly: F07599 shows a $>$25000 \kms\ wide \nv\ BAL accompanied by narrower absorbers at smaller velocities, whereas F23060 shows relatively narrow and shallow \nv\ absorption features on top of the \nv\ emission. Overall, the absorption features in F07599 and F01004 fall in the BAL category (velocity width $>$ 2000 \kms), while all other absorption features are classified as narrow absorption lines (NAL; velocity width $<$ 500 \kms). The object-by-object description of these absorption features is given in Appendix \ref{A3}.

To quantify the strength of these absorbers, we follow the same procedure as in Paper I. First, we fit the continuum and/or broad emission lines (\lya, O~VI, N~V) with low-order polynomials or Gaussian profiles. After the spectra are normalized by the best-fits from the continuum and/or broad emission line fits, these absorbers are quantified using a non-parametric approach, where we measure the total velocity-integrated EWs of the outflowing absorbers in the object’s rest frame,
\begin{eqnarray}
W_{eq} = \int [1-f(\lambda)] d\lambda, 
\label{eqn:weq} 
\end{eqnarray}
the weighted average outflow velocity
\begin{eqnarray}
v_{wtavg} = \frac{\int v[1-f(v)]dv}{W_{eq}}, 
\label{eqn:vavg} 
\end{eqnarray}
and the weighted outflow velocity dispersion,
\begin{eqnarray}
\sigma_{wtavg} = \{\frac{\int (v-v_{wtavg})^2[1-f(v)] dv}{W_{eq}}\}^{\frac{1}{2}}.
\label{eqn:savg} 
\end{eqnarray}
The results are summarized in Table \ref{tab:abs}.

\begin{deluxetable*}{cccc ccccc}
\tablecolumns{8}
\tabletypesize{\scriptsize}
\tablecaption{Properties of the O~VI and N~V Absorption Features \label{tab:abs}}
\tablehead{
\colhead{Name} & \colhead{\weq} & \colhead{\vavg} & \colhead{\savg} & log(N$_{ion,d}$)  & \colhead{C$_{f,d}$} & \colhead{\# comp.}  &  log(N$_{ion}$) & \colhead{\cf}\\
& [\AA] & [\kms] & [\kms] & [cm$^{-2}$] &  & & [cm$^{-2}$] &  }
\colnumbers
\startdata
F01004, O~VI & 10.59  $\pm{   0.26  }$ &   -2720  $\pm{    110  }$  &  1470  $\pm{    120  }$ &   ...   & ... & 1      &  $>$17.1                                                    &   ...  \\
F01004, N~V & $<$0.14  &   ...  &    ... & ... & ... &   ...  &    ... & ... \\
Mrk~1014, O~VI  & $<$0.09 &   ...  &   ...  &    ... & ... &   ...  &    ... & ... \\
Mrk~1014, N~V  & $<$0.13  &   ...  &   ...  &    ... & ... &   ...  &    ... & ... \\
F05189, N~V & $<$0.07     &   ...  &   ...  &    ... & ... &   ...  &    ... & ... \\
F07599, N~V   &   55.39  $\pm{   0.07  }$ &  -12690  $\pm{     20  }$ &    4620  $\pm{     10  }$ & ...  & ...   & 1  &  $>$17.2   &   ...     \\
Z11598, O~VI  &  2.03  $\pm{   0.12  }$ &    -190  $\pm{     30  }$ &      90  $\pm{     40  }$ & $>$16.4 & 0.80$\pm{0.05}$  & 2  & $>$16.3, 14.5$\pm{0.3}$,                                  & 0.90$\pm{0.03}$, 1(f)  \\
Z11598, N~V   &     2.07  $\pm{   0.12  }$ &    -170  $\pm{     30  }$ &     100  $\pm{     30  }$ & $>$16.1 & 0.57$\pm{0.22}$ & 1  & $>$15.9                                                   & 0.82$\pm{0.03}$ \\
3C~273, O~VI & $<$0.04 &  ...  &  ...  &  ... & ... &   ...  &    ... & ... \\
3C~273, N~V & $<$0.04 &   ...  &  ...  &  ... & ... &   ...  &    ... & ... \\
 Mrk~231 N~V & $<$0.18 &  ...  &  ...  &  ... & ... &   ...  &    ... & ...  \\
F13218, O~VI  & 2.23  $\pm{   0.18  }$ &    -250  $\pm{     40  }$ &     150  $\pm{     50  }$ & $>$15.6 & 0.66$\pm{0.12}$ & 3   & $>$15.7, $>$15.2, $>$15.9                                                 & 0.79$\pm{0.05}$, 0.79$\pm{0.09}$, 0.55$\pm{0.04}$ \\
 F13305, N~V  & $<$0.07 &   ...  &    ... & ... &   ...  &    ... & ... \\
 F21219, O~VI  &  1.17  $\pm{   0.05  }$ &   -4400  $\pm{    190  }$ &     180  $\pm{     20  }$ & ... & ...   & 4  & 14.1$\pm{0.2}$, 14.5$\pm{0.2}$,   & 1(f), 1(f), 1(f), 1(f) \\
 & & & & & & &  14.2$\pm{0.1}$, 14.2$\pm{0.1}$ & \\
 F21219, N~V   & 1.28  $\pm{   0.05  }$ &   -4430  $\pm{    300  }$ &     190  $\pm{     30  }$ & 14.8$\pm{0.2}$  & 0.78$\pm{0.20}$   & 4 & 14.4$\pm{0.1}$, 13.6$\pm{0.1}$, & 1(f), 1(f), 1(f), 1(f) \\
& & & & & & &  14.2$\pm{0.1}$, 13.5$\pm{0.1}$ & \\
F23060, N~V &    1.34  $\pm{   0.24  }$ &    -610  $\pm{ 150  }$ &     160  $\pm{    130  }$   & ... & ...  & 2     &  14.2$\pm{1.6}$, 14.1$\pm{1.6}$                             &   1(f), 1(f)
\enddata
\tablecomments{Column (1): Object and absorption feature name; Column (2)--(4): Velocity-integrated EW, \weq\ (eq. \ref{eqn:weq}), average depth-weighted velocity \vavg\ (eq. \ref{eqn:vavg}), and average depth-weighted velocity dispersion \savg\ (eq. \ref{eqn:savg}) of the absorption lines; Column (5): Ion column densities obtained from the analysis of the absorption doublet with partial covering model as described in Sec. \ref{331}; Column (6): Velocity-weighted covering fractions calculated from the same analysis for ion column densities in Column (5); Column (7): Number of components in the best-fits from the Voigt profile fits as described in Sec. \ref{332}. Notice that the O~VI absorption in F01004 and the N~V absorption of F07599 are BAL, and the 1-component fit is only tentative/experimental, with the aim to capture the overall absorption profile. Column (8): Ion column densities from the Voigt profile fits. The values for individual components from the best-fit model are separated by comma. This is also True for Column (9); Column (9): Covering fractions from the Voigt profile fits. The flag ``f'' in parenthesis indicates that the covering fraction is fixed to the corresponding value in the fits.}
\end{deluxetable*}

\subsubsection{Evidence of Partial Covering for the O~VI and N~V Doublets} \label{331}

The profiles of the resolved O~VI and N~V doublets may be used to derive the basic characteristics of the absorbing cloud -- background source system. For the absorption features in Z11598 and F13218, there is evidence of partial covering: the optical depth ratio of the spectrally resolved doublet deviates from the theoretical expectation from a simple model where the foreground cloud is illuminated by the background point source with a 100\% covering fraction.

For the cases where the optical depths of the doublets (proportional to $\lambda f_{osc}$, the product of wavelength and oscillator strength) differ by a factor of $\sim$2 (like \ovi\ and \nv), and the continuum intensity is normalized to unity, the coverage fraction (or covering factor, \cf) as a function of velocity may be obtained. Following \citet{Hamann1997a}, in the simple situation where the two transitions of the doublet do not overlap with each other, we have
\begin{eqnarray}
C_f(v) = \frac{I_1(v)^2-2I_1(v)+1}{I_2(v)-2I_1(v)+1};\ \ \ I_1>I_2\geq I_1^2 \\
C_f(v) = 1;\ \ \ I_2 < I_1^2 \\
C_f(v) = 1-I_1(v);\ \ \ I_2 \geq I_1
\label{eqn:cf} 
\end{eqnarray}
$I_1$ and $I_2$ are the normalized intensities of the weaker and stronger absorption lines, respectively, and \cf\ is the covering factor.

Then the optical depth as a function of the velocity can be written as
\begin{eqnarray}
\tau_1(v) = ln(\frac{C_f(v)}{I_1(v)+C_f(v)-1}) \\
\tau_2(v) = 2\tau_1(v)
\label{eqn:tau} 
\end{eqnarray}
The column density of the ion can then be obtained by integrating the optical depth over the velocity adopting
\begin{eqnarray}
N_{ion} = \frac{m_ec}{\pi e^2f\lambda}\int{\tau(v)dv}
\end{eqnarray}
The resulting values of $N_{ion}$ are listed in Table \ref{tab:abs}.

\subsubsection{Voigt Profile Fitting of O~VI and N~V Absorbers} \label{332}

A popular approach to quantify the absorption features is to fit them with the product of individual components assuming Voigt profiles for the optical depth distribution in frequency (or velocity, wavelength) space. As discussed in Sec. \ref{331}, there is evidence for partial covering in a few objects. To account for this, we also include a constant covering fraction parameter to each component of the model. The final model of the normalized intensity can then be written as
\begin{eqnarray}
I(\nu) = \prod\{1 - C_f [1-e^{-\tau(\nu)}]\} \\
\tau(\nu | N, b, z)  = N \sigma_0 f_{\rm osc} \Phi(\nu | b, z)
\label{eqn:absfit} 
\end{eqnarray}

%Collectively, the set of atomic constants $\{e, m_e, f_{\rm osc}, \nu_0, \Gamma\}$ are the charge and mass of the electron, oscillator strength, rest-frame frequency and damping coefficient of an atomic transition. 

\cf, $\tau$, $\nu$, $N$, $b$, $f_{\rm osc}$, and $\sigma_0$ are the covering fraction parameter, optical depth, frequency, ion column density, Doppler parameter, oscillator strength, and cross section, respectively. $\Phi(\nu | b, z)$ is the normalized Voigt profile. The adopted atomic parameters are taken from \citet{Morton2003}. 

In the fitting procedures, the model described above is further convolved with the line-spread function of HST/COS tabulated on the HST/COS website\footnote{\url{https://www.stsci.edu/hst/instrumentation/cos/performance/spectral-resolution}}. We adopt a customized software built on the non-linear least-squares fit implemented in LMFIT to search for the best-fit model.  In our software, a velocity component is added to the model if the Bayesian Information Criterion (BIC) \citep{bic} decreases, and this process stops when the minimum BIC value is found. The model with the lowest BIC value is then chosen as the best-fit to the data, which is also confirmed by a visual inspection. In addition, we have tested this by manually fitting the absorption line profile with n$+$1 components when n components are required by the best fit. The change in total column density is in general $\lesssim$0.1 (in logarithm), within the uncertainty of total column density derived from the best-fit model. The uncertainties of the best-fit parameters are calculated from the 1-$\sigma$ (68.3\%) confidence interval adopting the conf\_interval function of LMFIT, which takes into account the covariances between blended absorption components.

\begin{figure*}[!htb]   
\centering
\begin{minipage}[t]{0.32\textwidth}
\includegraphics[width=\textwidth]{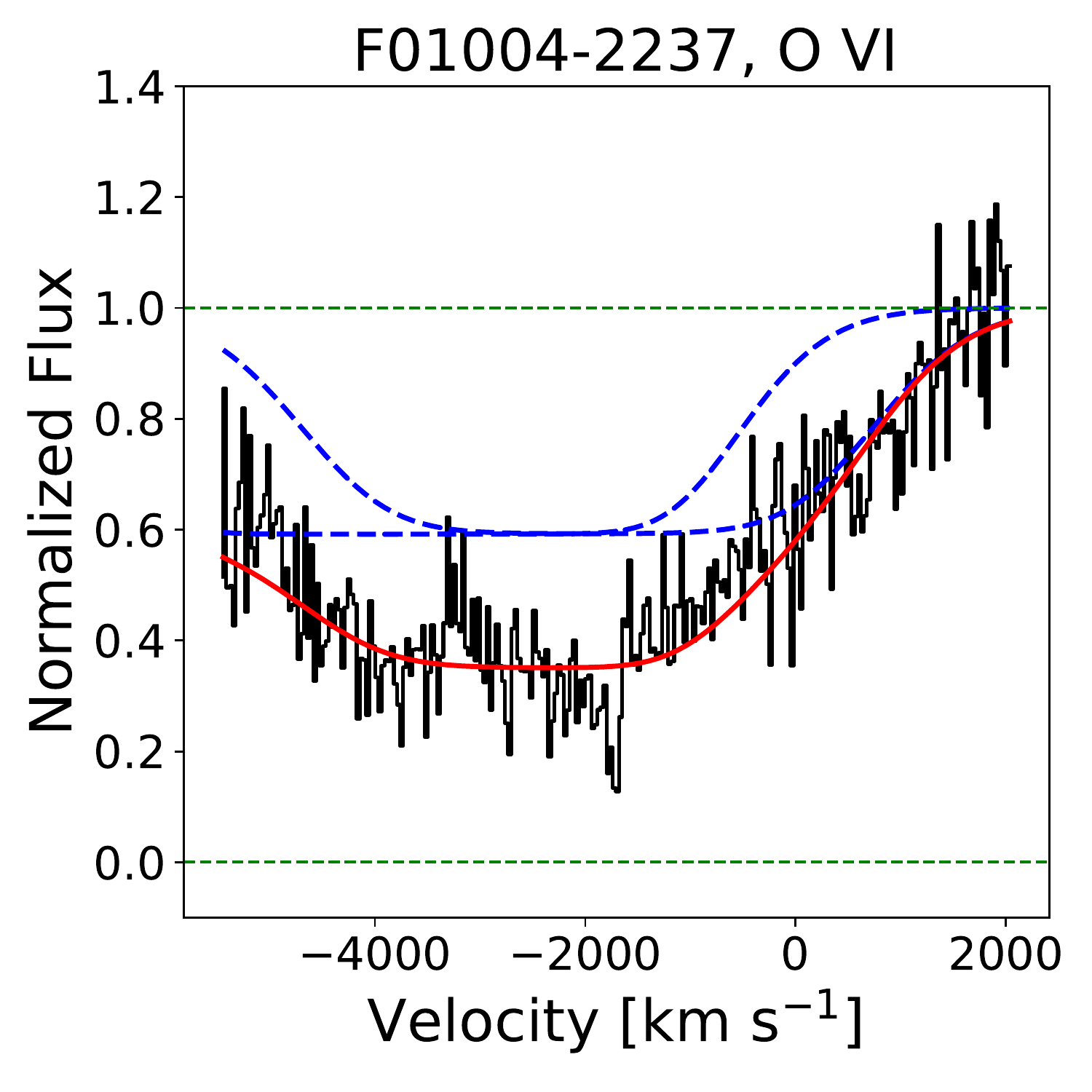}
\end{minipage}
\begin{minipage}[t]{0.32\textwidth}
\includegraphics[width=\textwidth]{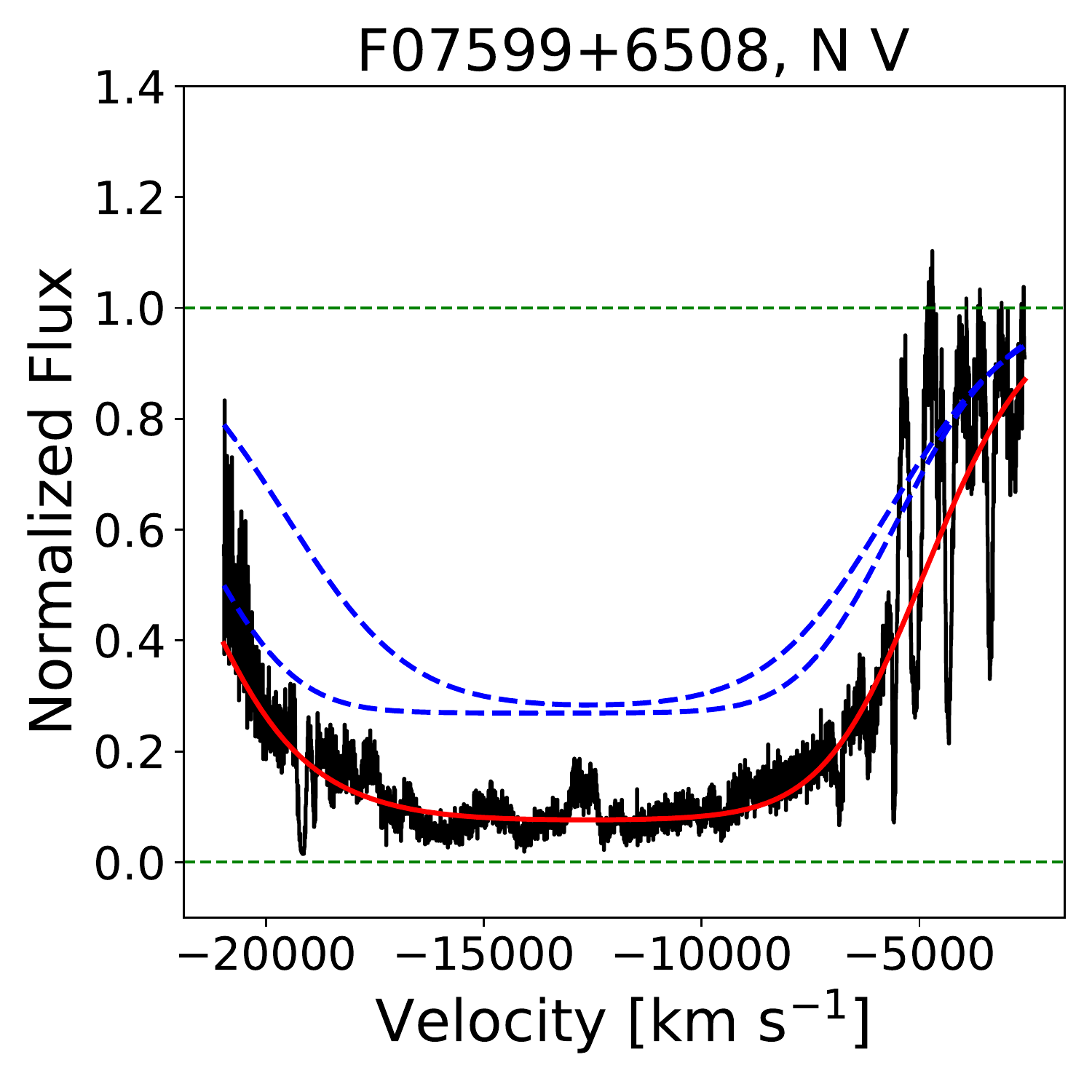}
\end{minipage}
\begin{minipage}[t]{0.32\textwidth}
\includegraphics[width=\textwidth]{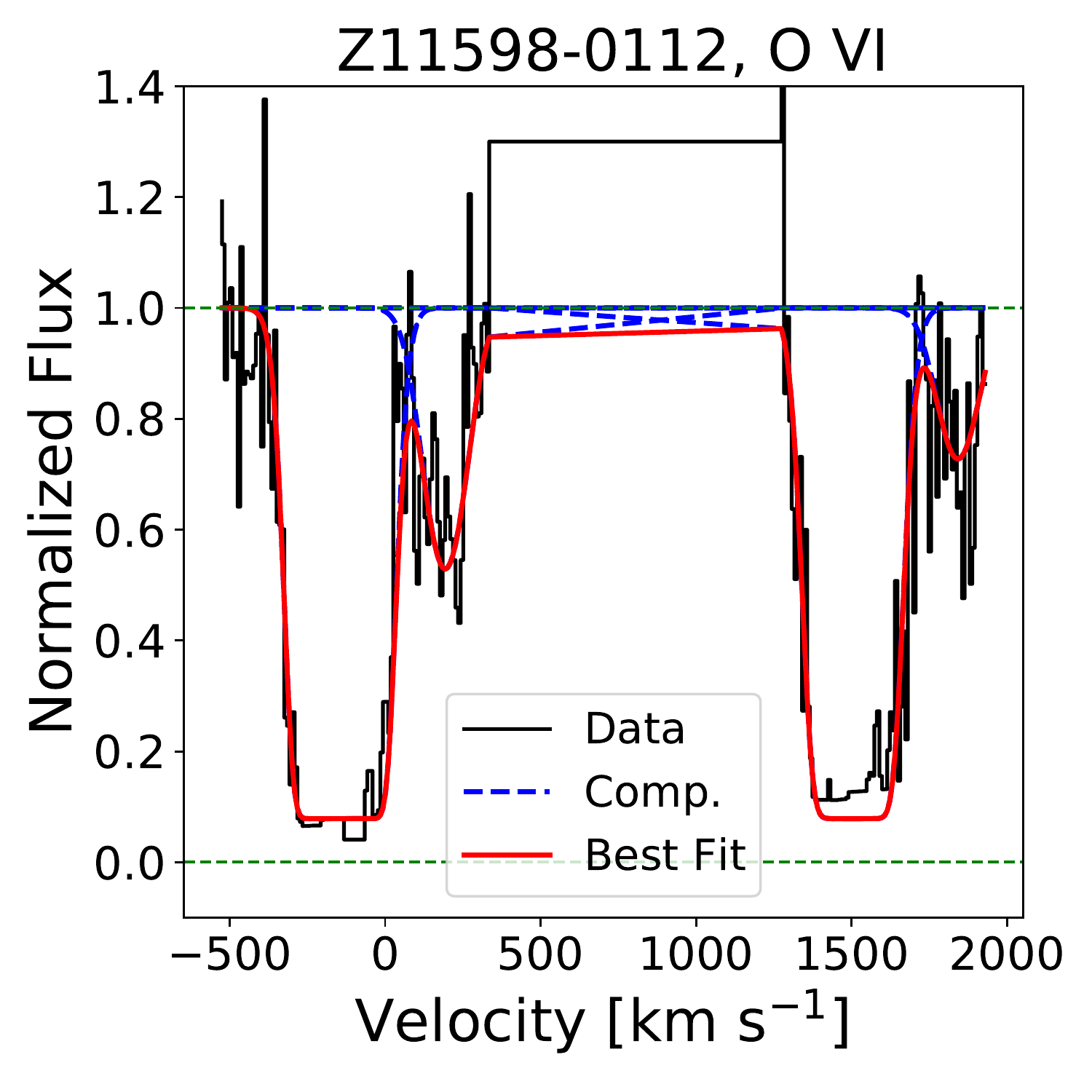}
\end{minipage}

\begin{minipage}[t]{0.32\textwidth}
\includegraphics[width=\textwidth]{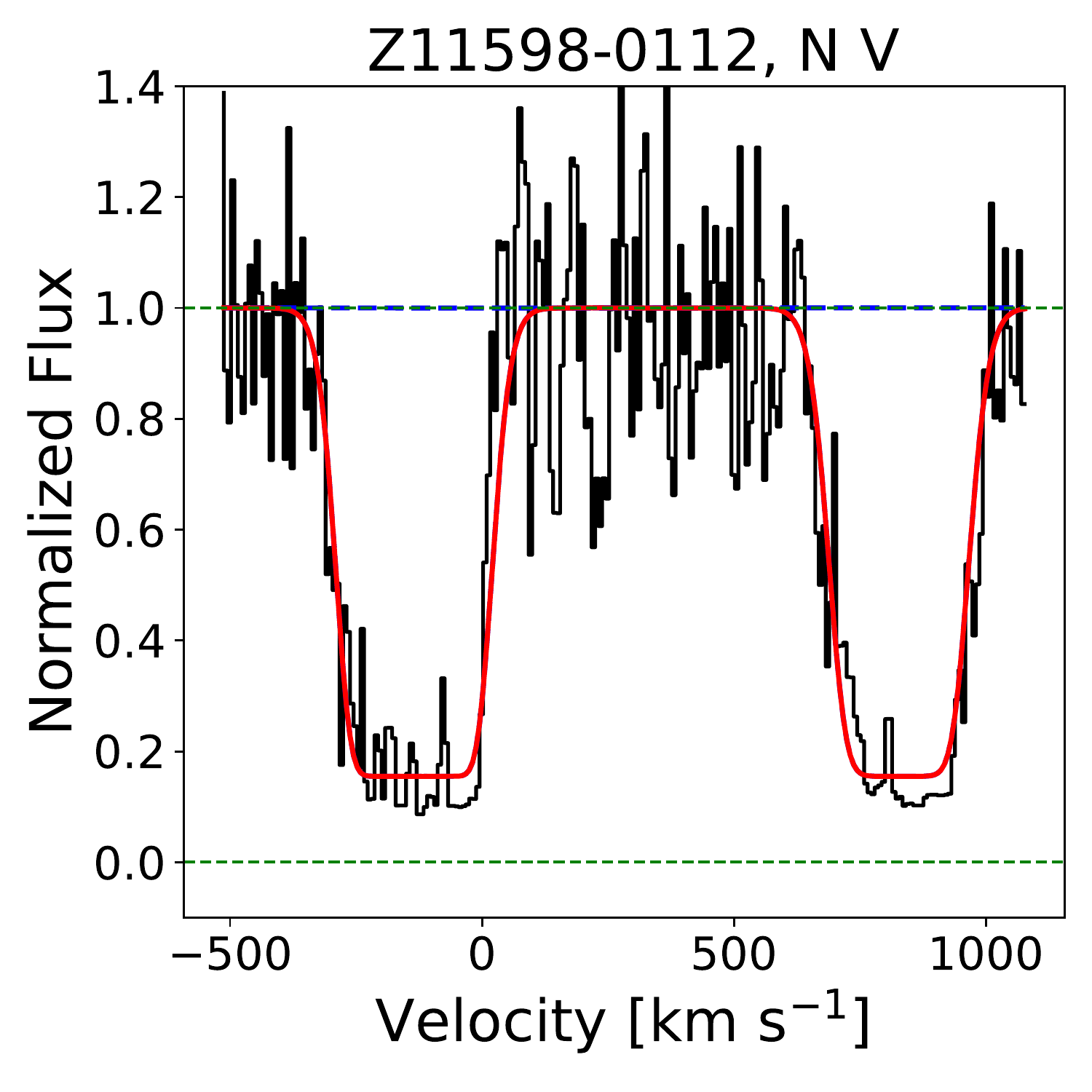}
\end{minipage}
\begin{minipage}[t]{0.32\textwidth}
\includegraphics[width=\textwidth]{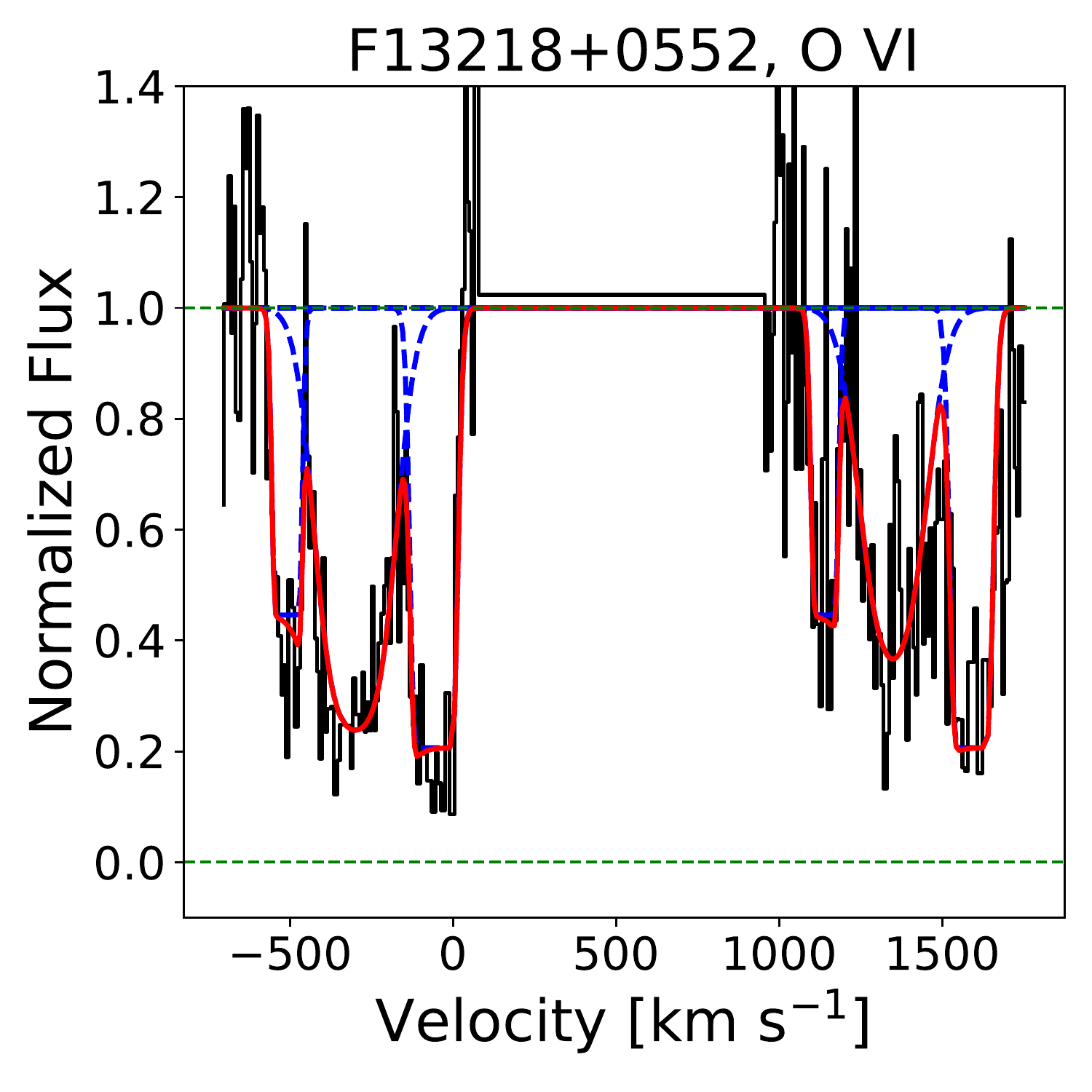}
\end{minipage}
\begin{minipage}[t]{0.32\textwidth}
\includegraphics[width=\textwidth]{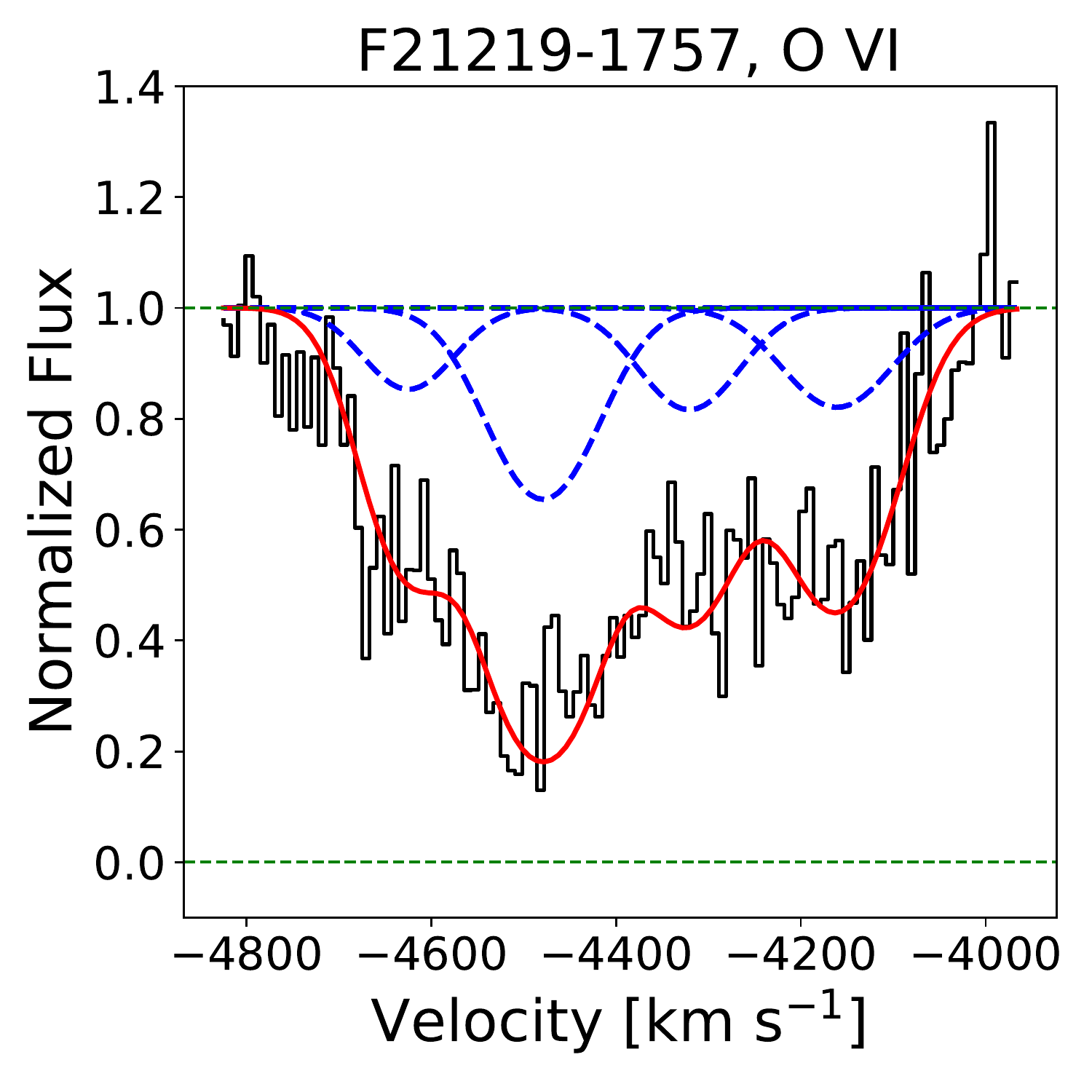}
\end{minipage}

\begin{minipage}[t]{0.32\textwidth}
\includegraphics[width=\textwidth]{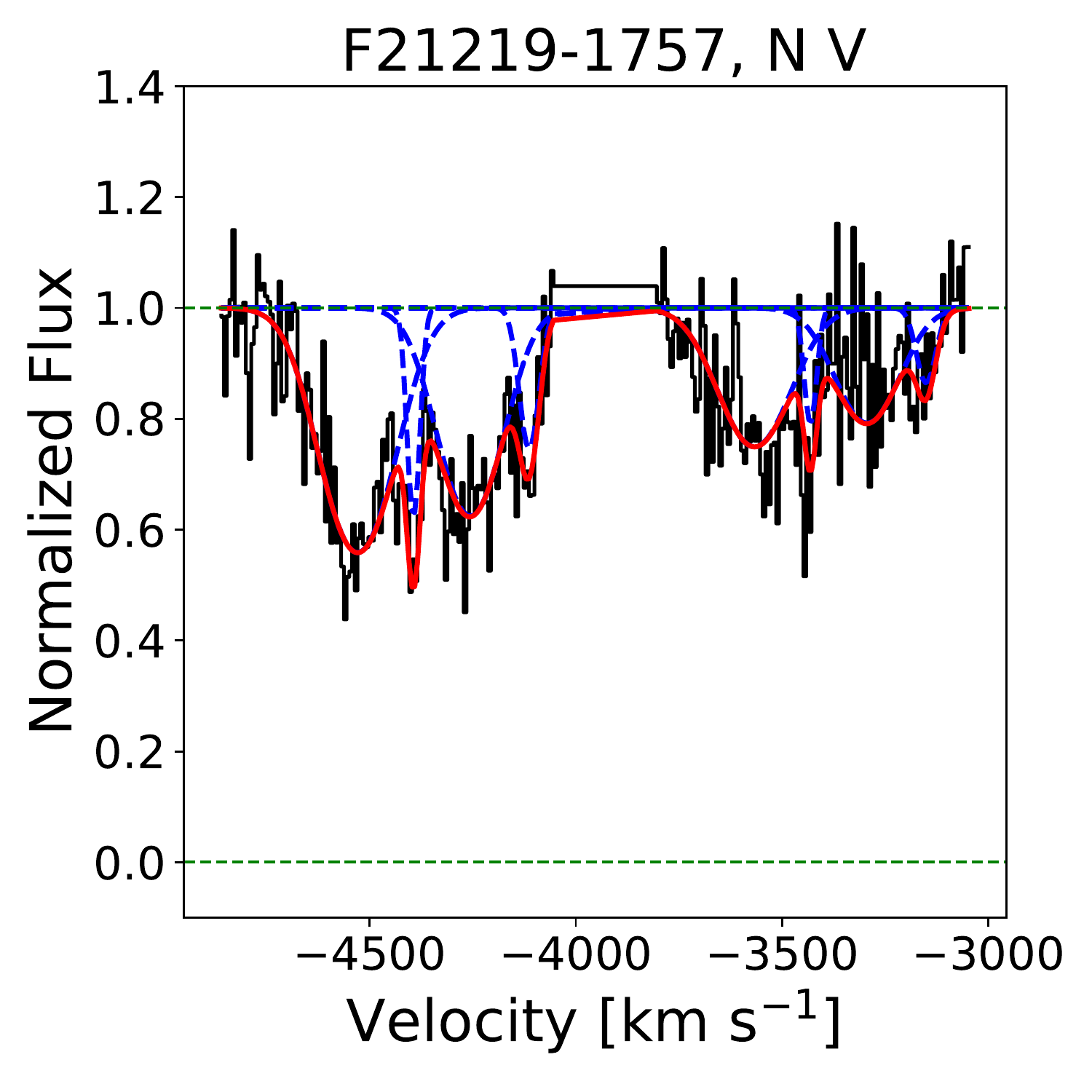}
\end{minipage}
\begin{minipage}[t]{0.32\textwidth}
\includegraphics[width=\textwidth]{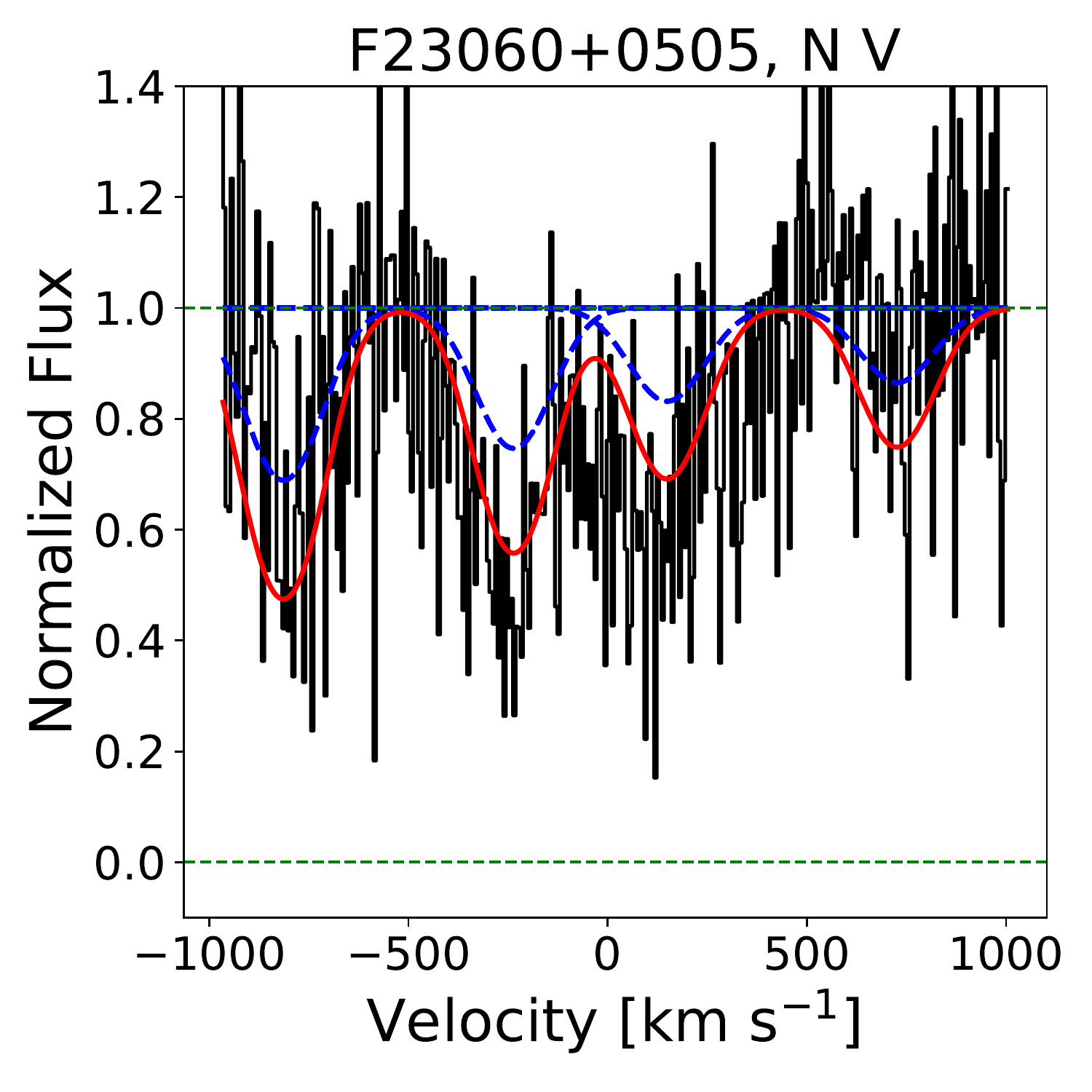}
\end{minipage}
\caption{Best fits to the \ovi\ and \nv\ absorption features detected in our sample of ULIRGs using multi-component Voigt profile fitting. Notice that for the BAL in F01004 and F07599, the single-component fits are tentative/experimental, only aiming to capture the overall shape of the BAL. In each panel, the normalized flux density is shown by the black solid curve. The overall best-fit model and individual components of this model are shown by the red solid curve and blue dashed curves, respectively. All spectra are normalized to the local continuum and shown in the rest-frame of the bluer transitions of corresponding objects.}
\label{fig:absfit}
\end{figure*}

The best fits for all absorbers clearly detected in our sample are shown in Fig. \ref{fig:absfit}. The covering factor, $C_f$, is fixed to unity in two objects: for F21219, the fit is unable to break the degeneracy between the covering factor and ion column density; for F23060 no evidence of partial covering is suggested by the data. For Z11598 and F13218, the ion column densities are reported as lower limits due to the saturation of the absorption features. Additionally, the fits for the N~V BAL in F07599 and O~VI BAL in F01004 are highly uncertain given the model parameter degeneracy caused by the saturation and smoothness of the absorption feature, and the large uncertainties in the continuum determination.

Following Paper I, our fitting scheme assumes that the velocity dependence of the optical depth can be parameterized as the sum of discrete independent components with Voigt profiles and constant covering fractions, and that the individual absorbers simply overlap with each other along the line-of-sight. In reality, the \cf\ is probably a more complex function of velocities \citep[][]{Arav2005,Arav2008,Arav2013}, and the absorbing material may completely overlap, partially overlap or not overlap each other. While our approach cannot account for these details, it is sufficient to meet our primary goal, which is to characterize the overall strength and kinematics of these absorbers, and put constraints on the ion column density.

The column densities and covering fractions from these fits are summarized in Table \ref{tab:abs}. For sources Z11598, F13218 and F21219, the results are consistent with those obtained from the analysis of the absorption doublets with partial covering model as described in Sec. \ref{331}. Further discussion based on these fits is postponed until Sec. \ref{5}.

\section{Origin of the \lya\ Emission}\label{4}

Intuitively, strong \lya\ emission is not expected in dusty ULIRGs due to the huge optical depth. The origin of \lya\ emission in our objects is therefore worth investigating. In this section, we focus on three factors that may help with the production and/or escape of \lya\ photons in our ULIRG sample:

(i) AGN: The AGN activity may intrinsically produce more \lya\ photons than what we have assumed. For example, the gas density may be so high in the broad line region (and perhaps also narrow line region) of AGN that collisional excitation becomes important in promoting \lya\ emission \citep[e.g.,][]{Dijkstra2017}. Radiation from the AGN may also ionize the gas and destroy the dust in the \lya-emitting regions and material along the line-of-sight, and therefore reduce the overall opacity to the \lya\ emission. (ii) Outflow: The blueshifted \lya\ emission may come from the outflowing gas. The velocity offset between the outflow and the interstellar medium decreases the optical depth to the \lya\ emission radiated from the fast-moving gas. The outflow can also create low opacity pathways for \lya\ photons by clearing out the gas and dust. In addition, the outflow may have broken out of the dusty ISM so that \lya\ photons can escape freely. (iii) Dust: The \lya\ photons are heavily affected by complicated, dust-related radiative transfer effects, which directly affect the observed properties of the \lya\ emission in ULIRGs.

In our analyses, we adopt Kendall tau correlation tests to examine potential correlations between the properties of \lya\ emission and those of the AGN, outflows, and dust reddening. Specifically, we adopt the method from \citet{Isobe1986} to compute the Kendall tau correlation coefficient. The $p$-value of null hypothesis (no correlation) is calculated to show the statistical significance of the correlation. This method can handle censored data, which is the case for \ewlya\ and \fesc. We use the implementation of pymccorrelation \citep{Privon2020}, which perturbs the data with Monte Carlo method to compute the error in the correlation coefficient \citep{curran2014}.
To expand the dynamic ranges of the variables in the analyses, by default, we also include the results of the starburst-dominated ULIRGs from M15, whenever possible. The results from these correlation tests are summarized in Table \ref{tab:lyastatistic}.

\begin{deluxetable}{cccccc}
\tablecolumns{6}
\tabletypesize{\scriptsize}
\tablecaption{Correlation Tests with the \lya\ Properties \label{tab:lyastatistic} }
\tablehead{\colhead{x} & \colhead{y} & sample & \colhead{N} & \colhead{$p$-value} & \colhead{r} }
\colnumbers
\startdata 
  \lagn & \ewlya    &        A          &                  20 & 0.002 & $0.49_{-0.14}^{+0.13}$   \\ 
  \lagn & \fesc   &          A          &            26 & 0.062 & $0.26_{-0.16}^{+0.16}$   \\ 
  \lagn & \vba  &            A          &           14 & 0.058 & $-0.38_{-0.15}^{+0.17}$  \\ 
  \lagn & \vwu  &            A          &           14 & 0.095 & $-0.34_{-0.15}^{+0.18}$  \\ 
  \lagn & \wba  &            A          &           14 & 0.066 & $0.37_{-0.18}^{+0.18}$ \\ 
  \fagn & \ewlya    &        A          &                  20 & 0.023 & $0.37_{-0.14}^{+0.13}$   \\ 
  \fagn & \fesc   &          A          &            26 & 0.068 & $0.25_{-0.14}^{+0.14}$   \\ 
  \fagn & \vba  &            A          &           14 & 0.470 & $-0.08_{-0.20}^{+0.19}$  \\ 
  \fagn & \vwu  &            A          &           14 & 0.429 & $-0.08_{-0.21}^{+0.22}$  \\ 
  \fagn & \wba  &            A          &           14 & 0.439 & $-0.08_{-0.22}^{+0.20}$  \\ 
  \vbao & \ewlya    &        A          &                 18 & 0.064 & $-0.32_{-0.16}^{+0.19}$   \\ 
  \vbao & \fesc    &         A          &             20 & 0.049 & $-0.32_{-0.16}^{+0.20}$   \\ 
  \vbao & \vba    &          A          &             18 & $<$0.001 & $0.72_{-0.11}^{+0.08}$   \\ 
  \vbao & \vba             & B &                   12 & 0.003 & $0.65_{-0.20}^{+0.14}$   \\ 
  \vbao & \vba         &        C  &                   8 & 0.008 & $0.77_{-0.23}^{+0.13}$   \\ 
  \vwuo & \ewlya  &          B          &               12 & 0.398 & $-0.06_{-0.24}^{+0.30}$   \\ 
  \vwuo & \fesc  &           B          &           13 & 0.381 & $0.04_{-0.26}^{+0.26}$   \\ 
  \vwuo & \vwu          &   B  &                   12 & 0.041 & $0.45_{-0.21}^{+0.18}$   \\ 
  \vwuo & \vwu       &         C   &                 8 & 0.013 & $0.72_{-0.26}^{+0.15}$   \\ 
  \wbao & \ewlya  &          B          &               12 & 0.439 & $-0.05_{-0.25}^{+0.25}$   \\ 
  \wbao & \fesc  &           B          &           14 & 0.228 & $0.23_{-0.22}^{+0.19}$   \\ 
  \wbao & \wba          &   B  &                   12 & 0.007 & $0.60_{-0.18}^{+0.14}$   \\ 
  \wbao & \wba       &         C  &                  8 & 0.083 & $0.49_{-0.27}^{+0.27}$   \\ 
  \ajiuyio & \ewlya  &       B          &                 12 & 0.285 & $-0.20_{-0.24}^{+0.28}$   \\ 
  \ajiuyio & \fesc    &      B          &               14 & 0.468 & $0.10_{-0.18}^{+0.16}$   \\ 
  \ajiuyio & \ajiuyi     &  B  &                       12 & 0.012 & $0.56_{-0.17}^{+0.13}$   \\ 
  \ajiuyio & \ajiuyi       &    C  &                     8 & 0.017 & $0.69_{-0.20}^{+0.17}$   \\ 
  \vbaha & \ewlya  &         B          &               14 & 0.174 & $-0.26_{-0.23}^{+0.29}$   \\ 
  \vbaha & \fesc  &          B          &           17 & 0.428 & $-0.06_{-0.20}^{+0.20}$   \\ 
  \vbaha & \vba          &  B  &                   14 & 0.001 & $0.64_{-0.13}^{+0.10}$   \\ 
  \vwuha & \ewlya  &         B          &               14 & 0.534 & $-0.03_{-0.18}^{+0.19}$   \\ 
  \vwuha & \fesc  &          B          &           15 & 0.316 & $0.13_{-0.26}^{+0.20}$   \\ 
  \vwuha & \vwu          &  B  &                   14 & 0.437 & $0.01_{-0.24}^{+0.22}$   \\ 
  \wbaha & \ewlya  &         B          &               14 & 0.203 & $0.23_{-0.28}^{+0.21}$   \\ 
  \wbaha & \fesc  &          B          &           16 & 0.411 & $-0.03_{-0.21}^{+0.23}$   \\ 
  \wbaha & \wba          &  B  &                   14 & 0.005 & $0.56_{-0.14}^{+0.13}$   \\ 
  \ajiuyiha & \ewlya  &      B          &                 14 & 0.373 & $0.00_{-0.26}^{+0.27}$   \\ 
  \ajiuyiha & \fesc  &       B          &             16 & 0.466 & $-0.10_{-0.16}^{+0.18}$   \\ 
  \ajiuyiha & \ajiuyi   &   B  &                       14 & 0.039 & $0.41_{-0.20}^{+0.16}$   \\ 
    \ebv & \fesc &           A          &       20 & 0.371 & $0.00_{-0.22}^{+0.20}$ 
\enddata
\tablecomments{Column (1): Independent variable (AGN and host galaxy property). Column (2): Dependent variable (\lya\ property). Column (3): Flag for the sample adopted in the analysis. A: Our ULIRG sample $+$ the ULIRG sample from M15; B: Our ULIRG sample alone; C: Sources with Type 2 AGN in our ULIRG sample. Column (4): Number of data points; Column (4) $p$-value of null hypothesis (no correlation) from the Kendall tau correlation test; Column (5) Kendall tau correlation coefficient r with the associated 1-$\sigma$ error.}
\end{deluxetable}

\subsection{Effect of the AGN} \label{41}

\begin{figure*}[!htb]   
\epsscale{1}
\plottwo{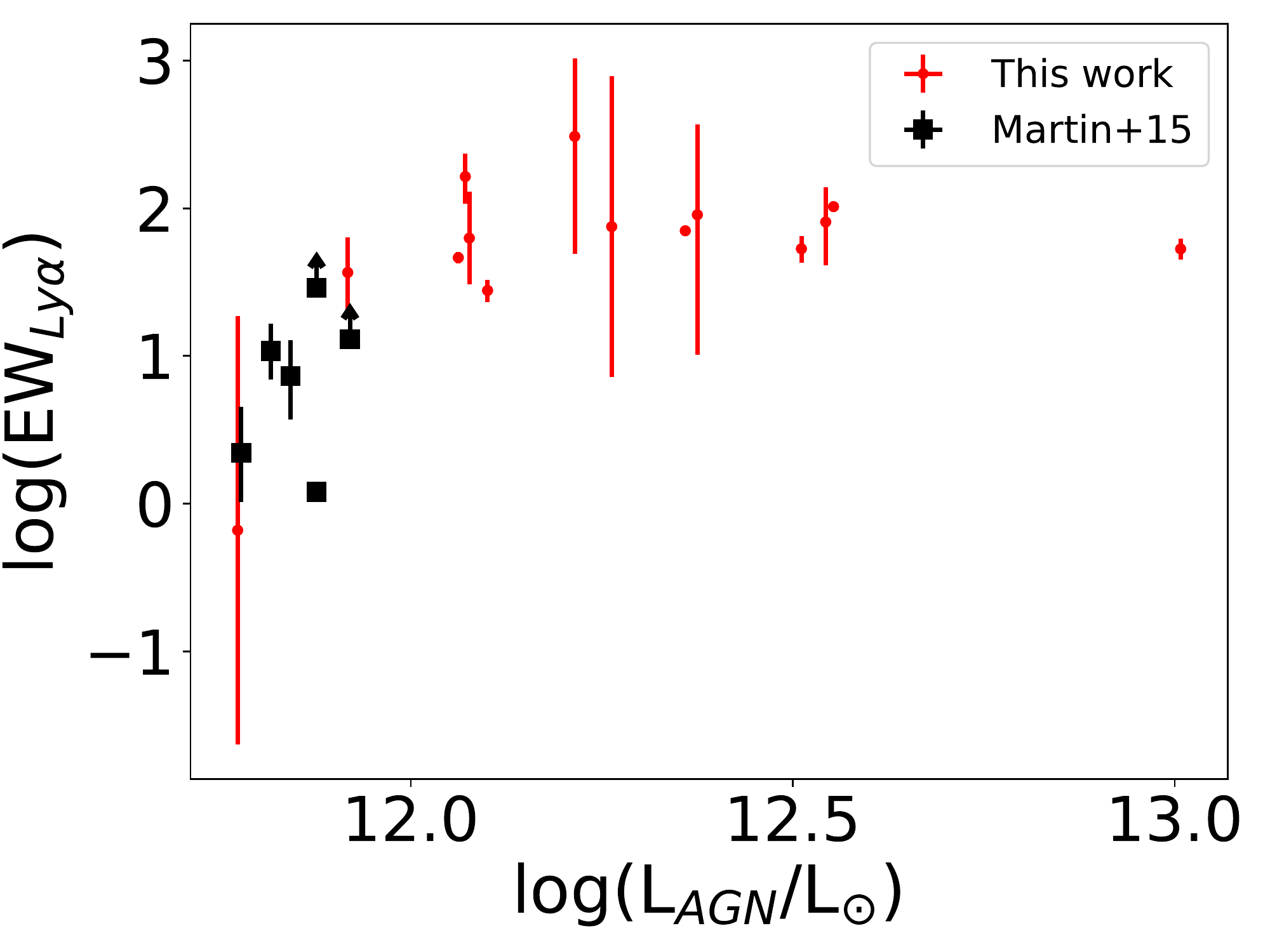}{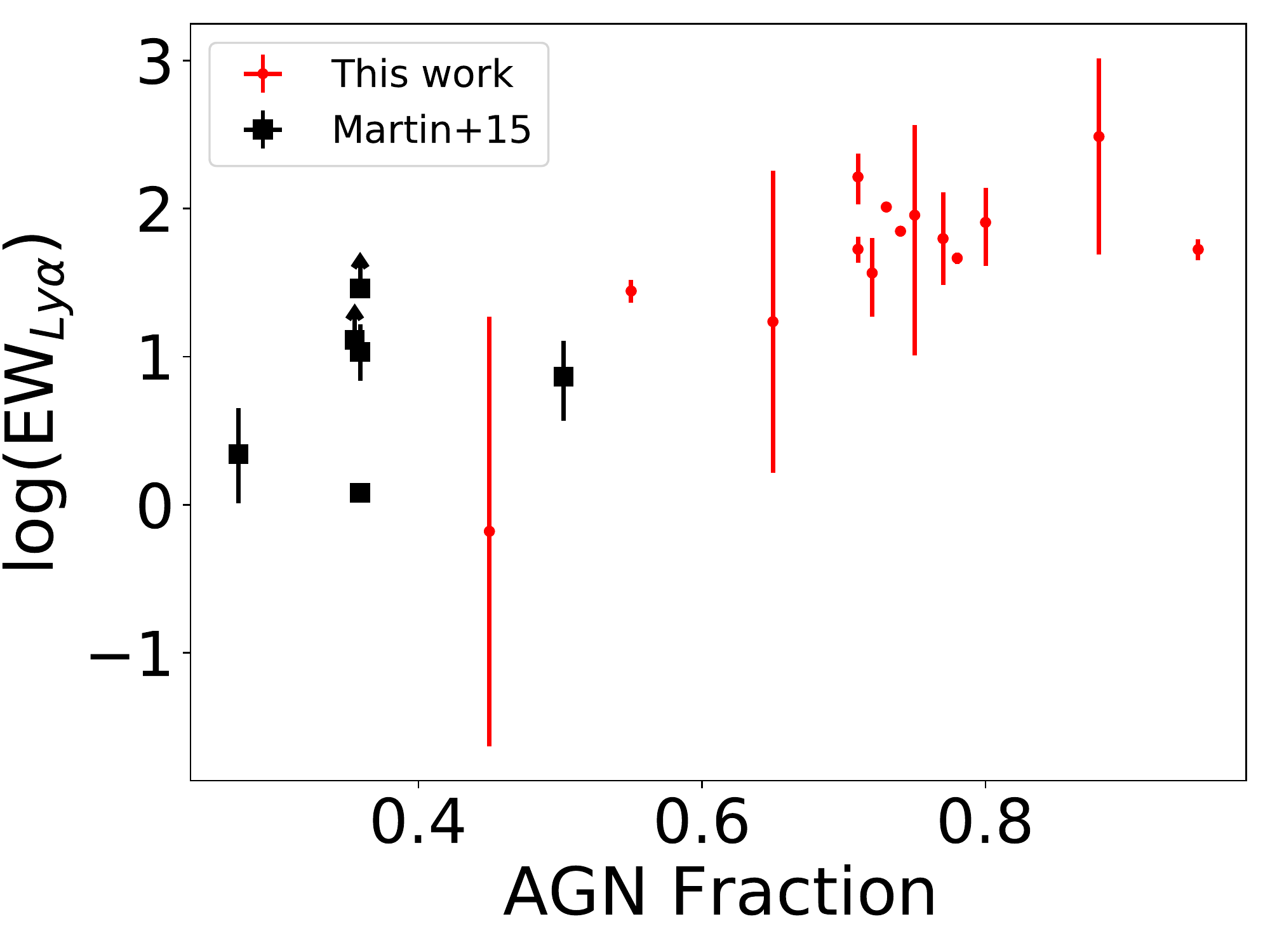}
\caption{The \lya\ EWs for the ULIRGs in our sample (red) and those from M15 (black) versus AGN luminosities (left) and AGN fractions (right). There are statistically significant positive correlations between \lagn\ and \ewlya, and between \fagn\ and \ewlya.}
\label{fig:ewlya_AGN}
\end{figure*}

\begin{figure*}[!htb]   
\epsscale{1}
\plottwo{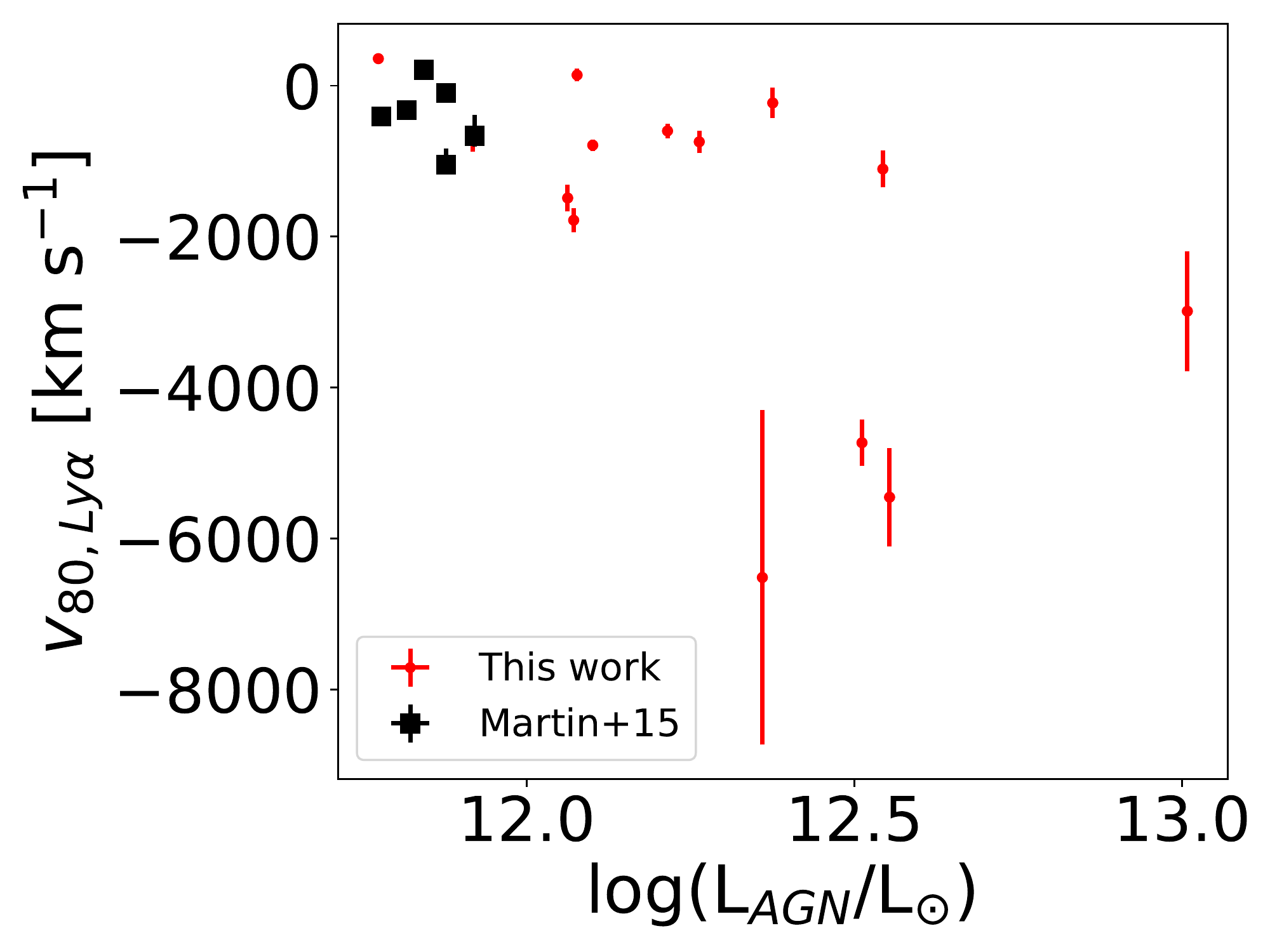}{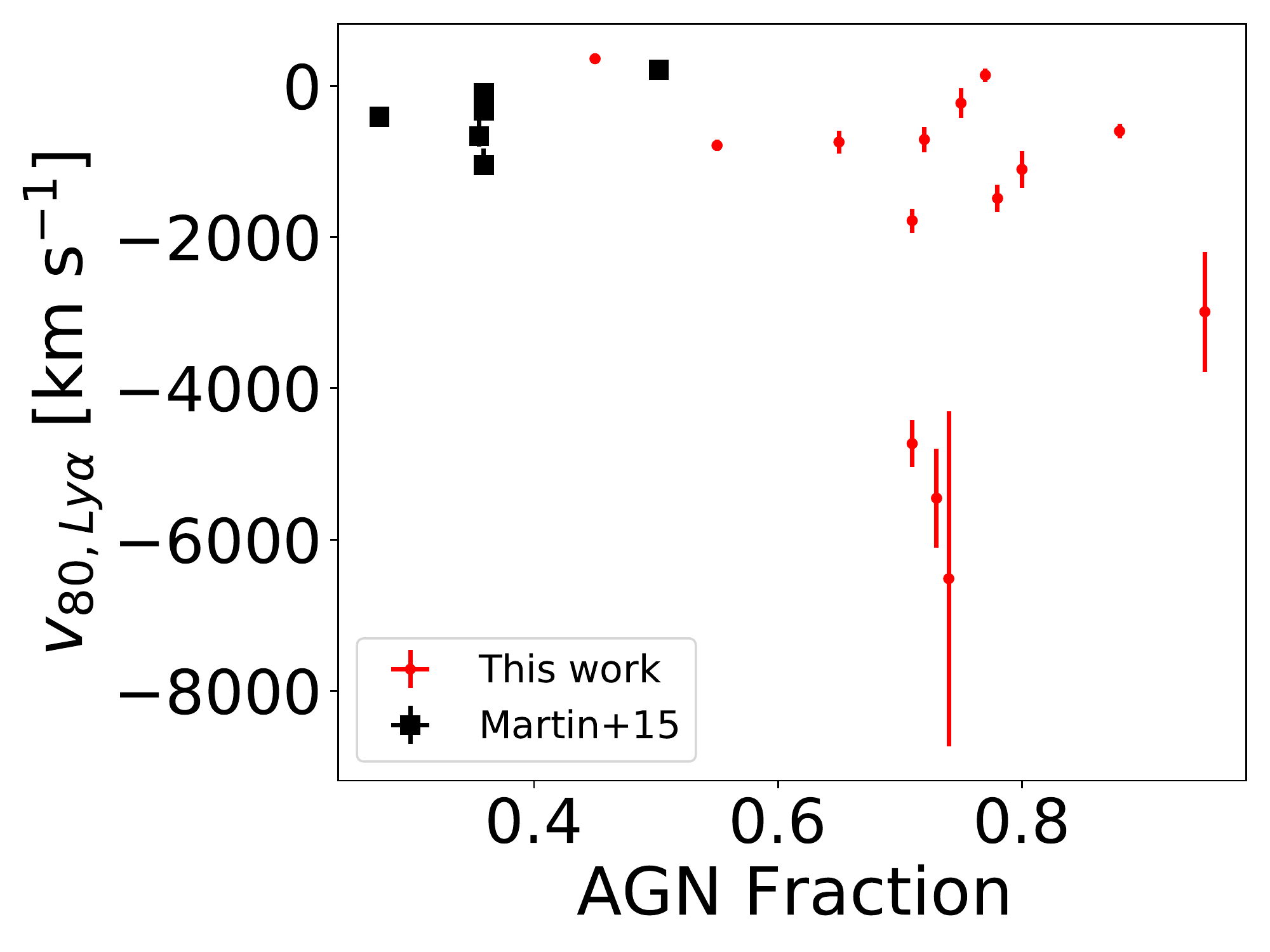}
\caption{The 80-percentile velocities \vba\ of \lya\ for the ULIRGs in our sample (red) and those from M15 (black) versus AGN luminosities (left) and AGN fractions (right). There are weak correlations between \lagn\ and \vba, and between \fagn\ and \vba, which are, however, mainly driven by the Type 1 sources with small \vba.}
\label{fig:v80lya_AGN}
\end{figure*}

First, we examine the possible link between the strength of the AGN (more specifically, the AGN luminosities, \lagn, and AGN fractions, \fagn; see Table \ref{tab:targets}) and the properties of \lya\ emission. As shown in Fig.\ \ref{fig:ewlya_AGN}, the EWs of \lya\ (\ewlya) increase with both \lagn\ and \fagn. The $p$-values are $\sim$0.002 and $\sim$0.023 based on the Kendall tau tests, respectively. In addition, \lagn\ may also correlate with the kinematic properties of \lya\ (\vba, \vwu\, \ajiuyi\, and \wba\ of \lya; $p$-values $\simeq$ 0.06--0.10). As an example, \vba\ of \lya\ are plotted against \lagn\ and \fagn\ in Fig. \ref{fig:v80lya_AGN}. These weak trends are mainly driven by the Type 1 sources with small \vba\ and large \wba: indeed, the correlations are not statistically significant ($p$-values $>$ 0.1) when the Type 1 sources are excluded.

\begin{figure*}[!htb]   
\epsscale{1}
\plottwo{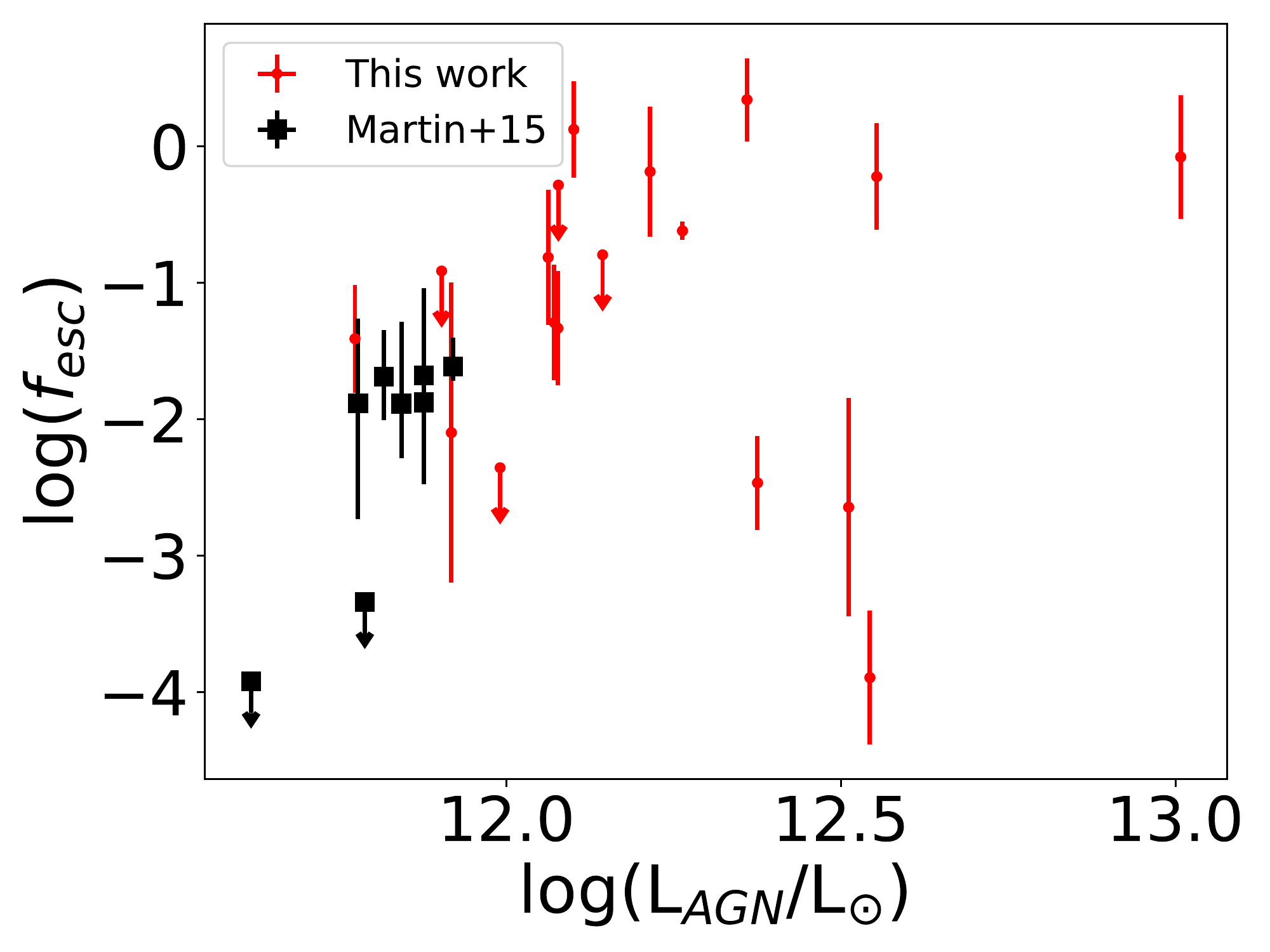}{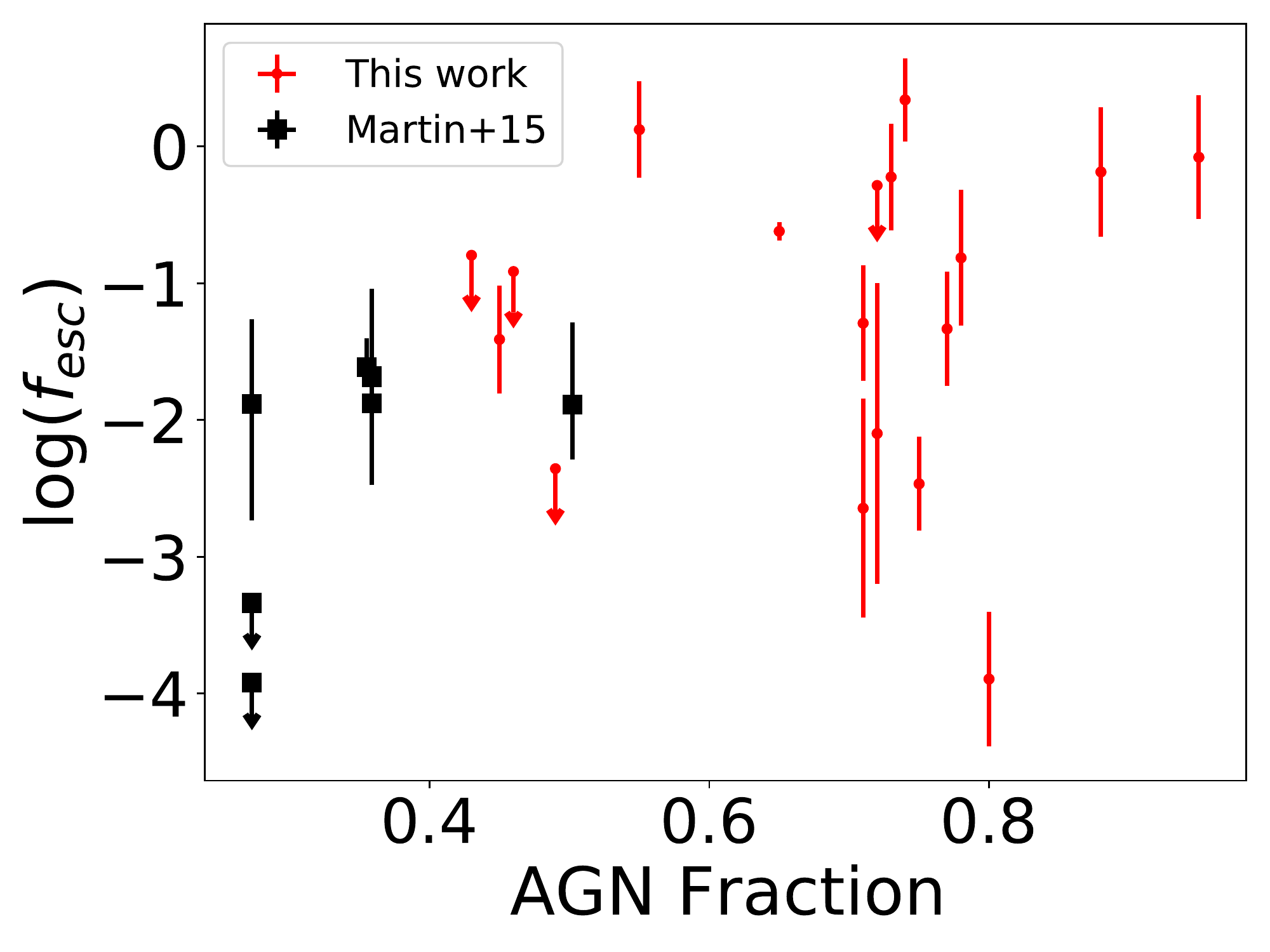}
\caption{The \lya\ escape fractions for the ULIRGs in our sample (red) and those from M15 (black) versus AGN luminosities (left) and AGN fractions (right). There may be weak correlations between \lagn\ and \fesc, and between \fagn\ and \fesc, when both the data from our sample and sources in Martin$+$15 are considered.}
\label{fig:fesc_AGN}
\end{figure*}

Next, we explore the behavior of the \lya\ escape fraction, \fesc, with the strength of the AGN, as shown in Fig. \ref{fig:fesc_AGN}. There are possible positive correlations between \lagn\ and \fesc\ ($p$-value $\simeq$ 0.062), and between \fagn\ and \fesc\ ($p$-value $\simeq$ 0.068). However, we note that these correlations are no longer significant ($p$-values $\simeq$ 0.2 and 0.4, respectively) if we exclude the data points of the starburst-dominated ULIRGs from M15.

\subsection{Effect of the Outflow}  \label{42}

The prevalence of blueshifted \lya\ emission line profiles in our sample, as stated in Sec. \ref{31}, hints at a potential link between outflows and the \lya\ emission in our objects. Therefore, we start by simply checking whether there is a correlation between the \ewlya\ and \vba\ of \lya, and confirm a positive result ($p$-value $\simeq$ 0.02). However, we note that this trend disappears ($p$-values $>$ 0.1) if we do not include the starburst-dominated objects from M15. Similarly, we have also checked the potential correlation between \fesc\ and \vba\ of \lya, but no statistically significant trend is present (p-value $\simeq$ 0.13).

\subsubsection{Connection with the Ionized Outflow in Emission}  \label{421}

\begin{figure*}[!htb]   
\plotone{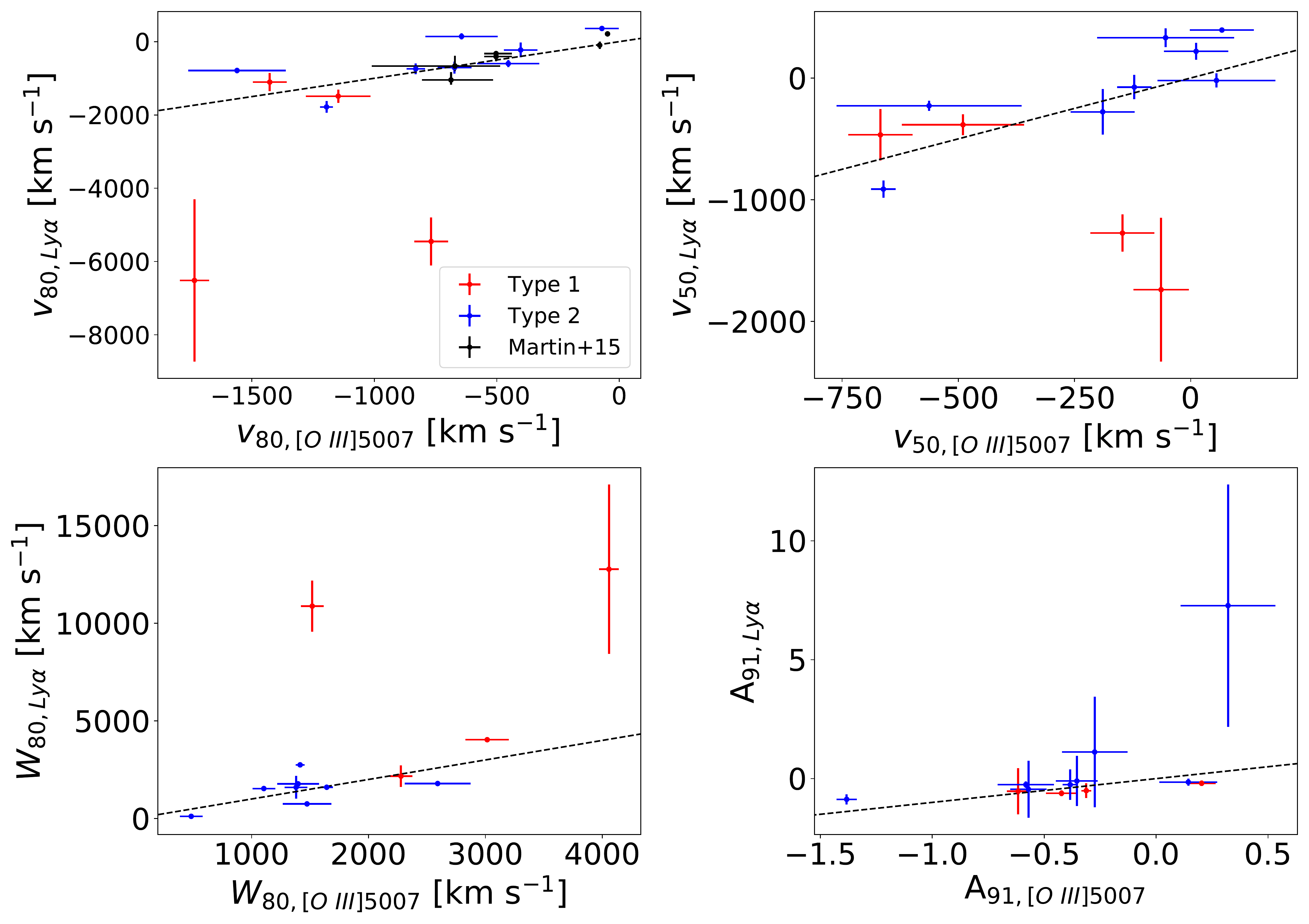}
\caption{Comparisons of the kinematic properties derived from \lya\ and \oiii.  From top left to bottom right, \vba, \vwu, \wba, and \ajiuyi\ are shown. The ULIRGs with Type 1 and Type 2 AGN in our sample are shown in red and blue, respectively. In the upper-left panel, the ULIRGs from M15 are shown in black. The black dashed lines indicate the 1:1 equality lines. The \vba\ of \lya\ and \oiii\ emission lines correlate with each other. Other kinematic properties of the \lya\ and \oiii\ emission lines are also correlated but less significantly.}
\label{fig:velo3lya}
\end{figure*}

The blueshift of the non-resonant, forbidden emission \oiii\ line in galaxies is strong  evidence for ionized gas outflows \citep{Veilleux2005}. To investigate the connection between the blueshift of the \lya\ emission and [O III] outflowing gas, it is thus natural to determine whether the kinematic properties derived from the \lya\ emission line are correlated with those based on the \oiii\ emission. This comparison is shown in Fig. \ref{fig:velo3lya}.

The most significant trend observed in our sample is the positive correlation between \vba\ of \lya\ and \oiii, with $p$-value $<$ 0.001. This trend still holds if the data from M15 are excluded ($p$-values $\simeq$ 0.003), or if only the type 2 sources are considered in the analysis ($p$-value $\simeq$ 0.008). The values of \vwu, \wba, and \ajiuyi\ of \lya\ also show positive correlations with those of \oiii, although they are statistically less significant ($p$-values of $\sim$0.007 -- 0.041 from the Kendall tau tests\footnote{As a reminder, in the analyses including \vwuo, \wbao, and \ajiuyio, we do not consider the sources from M15 as the corresponding measurements are not publicly available.}; see Table \ref{tab:lyastatistic}).

\begin{figure*}[!htb]   
\epsscale{1}
\plottwo{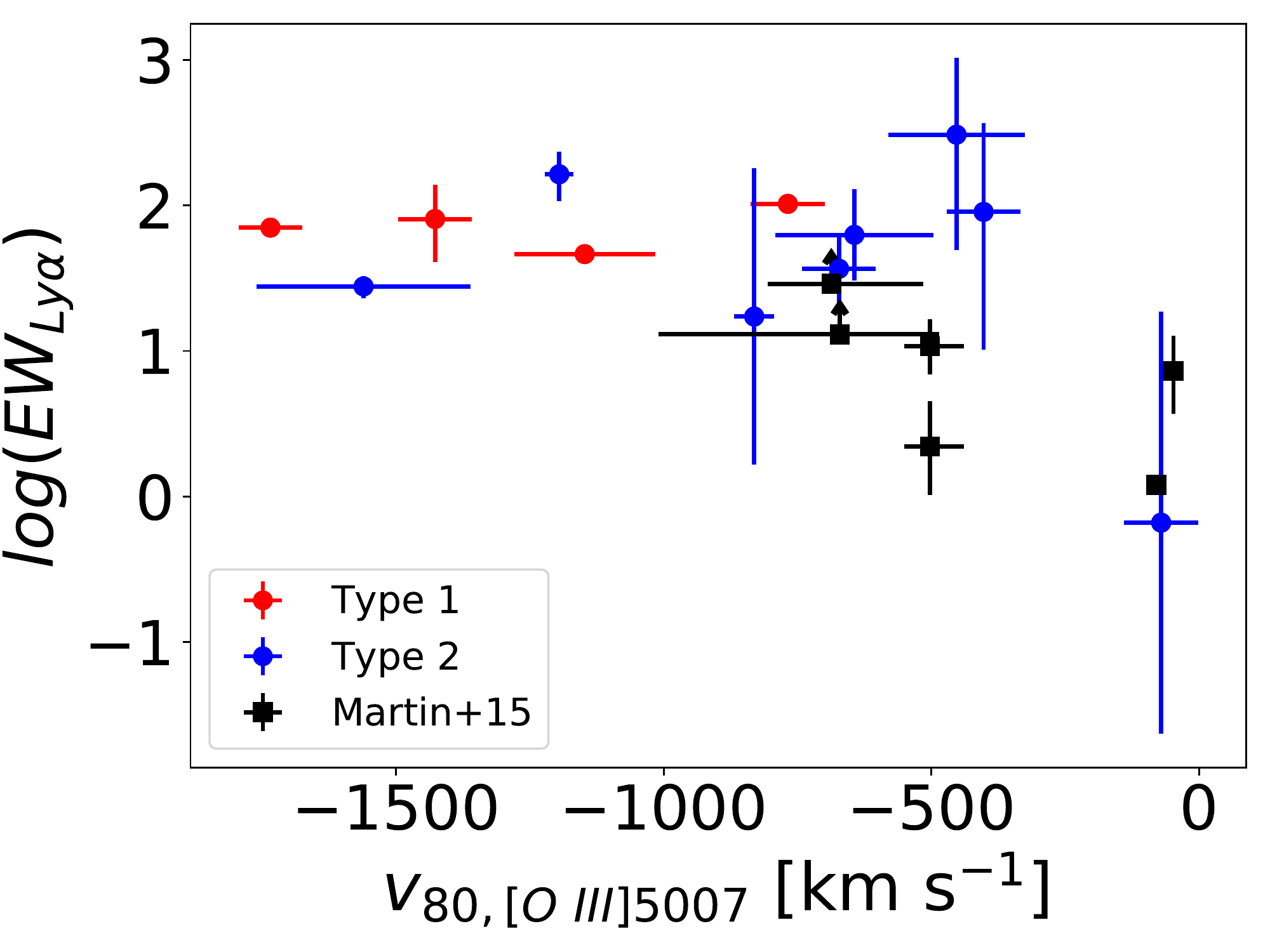}{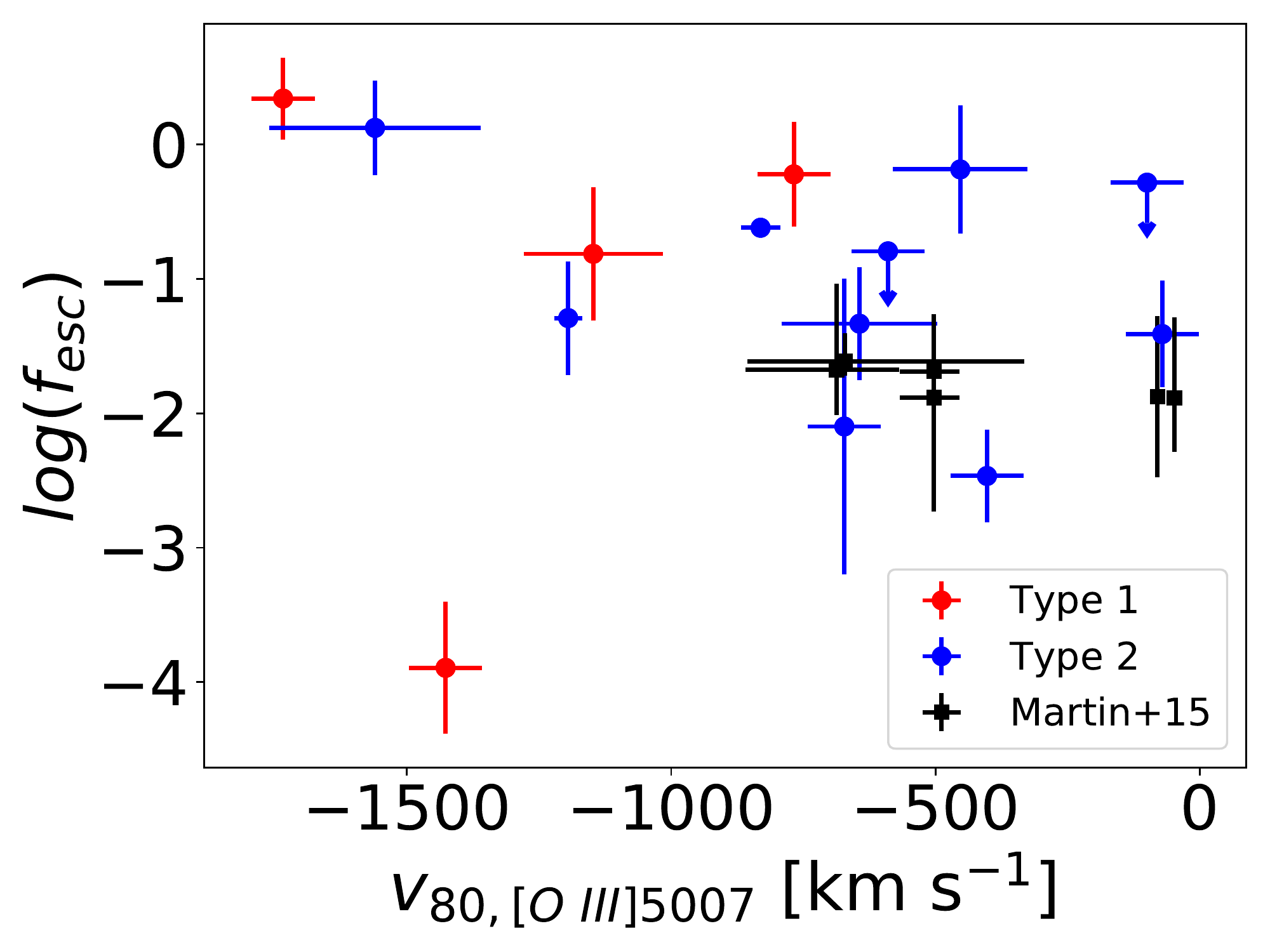}
\caption{The \lya\ EWs (left) and \lya\ escape fractions (right) for the Type 1 (red) and Type 2 (blue) ULIRGs in our sample and the starburst-dominated ULIRGs from M15 (black) as function of the 80-percentile velocities \vba\ of \oiii\ emission lines. There are weak correlations between \vbao\ and \ewlya, and between \vbao\ and \fesc, when both the data from our sample and sources in Martin$+$15 are considered.}
\label{fig:v80o3_ewlya}
\end{figure*}

Additionally, as shown in Fig.\ \ref{fig:v80o3_ewlya}, there may be weak correlations between \vbao\ (notice that this quantity is negative in the case of outflows) and \ewlya\ ($p$-value $\simeq$ 0.064), and between \vbao\ and \fesc\ ($p$-value $\simeq$ 0.049). This suggests that, to some extent, outflows may help clear the path for \lya\ photons to escape. However, these weak trends are no longer significant when the objects from M15 are excluded. Moreover, no statistically significant trends are visible when considering the other kinematic properties of \oiii\ (\vwuo, \wbao, and \ajiuyio).

For the sake of completeness, we also briefly discuss the strong \ha\ line emission in these objects. We note that \vba, \wba, and \ajiuyi\ of \ha\ are positively correlated with those of \lya\ ($p$-value $\simeq$ 0.001--0.039; see Table \ref{tab:lyastatistic}), whereas no correlation is seen between \vwu\ of \ha\ and \lya\ ($p$-value $\simeq$ 0.437). While blueshifted \ha\ emission may indicate outflowing gas, \ha\ is likely dominated by the emission from the broad emission line region (BELR) in Type 1 sources (e.g. 3C~273, Mrk~1014; see Fig. \ref{fig:HaloopAGN}). In addition, the nearby \niiab\ emission lines add uncertainty to the kinematic measurements based on \ha. Therefore, we do not use the \ha-based kinematics to examine the link between the blueshifts of \lya\ and the ionized outflows in the remainder of the paper.

\subsubsection{Connection with the O~VI and/or N~V Outflows}  \label{422}

As discussed in Sec. \ref{33}, highly ionized O~VI and/or N~V outflows are detected in the HST/COS spectra of our ULIRGs. Among the 9 objects with both \lya\ measurements and continuum S/N high enough to allow for relatively solid detection of O~VI and/or N~V absorption features, 4 of them show O~VI and/or N~V outflows. There is no clear difference in the mean values and overall ranges of the \lya\ escape fractions between the objects with and without O~VI and/or N~V outflows. The mean value of the \lya\ EWs in the objects without outflow detection in O~VI and/or NV is higher than those with outflows by $\sim$60\%. However, these results are based on a very small sample, so no statistically robust conclusion may be drawn.

\subsubsection{Connection with the Neutral Phase Outflows}  \label{423}

\begin{figure*}[!htb]   
\centering
\begin{minipage}[t]{0.31\textwidth}
\includegraphics[width=\textwidth]{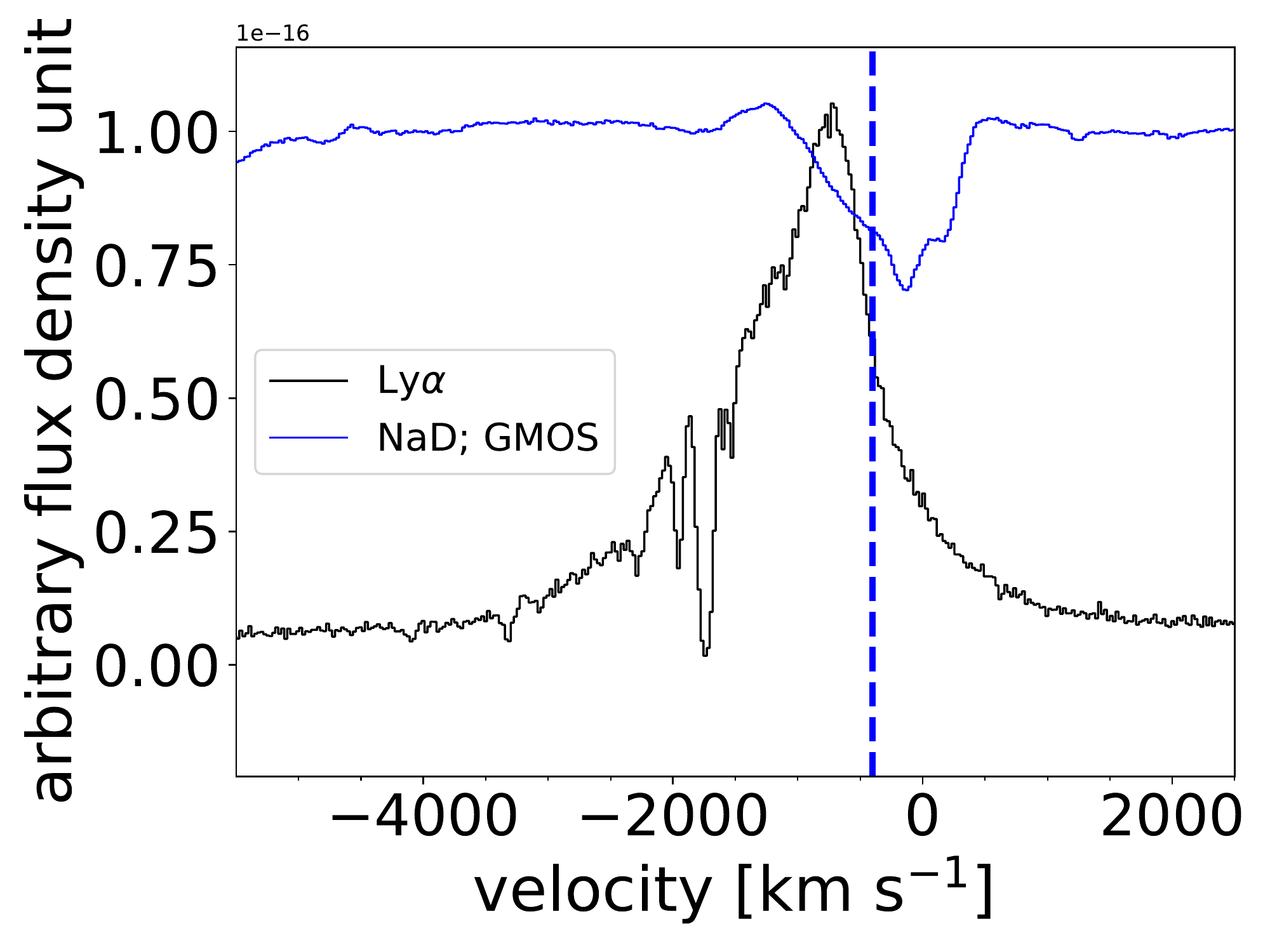}
\end{minipage}
\begin{minipage}[t]{0.31\textwidth}
\includegraphics[width=\textwidth]{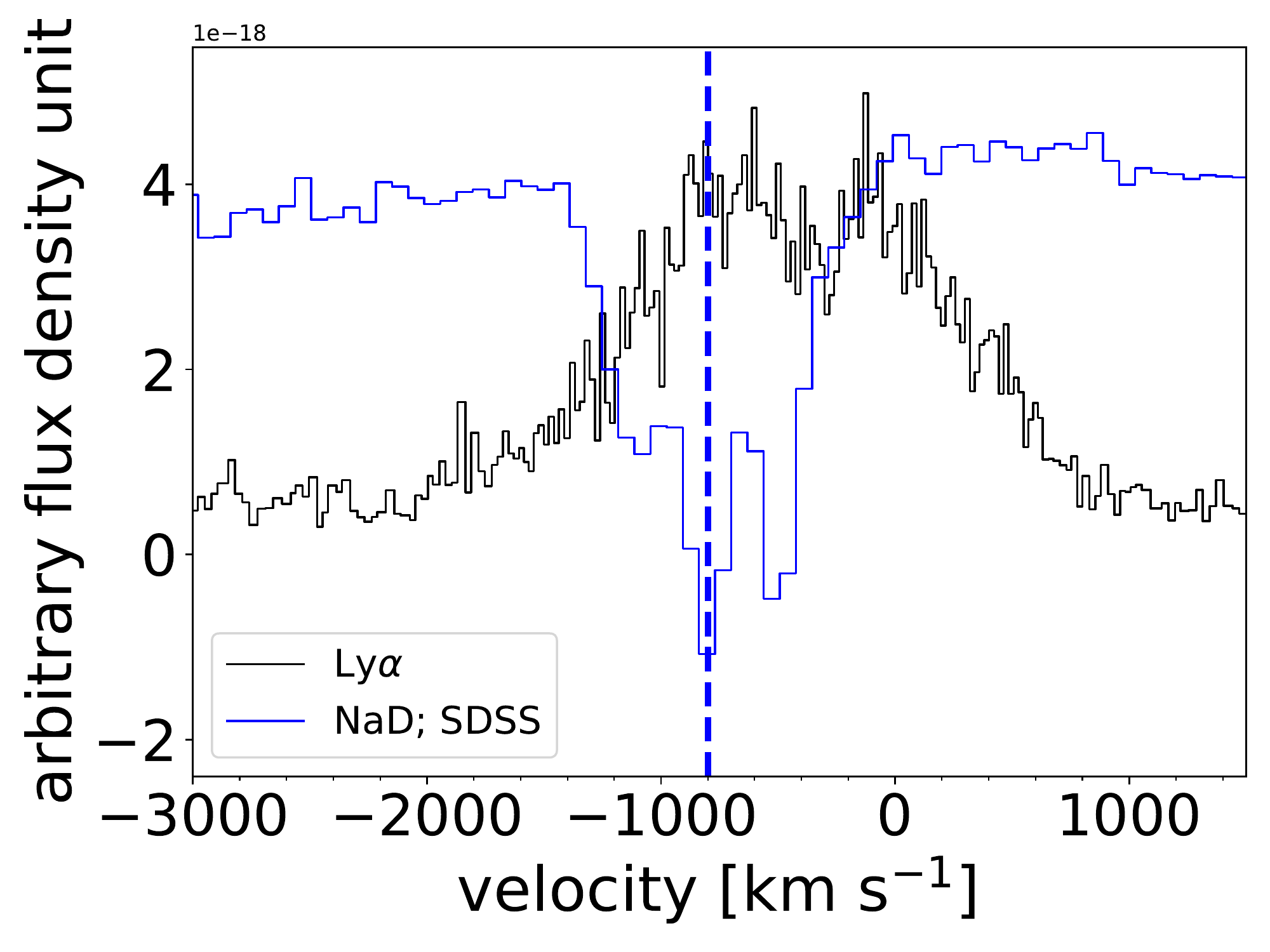}
\end{minipage}
\begin{minipage}[t]{0.31\textwidth}
\includegraphics[width=\textwidth]{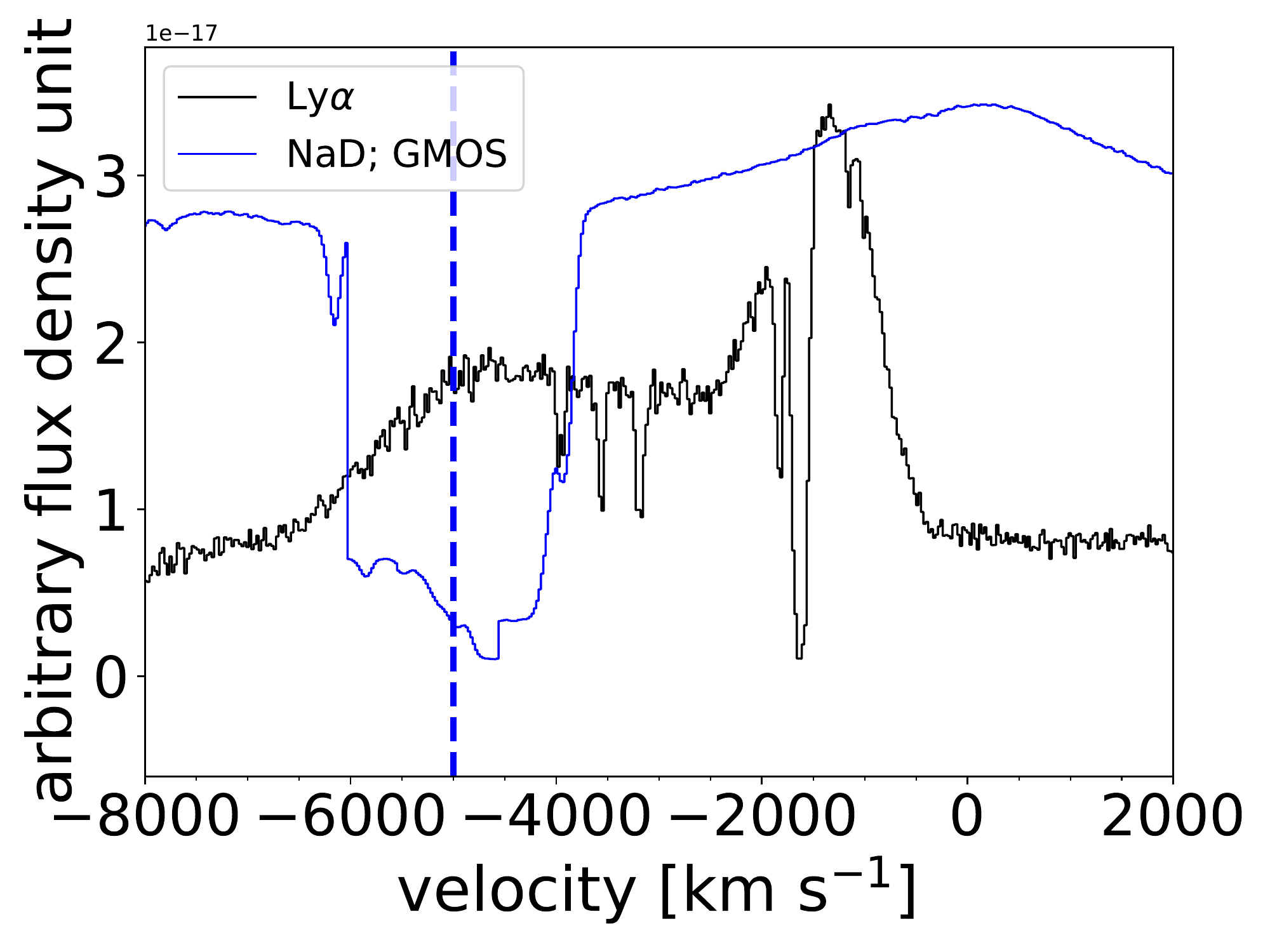}
\end{minipage}
\caption{Comparison between the \lya\ (black) and \nad\ (blue) profiles for sources F05189 (left), F11119 (middle) and Mrk 231 (right). The blue dashed lines indicate the \vwu\ of the outflowing \nad\ component from \citet[][excluding the systemic component in F05189]{Rupke2005c}. The spectra are scaled for display purposes and the $y$ axes are in arbitrary flux density units.}
\label{fig:nad}
\end{figure*}

Neutral gas outflows, traced by the blueshifted \nad\ absorption features, are often detected in ULIRGs \citep[e.g.,][]{Rupke2005c,RupkeVeilleux2011,RupkeVeilleux2013a,Rupke2017}. As shown in Fig. \ref{fig:nad}, three of our objects have both blueshifted \lya\ emission and blueshifted interstellar \nadb\ absorption features with similar kinematics.  Specifically, for F05189, \vwu\ of \lya\ emission is similar to that of the blueshifted component of the \nadb\ absorption ($\sim$$-$400 \kms). For F11119, the blueshifted peak of the \lya\ emission and \vwu\ of \nadb\ absorption have a similar velocity of $\sim$$-$800 \kms. For Mrk~231, \vwu\ of \nadb\ absorption ($\sim$$-$5000 \kms) is close to the velocity of the peak of the blueshifted wing of \lya\ emission. Overall, if the blueshifted \lya\ emission is indeed tracing the outflowing gas, as argued in Section \ref{421}, 
the results above hint at a possible connection between the outflowing gas traced by blueshifted \lya\ and the outflowing neutral gas in these three objects. For instance, \lya\ could be scattered off of the outflowing neutral gas traced by \nadb. However, given the very small number of objects where the \lya\ $-$ \nadb\ comparison was possible, these results may not apply to all objects in the sample.

\subsection{Effect of Dust}
\label{44}

Complex dust-related radiative transfer processes may shape the observed \lya\ emission. In the following, we explore how dust and its distribution within the galaxy may affect the escape of \lya\ photons qualitatively, by examining the relation between the nebular line color excess, \ebv\ (reflecting dust reddening of the line-emitting gas) and \fesc. A more quantitative analysis on this issue requires a careful modelling of the radiative transfer processes, which is beyond the scope of this paper.

\begin{figure}[!htb]   
\epsscale{1.2}
\plotone{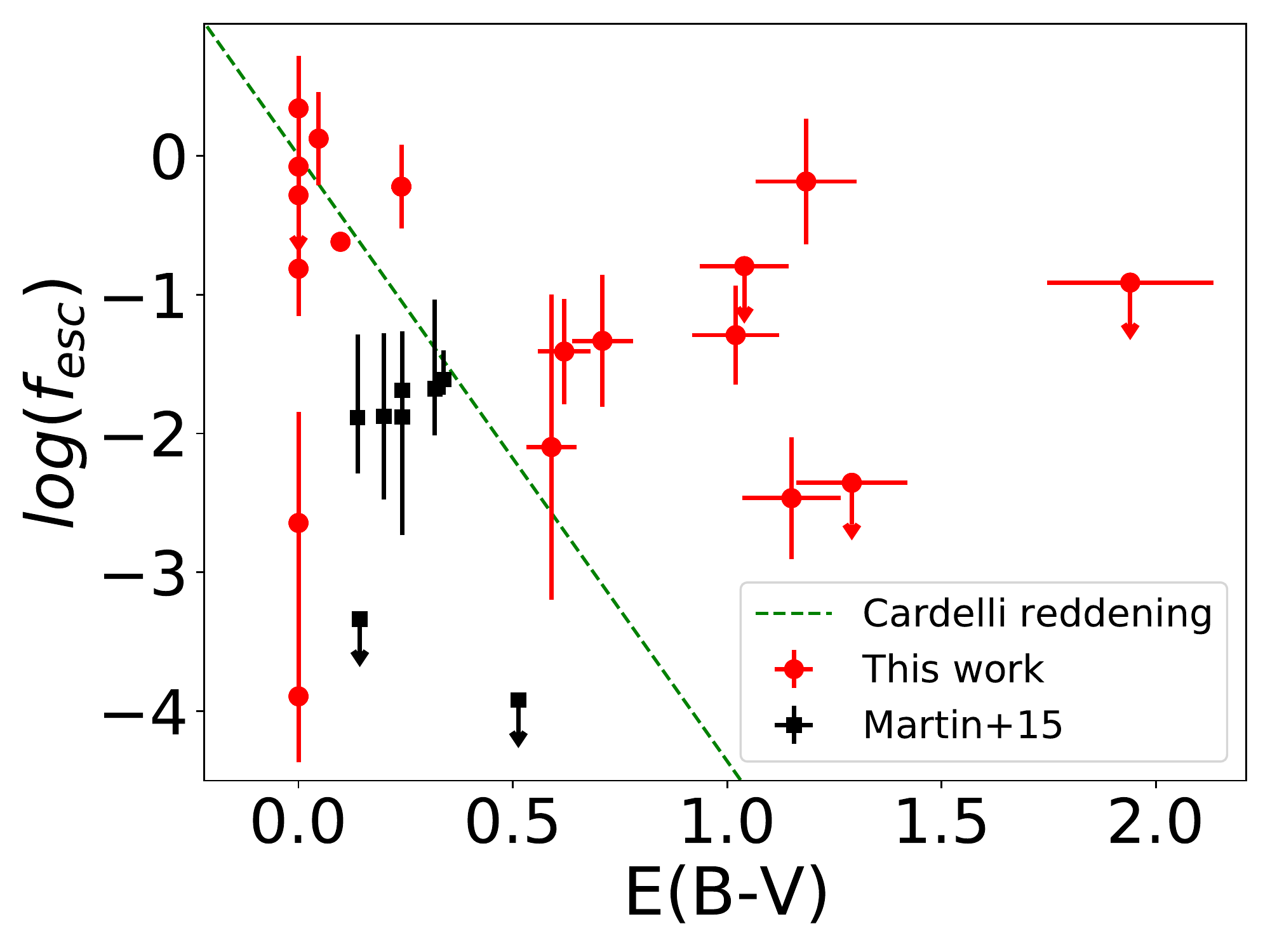}
\caption{\lya\ escape fractions, \fesc, for the AGN-dominated ULIRGs in our sample (red: Type 1 AGN; blue: Type 2 AGN) and the starburst-dominated ULIRGs from M15 (black) as a function of the color excess \ebv. The green dashed line indicates the continuum attenuation at the wavelength of \lya\ transition as given by a \citet{Cardelli1989} reddening curve and the color excess (measured from the Balmer decrement).}
\label{fig:ebvfesc} 
\end{figure}

\begin{figure}[!htb]   
\epsscale{1.2}
\plotone{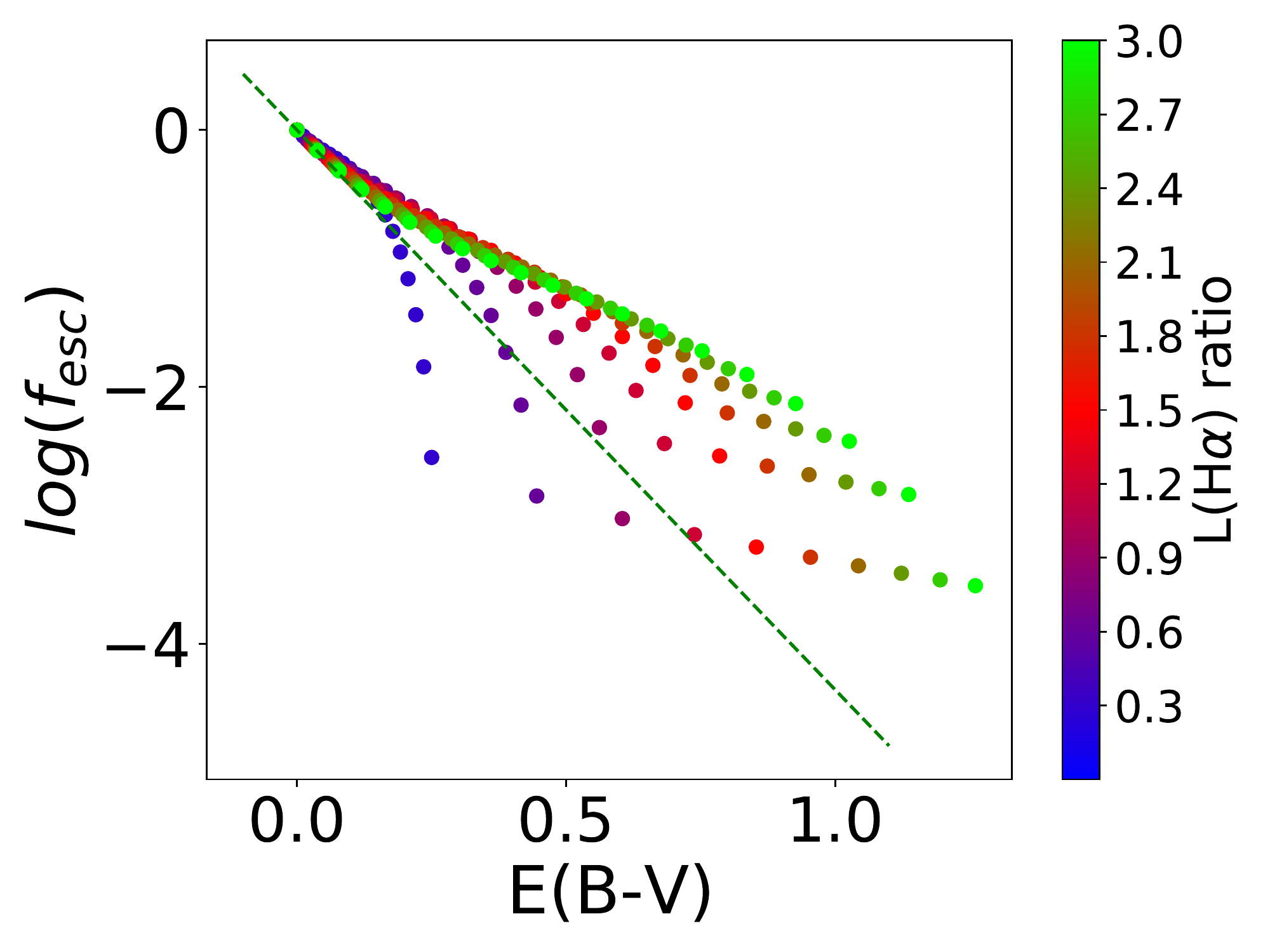}
\caption{\fesc\ as a function of \ebv\ from a simple model reflecting the effect of mixing two ionized gas clouds: one cloud with no dust content and unity \ha\ luminosity, and the other one with increasing dust reddening and \ha\ luminosity. The \ha\ luminosity ratio of the dusty cloud to the dust-free cloud is indicated by the color of the data point. Along each track of data points with the same color, the \ebv\ of the dusty cloud increases from left to right. The green dashed line indicates the same continuum attenuation as shown in Fig. \ref{fig:ebvfesc}.}
\label{fig:ebvsim} 
\end{figure}

In Fig. \ref{fig:ebvfesc}, we plot \fesc\ as a function of \ebv\ for both our objects and the starburst-dominated ULIRGs from M15. Also shown in the figure is the expected \fesc\ given the continuum attenuation at the \lya\ transition derived from the values of \ebv\ based on the \citet{Cardelli1989} reddening curve. 
In the figure, there is a lack of correlation between the \ebv\ and \fesc\  ($p$-value $\simeq$ 0.371 from the Kendall tau test, see Table \ref{tab:lyastatistic}), which is inconsistent with the naive expectation that \fesc\ should decrease with increasing \ebv. In addition, the \lya\ emission in 6 out of the 14 ULIRGs with clear \lya\ detections show \fesc\ much higher than the values expected from the reddening curve. Similar phenomena have been seen in nearby \lya-emitting galaxies with \ebv\ $\gtrsim$ 0.3, where \lya\ emission is on average several times stronger than expected \citep{Scarlata2009,Atek2014,Hayes2014}.

The two phenomena described above may be caused by the fact that our observations reflect the integrated properties of regions with various dust extinction within one galaxy. The observed emission lines come from both the dusty regions (more likely located in the central parts of the systems) and dust-free/less dusty regions such as broad line region of the AGN, off-nuclear H II region, diffused ionized gas, tidally stripped gas during the merging process or even outflowing gas at large distance from the nucleus. The observed \lya\ flux and \ebv\ are the integrated values of all these regions with different weights, which may therefore lead to large scatters in the relation between \ebv\ and \fesc. The higher \fesc\ at a given \ebv\ may be explained if dusty regions bias the overall \ebv\ to a value higher than those of less dusty regions where most of the observed \lya\ emission comes from \citep[this scenario has been pointed out in previous work; e.g.,][]{Atek2014}. To illustrate this point, we build a simple model that considers the total \lya\ and Balmer emission lines from two ionized gas clouds: one with zero dust reddening, and one with increasing dust reddening. The \ha\ luminosity ratios of the two regions are also varied. The overall \ebv\ and \fesc\ integrated over the two gas clouds are shown in Fig. \ref{fig:ebvsim}. At large \ebv, the \fesc\ can be easily larger than the expected values based on the dust extinction curve, adopting case B. In addition, the overall distribution of the data points can also mimic the scatter/non-correlation seen in Fig. \ref{fig:ebvfesc}.

Similarly, the enhancement of \fesc\ may also be caused by the internal geometry of the ISM and the dust distribution within it that, together, change the behavior of \lya\ photons with respect to dust attenuation. At least two possible solutions have been proposed by previous works: 
(i) \citet{Atek2009a} and \citet{Finkelstein2009} have invoked the \citet{Neufeld1991} geometry where dust is
embedded within the H~I clumps of a multiphase ISM, and the scattering of \lya\ photons prevents them from encountering
dust. However, radiative transport simulations show that this
effective ``boost'' of \lya\ is very difficult unless parameters are carefully fine-tuned \citep{Laursen2013,Duval2014}. The predicted increase of \lya\ EW with measured attenuation is not observed in our sample. (ii) Instead, \citet{Scarlata2009} argue for a scenario that is also built upon a clumpy dust distribution. It neither requires preservational scattering as in scenario (i) nor predicts that \lya\ EW increase with \ebv.

Alternatively, the higher-than-expected \fesc\ in the 6 sources may be caused by higher intrinsic \lya\ emission. Such deviation from case B may be caused by the high density gas within the BELR of the AGN, where the collisional excitation enhances the intrinsic \lya\ emission. Nevertheless, following this logic, it may be odd that the higher-than-expected \fesc\ are mostly seen in Type 2 sources rather than Type 1 sources where the \lya\ enhancement due to the dense BELR should be more prominent. For example, in the Type 1 source 3C~273, while the broad \lya\ emission line resembles the broad \ha\ emission line (see Fig. \ref{fig:HaloopAGN}) and is thus likely originated from the BELR, the \fesc\ in 3C~273 is close to the expected value under case B condition.

\subsection{Overall Trends}

\label{45}

In short, the analyses above indicate that the EWs of \lya\ are more closely related to the strength of AGN activity (e.g., \lagn, \fagn), while the blueshifts of \lya\ emission are more closely linked to those of non-resonant optical emission lines tracing ionized gas outflows. It is likely that the AGN activity governs the overall production of \lya\ emission and the outflowing gas generates the blueshifted \lya\ emission.

Additionally, the \lya\ escape fractions, \fesc, tend to be slightly higher in sources with stronger AGN and faster outflows. Nevertheless, the \fesc\ does not correlate with the dust reddening, \ebv, and 6 out of the 14 objects show \fesc\ higher than the expectation from attenuation adopting case B conditions and the extinction curve
from \citet{Cardelli1989}.

\section{O~VI and N~V Absorbers in the ULIRGs}  \label{5}

\subsection{Origin of the O~VI and N~V Absorbers}  \label{51}

We now turn our attention to the O~VI and N~V absorption features. Given the general blueshifts of these features, they are most likely tracing gas driven out of these galaxies by the starburst and/or AGN. Other possible origins include tidal debris from the galaxy merger, intervening circumgalactic medium (CGM), and stellar absorption.

Following Paper I, the characteristics of O~VI and N~V absorption features that indicate a quasar-driven wind origin include (1) line profiles that are broad and smooth compared to the thermal line widths (10--20 \kms\ for N$^{4+}$ and O$^{5+}$) ions at temperature T $\simeq$ 10$^{4.5}$ -- 10$^{5.5}$ K, (2) line ratios of the doublets O~VI $\lambda$1032/$\lambda$1038 and N~V $\lambda$1238/$\lambda$1243 that imply partial covering of the quasar emission source, and perhaps (3) large column densities in these high-ionization ions and/or high O VI/H I column density ratio \citep{Hamann1997b,Hamann1997a,Tripp2008,Hamann2019b}.
 
All blueshifted O~VI and N~V absorbers detected in the 6 ULIRGs meet criterion (1) except for the 2 narrow ($\sigma \sim 30\ km\ s^{-1}$) components out of the 3 components in the best-fit O~VI profile of F13218 and the 2 narrow ($\sigma \sim 20-30\ km\ s^{-1}$) components out of the 4 components in the best-fit N~V profile of F21219\footnote{The narrow line widths of these components do not rule out their outflow origin. AGN outflows can also show combinations of smooth broad components with narrow comps superimposed \citep[e.g.,][]{Yuan2002}}. In addition, the O~VI and N~V absorption features in Z11598 and F13218 meet criterion (2), and the deep BAL features in F01004 and F07599 suggest potentially high column densities in agreement with criterion (3). So, in the end, all 6 ULIRGs show certain O~VI and N~V absorption features consistent with quasar-driven winds. This translates into an apparent outflow incidence rate of $\sim$50\%, close to that of the quasar sample in Paper I. We will expand on this topic in Sec. \ref{6}.

In contrast, the redshifted component of the \ovi\ absorption feature in Z11598 may arise from tidal debris or infalling gas. It is not likely associated with the CGM since such relatively strong N~V absorption line is rarely found in CGM studies at the low redshift \citep[][]{Werk2016}, but with several notable exceptions \citep[][]{Ding2003,Lehner2009,Savage2010,Tripp2011,Muzahid2015,Rosenwasser2018,Gatkine2019,Zahedy2020}. In addition, we cannot rule out the possibility that the two narrow components ($|v|\lesssim 500$ \kms, $\sigma \sim 30\ km\ s^{-1}$) in the O~VI profile of F13218 come from the turbulent ISM and/or CGM of the system.

\begin{deluxetable}{cccc}
\tablecolumns{4}
\tabletypesize{\scriptsize}
\tablecaption{Estimated Outflow Energetics for O~VI and N~V Absorptions} \label{tab:energetics}
\tablehead{
\colhead{Name}  & \colhead{log(d$M$/d$t$)} & \colhead{log(d$p$/d$t$)} & \colhead{log(d$E$/d$t$)}  \\
& [\msunyr] & [g cm s$^{-2}$] & [erg s$^{-1}$] 
}
\colnumbers
\startdata
F01004, O~VI  & $>$-2.6 & $>$32.3 & $>$40.6 \\
F07599, N~V   & $>$-1.0 & $>$34.4 & $>$43.3  \\
Z11598, O~VI  & $>$-3.9 & $>$29.7 & $>$36.7 \\
Z11598, N~V   & $>$-4.0 & $>$29.7 & $>$36.9 \\
F13218, O~VI  & $>$-4.7 & $>$29.3 & $>$36.7 \\
F21219, O~VI  & $>$-4.7 & $>$30.3 & $>$38.6 \\
F21219, N~V   & $>$-4.0 & $>$31.0 & $>$39.3 \\
F23060, N~V    & $>$-5.1 & $>$29.1 & $>$36.6
\enddata
\tablecomments{Column (1): Object and absorption feature names; Column (2)--(4): Estimated lower limits of mass, momentum, and kinetic energy outflow rates as described in Sec. \ref{52}. A radial distance of 0.1 pc, an ionization correction factor of 0.2 \citep[e.g.,][]{Tripp2000}, and a solar metallicity are adopted in the calculation. The absorption features in F01004 and F07599 are broad absorption lines (BAL), and the lower limits 
listed are just a rough estimation from a tentative/experimental single component fit to the BAL.}
\end{deluxetable}

\subsection{Location and Energetics of the Outflows}  \label{52}

Here we start with the constraints on the location of the O~VI and N~V absorbers detected in our ULIRG sample. In Z11598 and F13218 (and perhaps also all other 4 ULIRG with detected absorbers), the O~VI and/or N~V absorption features are deeper than the FUV continuum level, implying that part of the O~VI and N~V emission produced in the BELR is also absorbed. These absorbers are thus located outside the BELR which has a scale of
\begin{eqnarray}
r_{BELR} \simeq 0.1(\frac{\lambda L_{\lambda}(1350 \text{\AA})}{2\times10^{46}\ {\rm erg~s}^{-1}})^{0.55}~{\rm pc}
\end{eqnarray}
\citep[e.g.,][]{Kaspi2005,Kaspi2007,Bentz2013}. The equation (16) above is based on the C~IV-traced BELR size luminosity relation from equation (2) in \citet{Kaspi2007}, and the denominator, $2\times10^{46}\ {\rm erg~s}^{-1}$, corresponds to the $\lambda L_{\lambda}(1350~\text{\AA})$ of 3C~273. This typical scale is also consistent with the VLTI/GRAVITY result on the size of the BELR of 3C~273 \citep[$r \simeq$ 0.12$\pm$ 0.03 pc;][]{GRAVITY2018}. 

As for the upper limit on the radial distance of the outflows, qualitative constraints exist for the two objects with evidence of partial covering, Z11598 and F21219. The distance of the outflows in these two sources cannot be significantly larger than the size of the region where the continuum radiation comes from. The evidence of partial covering in these two objects also sets interesting limits on the size of the absorbing cloud: If the absorbing material is a single uniform cloud with a 100\% filling factor, the cloud size should thus be $\lesssim$0.1 pc. However, if the absorbing material is made of multiple clouds, the partial covering may instead reflect the small filling factor of the clouds, and the sizes of individual clouds may be much smaller than 0.1 pc.

The mass, momentum, and kinetic energy outflow rates of the highly ionized outflow detected in our sample are estimated using the following equations:

\begin{eqnarray}
\frac{dE}{dt} = 3.3\times10^{39} (\frac{Q}{0.15})(\frac{N_{H}}{10^{22} cm^{-2}})(\frac{R}{0.1 pc})v_{1000}^3 \\ 
\frac{dp}{dt} = 6.7\times10^{32}(\frac{Q}{0.15})(\frac{N_{H}}{10^{22} cm^{-2}})(\frac{R}{0.1 pc})v_{1000}^2 \\  
\frac{dm}{dt} = 0.003(\frac{Q}{0.15})(\frac{N_{H}}{10^{22} cm^{-2}})(\frac{R}{0.1 pc})v_{1000}    
\label{eqn:KE2} 
\end{eqnarray}

In the equations above, $Q$ is an approximate global outflow covering factor quoted from Paper I, based on the incidence of mini-BALs in the SDSS quasars \citep{Trump2006,Knigge2008,Gibson2009b,Allen2011}; $R$ is the radial distance of the outflow and the value of 0.1 pc is a place-holder adopted for illustrative purposes; $N_H$ is the column density of hydrogen, and $v_{1000}$ is the outflow velocity in units of 1000 \kms\ obtained from the Voigt profile fits described in Sec. \ref{332}.

Note that the Voigt profile fits are highly uncertain for the O~VI BAL in F01004, and the N~V BAL in F07599, as stated in Sec. \ref{332}. Therefore, only rough estimations of the ion column densities may be obtained from the fits. For Z11598 and F13218, the O~VI and N~V absorption features are also saturated despite their much narrower line widths, so the corresponding ion column densities from the fits are also uncertain. Nevertheless, the results derived from the Voigt profile fits are consistent with those derived from analyzing the absorption doublet with partial covering model, as described in Sec. \ref{331} (see Table \ref{tab:abs}). For the N~V absorption in F21219, the ion column density acquired from the analysis with partial covering model is a bit higher than that from the Voigt profile fits, which is expected as the partial covering factor, while fixed to unity in the Voigt profile fit, should be less than unity.

Next, the ion column densities are converted into hydrogen column densities. The metal abundance and ionization correction factor needed for the conversion can, in principle, be determined from elaborate photoionization modeling when multiple absorbers from both high- and low-ionization species are present \citep[e.g.,][]{Arav2013,Haislmaier2021}, but this information is not available for our objects. For simplicity, we adopt a solar abundance \citep[while super-solar metallicities are also reported in the literature; e.g.,][]{Moe2009} and a ionization correction factor of 0.2 \citep[which is a conservative upper limit reported in the literature; e.g.,][]{Tripp2000} in the calculations. As discussed at the beginning of this section, the radial distances of these absorbers are largely unconstrained, other than the fact that the absorbers in Z11598 and F13218 are located outside of the BELR ($r\gtrsim0.1$ pc). Adopting these aforementioned values, the obtained mass, momentum, and energy outflow rates are likely lower limits as reported in Table \ref{tab:energetics}. In general, they are modest ($\lesssim$1\%) compared with the star formation rates, AGN radiation momenta, and AGN luminosities of the systems. However, for the highly saturated O~VI BAL in F01004 and N~V BAL in F07599, the column density of the outflowing gas and thus the outflow energetics are probably severely underestimated.

\section{O VI and N V Absorbers in the Combined ULIRG $+$ Quasar Sample}  \label{6}

In this section, we explore the properties of the highly ionized O~VI and N~V absorbers along the ULIRG-QSO merger sequence \citep[e.g., see][and references therein]{Sanders1988,Hopkins2009,Veilleux2009} by combining the 11 ULIRGs with high enough continuum S/N to allow for O~VI and N~V absorption detections (ULIRG F15250 is excluded from the analysis given that the N~V absorption feature is highly uncertain due to contamination from geo-coronal emission) and the 30 quasars presented in Paper I (3 of the 33 quasars in Paper I overlap with the ULIRG sample and are thus already included). In total, 6 ULIRGs (see Sec. \ref{33}) and 17 quasars (see Section 7.1 in Paper I) show O~VI and N~V absorbers indicative of quasar-driven outflows. 
In addition, to be consistent with the analyses in Paper I and to maximize the sample size for better statistics, we also include the narrow O~VI and/or N~V absorption features in the 3 quasars from Paper I that do not meet our criteria for quasar-driven outflows. The absorption features in these three objects are relatively narrow (\savg $\simeq$ 10--30 \kms) and are redshifted in two of these objects.

\subsection{Overall Incidence Rates and Regressions}
\label{61}

Based on the measurements listed in Table \ref{tab:abs}, while quite uncertain due to the small sample size, we can estimate the incidence rate of absorption features in the ULIRG-only sample. Adopting the $\beta$ distribution \citep{Cameron2011} used in Paper I, we obtain an incidence rate of $\sim$55\% (1-$\sigma$ range: $\sim$40\%--68\%) for the detection of O~VI or N~V or both absorption features. This is similar to the rate of $\sim$61\% (1-$\sigma$ range: $\sim$52\%--68\%) obtained in Paper I, which is based on the quasar sample alone. In the combined ULIRG $+$ quasar sample, the overall incidence rate of O~VI or N~V or both absorption features is $\sim$63\% (1-$\sigma$ range: $\sim$55\%--70\%).

Next, we explore how the incidence rates and properties (velocity-integrated EWs, \weq, depth-weighted velocities, \vavg, and velocity dispersions, \savg) of the O~VI and N~V absorption features depend on the AGN and host galaxy properties of our objects, adopting the $\beta$\ distribution above and regressions described below. 
Following Paper I, we apply linear regressions adopting the Bayesian model in LINMIX\_ERR \citep{Kelly2007}. We use the Metropolis-Hastings sampler and a single Gaussian to represent the distribution of the parameters. LINMIX\_ERR allows censored values for dependent variables ($y$), which is the case for \weq. The only exceptions involve \NHxray, where both $x$ and $y$ values are censored, in which case we adopt the Kendall tau correlation test described at the beginning of Sec. \ref{4}. 
For both methods, we calculated the correlation coefficients $r$ and their 1-$\sigma$ errors. A perfect correlation gives $r=1$ and a perfect anti-correlation gives an $r=-1$. A sample with no correlation at all gives $r=0$. In addition, we have computed the significance of a correlation, P($r$), as the fraction of correlation coefficients $r \le 0$ ($r \ge 0$) for a positive (negative) correlation\footnote{From LINMIX\_ERR, it is technically difficult to calculate the classic $p$-value in null hypothesis significance testing. Nevertheless, our definition of P($r$) can describe the significance of the correlation similarly. Like $p$-value $<0.05$, P($r$) $<0.05$ also indicates a statistically significant correlation: it suggests that the possibility is 95\% for the correlation coefficient $r$ to be larger (smaller) than 0, in the case of a positive (negative) correlation. Note that this P($r$) was called $p$-value in Paper I, which is abandoned in this Paper II to avoid ambiguity.}. For LINMIX\_ERR, the distribution of $r$ are acquired from the posterior distribution, while for pymccorrelation, they are obtained from the Monte Carlo perturbations.

For the regressions, we do not take O~VI and N~V data as independent measurements. Therefore, when measurements are available for both doublets in a given source, we take the average of the measurements (either detection or limit) from the two lines. If only one line is detected, we use the measurement for the detection. In addition, when multiple X-ray measurements exist for a source, we take the average of them. Errors in \lbol, \lir, $L_{\rm IR}/L_{\rm BOL}$, $L_{\rm FIR}/L_{\rm BOL}$, and \aux\ are unknown or largely uncertain, so for the regressions we fix their errors to $\pm{0.1}$~dex \footnote{Note that the results of the regression analysis do not change even if we adopt larger errors of up to $\pm{0.5}$ dex.}. For \luv, we ignore the negligible statistical measurement errors. As examples, the final data points adopted for the regressions are shown in the inset panels of Fig. \ref{fig:aux} and \ref{fig:logNH}.

\begin{deluxetable}{cc cc}
\tablecolumns{4}
\tablecaption{Incidence Rates of O VI and N V Absorbers in the Combined ULIRG $+$ Quasar Sample \label{tab:incidence}}
\tablehead{\colhead{Line} &  \colhead{Detection} & \colhead{Total} & \colhead{Fraction (1-$\sigma$ range)} }
\colnumbers
\startdata
\multicolumn{4}{c}{All Quasars and ULIRGs}   \\     \hline
O VI & 17 & 25 & 0.68(0.58--0.76)  \\ 
N V & 17 & 34 & 0.50(0.42--0.58)  \\ 
Both & 8 & 18 & 0.44(0.34--0.56)  \\ 
Any & 26 & 41 & 0.63(0.55--0.70)  \\ 
\hline
\multicolumn{4}{c}{\lagn\ $<$ 12.1}   \\           
\hline
O VI & 9 & 15 & 0.60(0.47--0.71)  \\ 
N V & 15 & 27 & 0.56(0.46--0.64)  \\ 
Both & 6 & 10 & 0.60(0.44--0.73)  \\ 
Any & 13 & 21 & 0.62(0.51--0.71)  \\ 
\hline
\multicolumn{4}{c}{\lagn\ $\ge$ 12.1}   \\     
O VI & 7 & 9 & 0.78(0.59--0.86)  \\ 
N V & 1 & 5 & 0.20(0.12--0.45)  \\ 
Both & 2 & 8 & 0.25(0.16--0.44)  \\ 
Any & 12 & 19 & 0.63(0.51--0.72)  \\ 
\hline
\multicolumn{4}{c}{\NHxray $<$ 10$^{22}$ cm$^{-2}$} \\
\hline
O VI & 6 & 14 & 0.43(0.31--0.56)  \\ 
N V & 5 & 16 & 0.31(0.22--0.44)  \\ 
Both & 1 & 8 & 0.12(0.08--0.32)  \\ 
Any & 3 & 12 & 0.25(0.17--0.41)  \\ 
\hline
\multicolumn{4}{c}{\NHxray $\ge$ 10$^{22}$ cm$^{-2}$} \\
\hline
O VI & 9 & 9 & 1.00(0.83--0.98)  \\ 
N V & 10 & 15 & 0.67(0.53--0.76)  \\ 
Both & 4 & 6 & 0.67(0.45--0.79)  \\ 
Any & 16 & 21 & 0.76(0.65--0.83)  \\ 
\hline
\multicolumn{4}{c}{X-ray to FUV Spectral Index \aux $<$ -1.3} \\
\hline
O VI & 10 & 13 & 0.77(0.62--0.85)  \\ 
N V & 9 & 17 & 0.53(0.41--0.64)  \\ 
Both & 4 & 8 & 0.50(0.34--0.66)  \\ 
Any & 15 & 17 & 0.88(0.76--0.92)  \\ 
\hline
\multicolumn{4}{c}{X-ray to FUV Spectral Index \aux $\ge$ -1.3} \\
\hline
O VI & 1 & 2 & 0.50(0.25--0.75)  \\ 
N V & 4 & 9 & 0.44(0.30--0.60)  \\ 
Both & 3 & 8 & 0.38(0.25--0.55)  \\ 
Any & 9 & 20 & 0.45(0.35--0.56)  \\
\enddata
\tablecomments{
Column (1): Feature(s) used in the statistical analysis. “Both” means both O~VI and N~V doublets and “Any” means either O~VI or N~V doublet or both; Column (2): Number of objects with detected absorption features; Column (3): Total number of objects with enough continuum S/N ratios to allow for detections of corresponding absorption features; Column (4): Fraction of objects with detected absorption features. The two numbers in parentheses indicate the 1-$\sigma$\ range (68\% probability) of the fraction of objects with detected absorption features, computed from the $\beta$\ distribution \citep{Cameron2011}.}
\end{deluxetable} 

\subsection{Dependence on the X-ray Properties} \label{62}

The dependence of the incidence rates on several primary AGN/host galaxy properties are listed in Table \ref{tab:incidence}. The regression results between the AGN/host galaxy properties and the properties of the absorption features (\weq, \vavg, and \savg) are listed in Table \ref{tab:regressions}. In brief, we find that the incidence rates of the absorption features do not depend on the AGN/host galaxy properties, such as the bolometric luminosities, AGN luminosities, AGN fractions, IR luminosities, FIR luminosities, FIR-to-bolometric luminosity ratios, and FUV luminosities (note that only a few key quantities are listed in Table \ref{tab:incidence}).  Similarly, no statistically significant (P($r$) $< 0.05$, $|r| >> 0$) trends are seen between these AGN/galaxy properties and the properties of the absorption features. These negative results are largely consistent with those in Paper I based on the quasar-only sample.

Nevertheless, as discussed in Paper I, the incidence rate and properties of the absorption features do correlate with several X-ray properties of the sources. We examine these trends with the ULIRG $+$ quasar sample below. In general, we find that the incidence rate of the absorption features is higher in the X-ray weak (relative to their UV luminosities, as quantified by \aux\ described in Sec. \ref{232}) or absorbed sources (Table \ref{tab:incidence}). We also find that the \weq, \vavg\ and \savg\ of the absorption features correlate with \aux\ (see Table \ref{tab:regressions}).

\begin{figure*}[!htb]   
\epsscale{4}
\begin{minipage}[t]{0.33\textwidth}
\includegraphics[width=\textwidth]{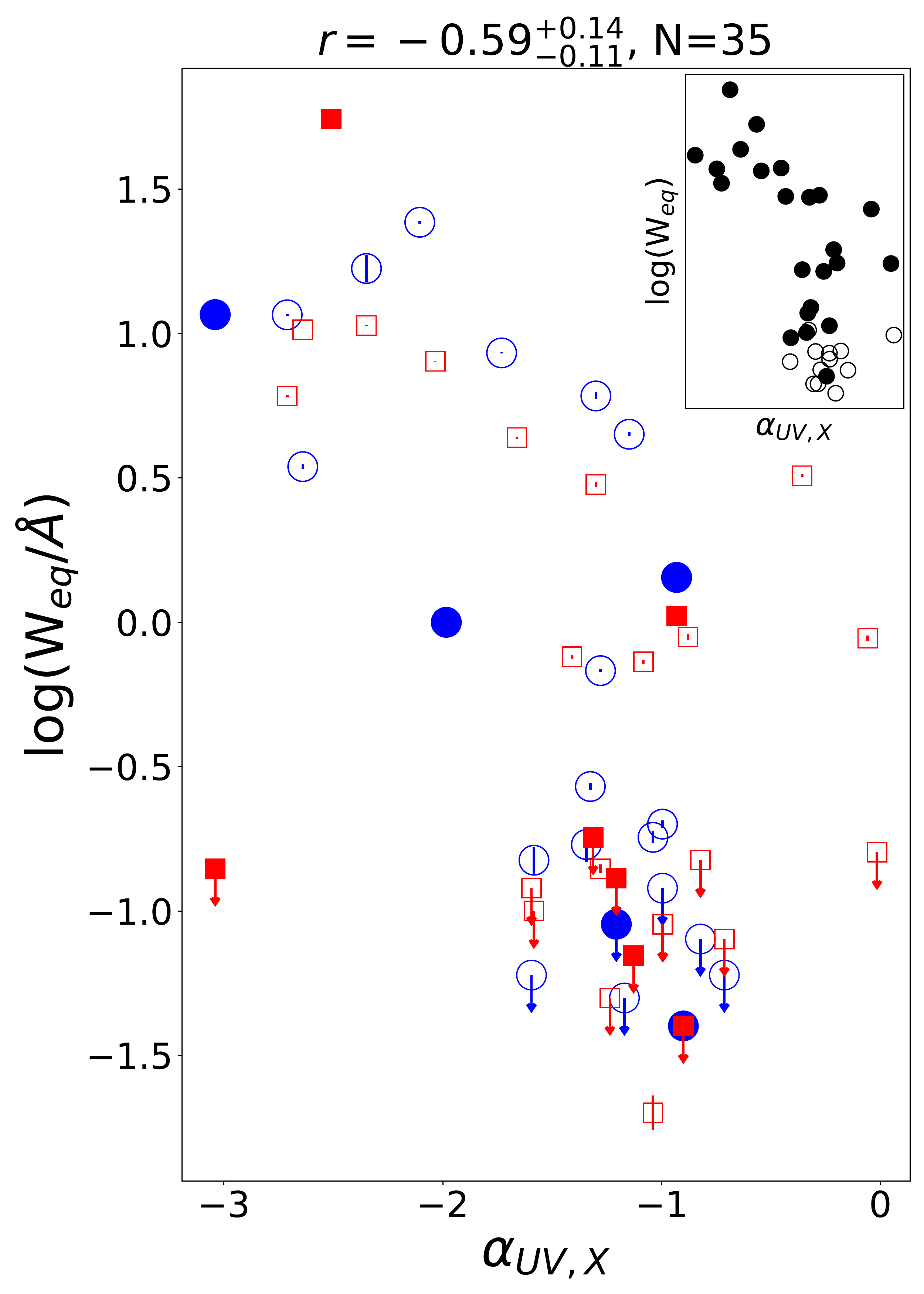}
\end{minipage}
\begin{minipage}[t]{0.33\textwidth}
\includegraphics[width=\textwidth]{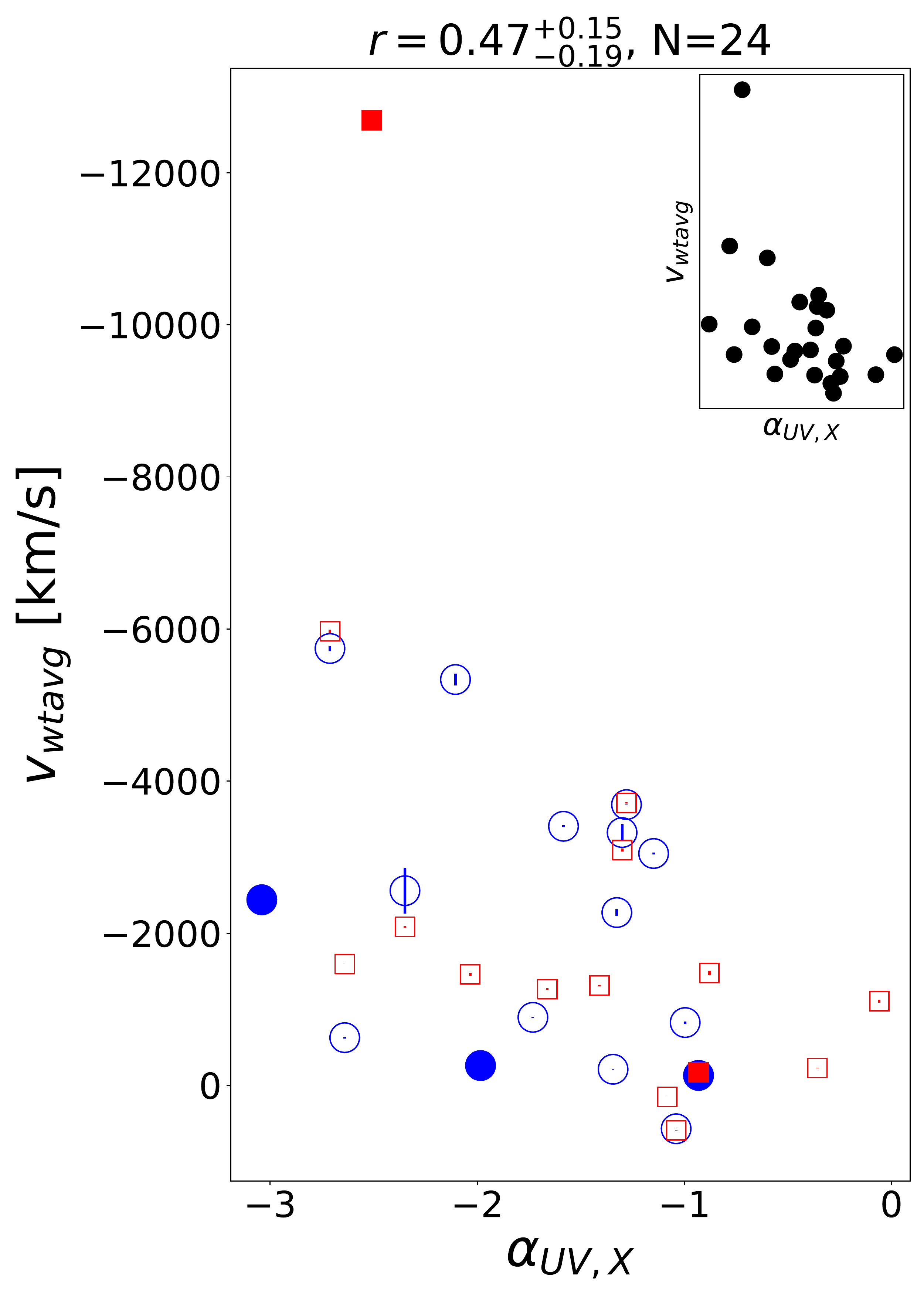}
\end{minipage}
\begin{minipage}[t]{0.33\textwidth}
\includegraphics[width=\textwidth]{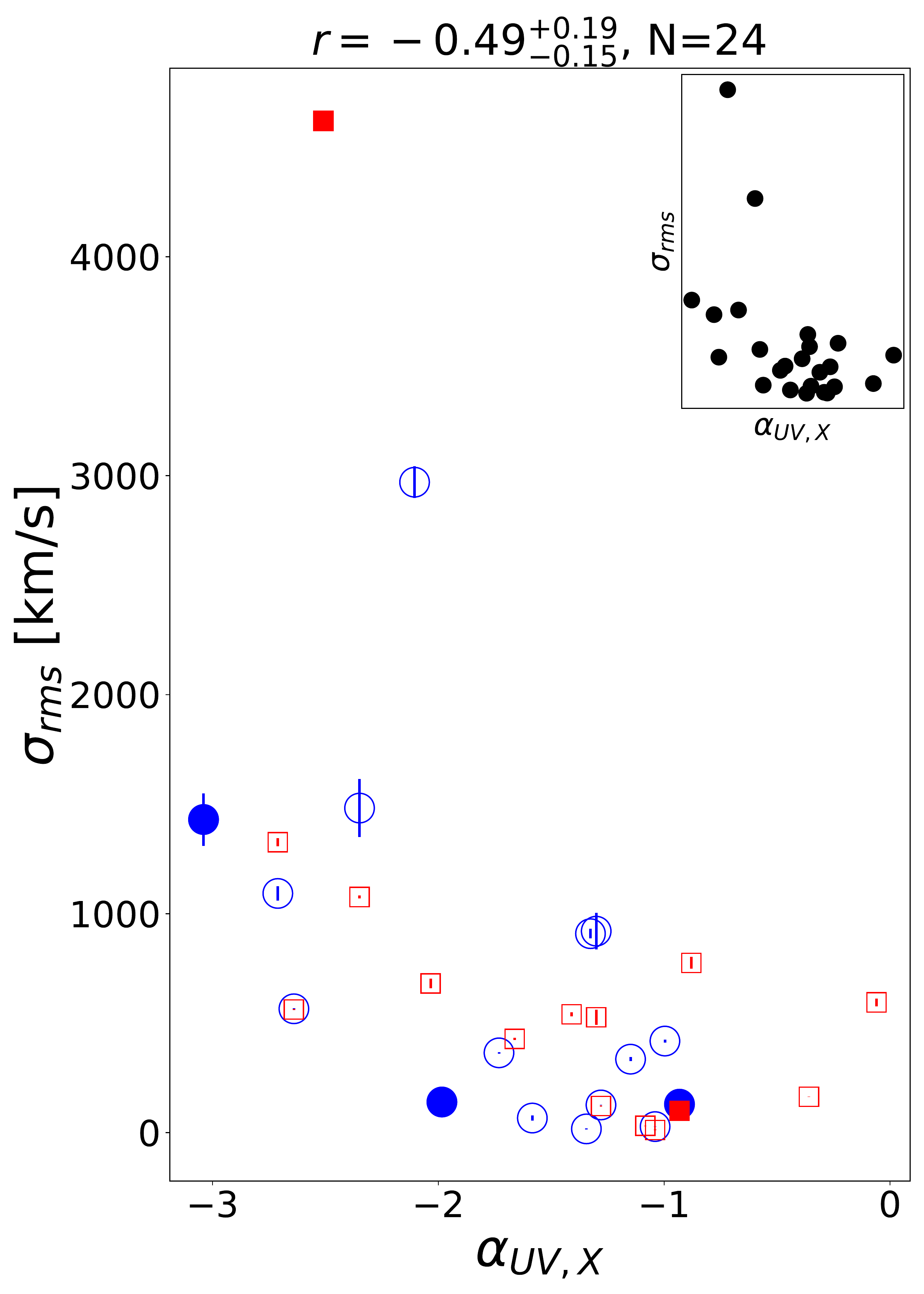}
\end{minipage}
\caption{The velocity-integrated EWs, \weq\ (left), depth-weighted velocities, \vavg\ (middle) and depth-weighted velocity dispersions, \savg\ (right) for the O~VI (blue circle) or N~V (red diamond) absorption features in ULIRGs (filled symbols) and quasars (hollow symbols) as function of \aux. The actual data points used in the regressions, in which O~VI and N~V quantities and/or X-ray measurements are averaged for a given source, are shown in each inset panel. The solid symbols are detections, while the open symbols are censored values in one or both quantities plotted. The regression results (correlation coefficients $r$ with 1-$\sigma$ errors, and number of data points in the inset panel, N; see the 3rd paragraph in Sec. \ref{61} for more information) are shown above each panel.}
\label{fig:aux}
\end{figure*}

The incidence rates of the absorption features (either O~VI or N~V or both) are 88\% (1-$\sigma$ range: 76\%--92\%) for the objects with \aux\ $<-$1.3\footnote{The value $-$1.3 is chosen since (1) it divides the sample into two groups with approximately equal numbers of objects; (2) it corresponds to an \aox\ of $\sim -$1.7 as adopted in Paper I.} and 45\% (1-$\sigma$ range: 35\%--56\%) for those with \aux\ $\ge -$1.3. Such difference is statistically significant with a $p$-value of $\sim$0.014, adopting the scipy.stats implementation of the Fisher exact test with a null hypothesis that galaxies with \aux\ $<-$ 1.3 and \aux\ $\ge-$1.3 are equally likely to show O~VI or N~V absorption features. Additionally, as shown in Fig. \ref{fig:aux}, the objects with lower \aux\ tend to have larger \weq, smaller \vavg, and larger \savg, where the regressions give correlation coefficients $r$ of $-0.59^{+0.15}_{-0.11}$, $0.48^{+0.15}_{-0.19}$ and $-0.50^{+0.18}_{-0.16}$, respectively. These results in general confirm or strengthen those from Paper I based on the quasar-only sample and \aox: the incidence rates in Paper I were found to be 75\% (1-$\sigma$ range: 59\%--83\%) versus 55\% (1-$\sigma$ range: 44\%--65\%) for sources with \aox$< -$1.6 and \aox$\geq -$1.6 (This difference was not considered significant since the $p$-value was $\sim$0.45, adopting the same Fisher test on the difference between the incidence rates), and the correlation coefficients $r$ for the trends between the \aox\ and the \weq, \vavg, and \savg, were $-0.62^{+0.17}_{-0.13}$, $0.31^{+0.21}_{-0.24}$ and $-0.55^{+0.20}_{-0.15}$, respectively.

\begin{figure*}[!htb]   
\epsscale{4}
\begin{minipage}[t]{0.33\textwidth}
\includegraphics[width=\textwidth]{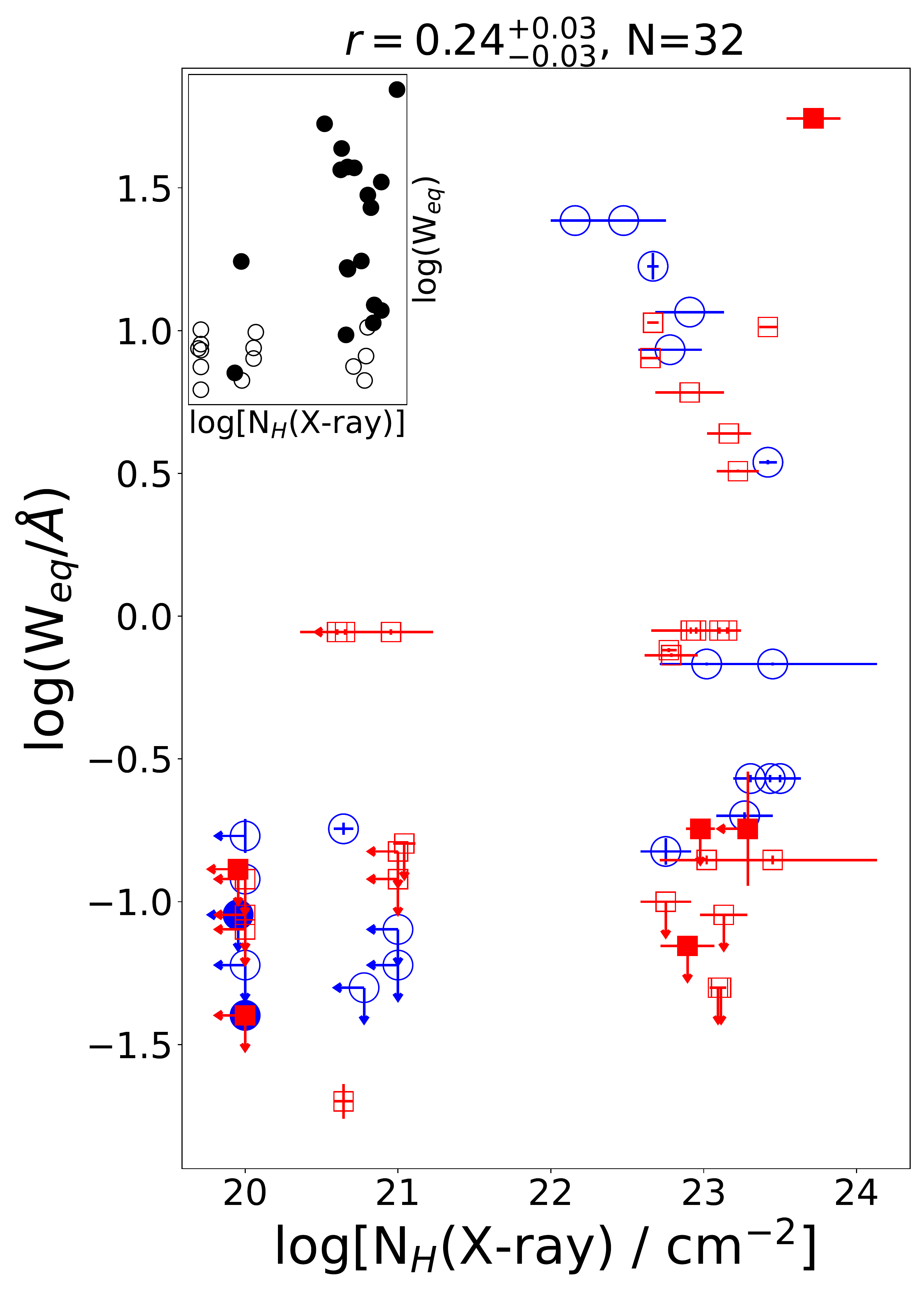}
\end{minipage}
\begin{minipage}[t]{0.33\textwidth}
\includegraphics[width=\textwidth]{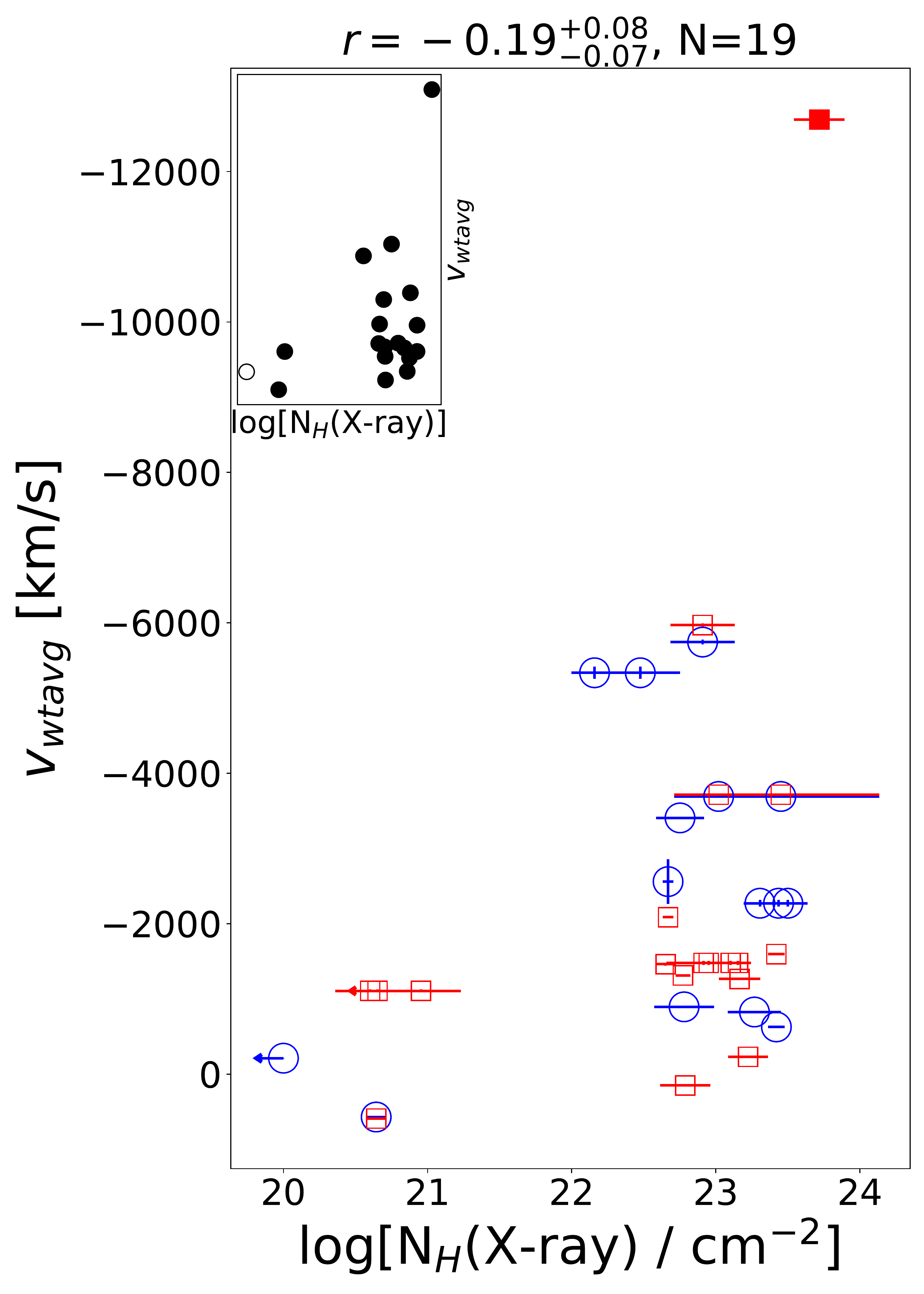}
\end{minipage}
\begin{minipage}[t]{0.33\textwidth}
\includegraphics[width=\textwidth]{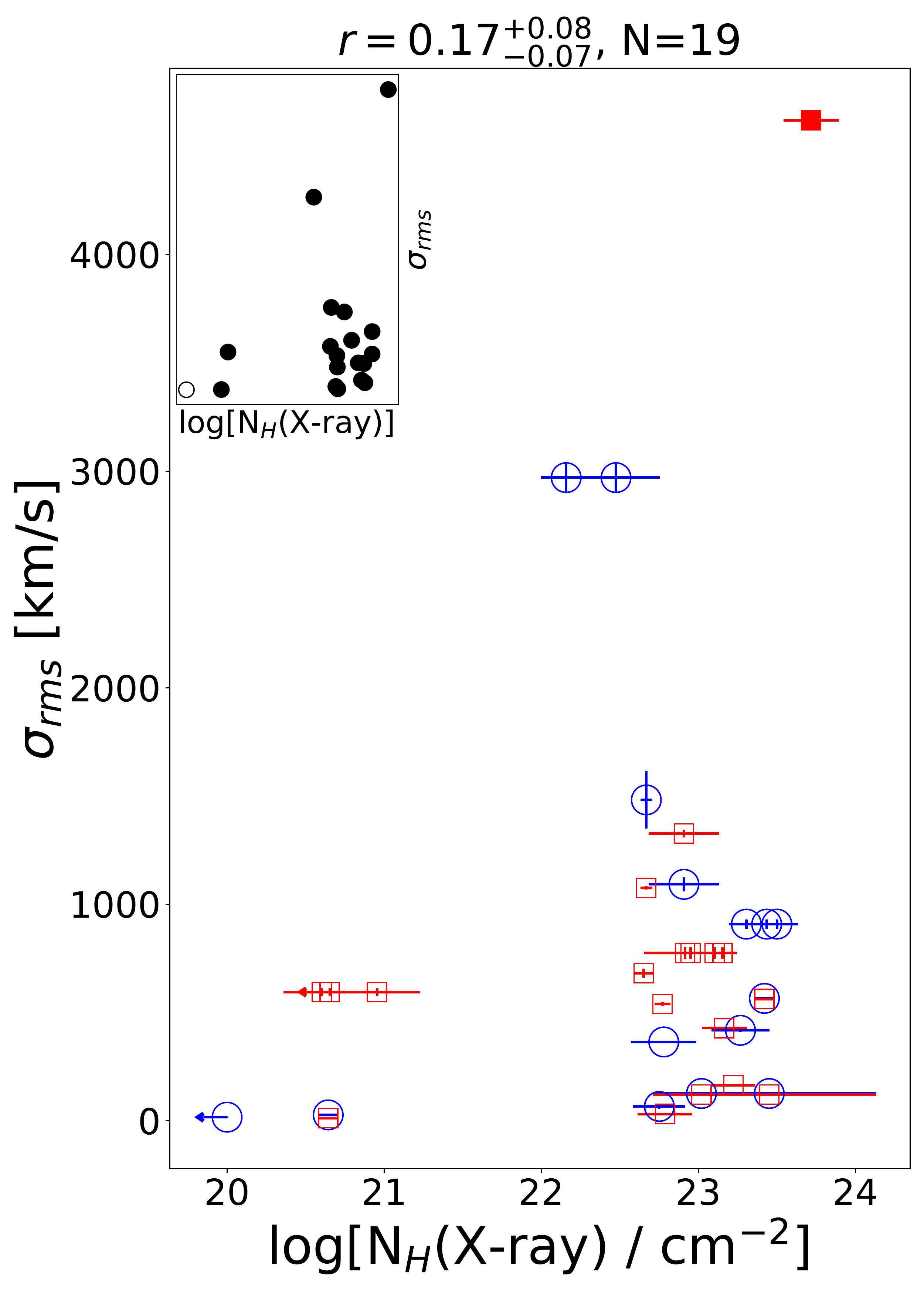}
\end{minipage}
\caption{Same as Fig. \ref{fig:aux} but as function of the logarithm of the X-ray absorbing column densities \NHxray.}
\label{fig:logNH}
\end{figure*}

Similarly, the incidence rate of absorption features for sources with \NHxray\ $>$ 10$^{22}$ cm$^{-2}$ is 76\% (1-$\sigma$ range: 65\%--83\%), whereas the rate for those with lower \NHxray\ is 25\% (1-$\sigma$ range: 17\%--41\%). This result is almost identical to that obtained in Paper I. Moreover, as shown in Fig. \ref{fig:logNH}, the \weq\ of the absorption features may be higher in objects with higher \NHxray, which is consistent with the result from Paper I: The correlation coefficients $r$ are $0.24^{+0.03}_{-0.03}$ for the ULIRG $+$ quasar sample and $0.19^{+0.03}_{-0.03}$ for the quasar-only sample, respectively. As for the \vavg\ and \savg\ of the absorption features, their lack of dependence on the \NHxray\ found in Paper I remains.

Furthermore, other trends seen among the quasars in Paper I that involve the X-ray luminosities are not statistically significantly anymore when the ULIRGs are included. These include the trend between \weq\ of the absorption features and the hard X-ray (2--10 keV) luminosities (r changes from $-0.51^{+0.20}_{-0.15}$ to $-0.09^{+0.24}_{-0.23}$), those between \vavg\ and \savg\ of the absorption features and the soft X-ray (0.5--2 keV) to ``total'' X-ray (0.5--10 keV) luminosity ratios (r change from $-0.69^{+0.13}_{-0.20}$ and $-0.54^{+0.24}_{-0.18}$ to $-0.18^{+0.26}_{-0.24}$ and $-0.06^{+0.27}_{-0.27}$, respectively), and those between \vavg\ and \savg\ and the X-ray to bolometric luminosity ratios (r change from $0.64^{+0.14}_{-0.20}$ and $-0.61^{+0.21}_{-0.15}$ to $-0.26^{+0.23}_{-0.21}$ and $0.32^{+0.20}_{-0.23}$, respectively).

\begin{deluxetable}{ccccc}
\tabletypesize{\tiny}
\tablecolumns{5}
\tablecaption{Regression Analysis on the O~VI and N~V Absorbers in the Combined ULIRG$+$Quasar Sample\label{tab:regressions}}
\tablehead{\colhead{$y$} & \colhead{$x$} & \colhead{$N$} &
  \colhead{P(r)} & \colhead{$r$} }
\colnumbers
\startdata
        $W_{\rm eq}$  &          {                                 log($L_{\rm BOL}/L_\sun$)}   &         39 &      0.157 &     $-0.18_{- 0.17}^{+ 0.18}$ \\
        $W_{\rm eq}$  &                              log[$\lambda L_{1125}/{\rm erg~s}^{-1}$]   &         36 &      0.210 &     $-0.16_{- 0.18}^{+ 0.20}$ \\
        $W_{\rm eq}$  &                                                          AGN fraction   &         38 &      0.126 &     $-0.40_{- 0.33}^{+ 0.35}$ \\
        $W_{\rm eq}$  &                                             log($L_{\rm AGN}/L_\sun$)   &         38 &      0.088 &     $-0.25_{- 0.17}^{+ 0.18}$ \\
        $W_{\rm eq}$  &\underline{                                         \aux            }    &         35 &      $<$0.001 &  $-0.59_{- 0.11}^{+ 0.15}$ \\
        $W_{\rm eq}$  &                                         log($L_{\rm IR}/L_{\rm BOL}$)   &         39 &      0.058 &     $ 0.33_{- 0.20}^{+ 0.18}$ \\
        $W_{\rm eq}$  &                                        log($L_{\rm FIR}/L_{\rm BOL}$)   &         38 &      0.225 &     $-0.13_{- 0.18}^{+ 0.18}$ \\
        $W_{\rm eq}$  &{                           log[$N({\rm H})/{\rm cm}^{-2}$]}   &         32 &      $<$0.001 &     $ 0.24_{-0.03}^{+0.03}$   \\            
        $W_{\rm eq}$  &                                                              $\Gamma$   &         34 &      0.202 &     $ 0.17_{- 0.20}^{+ 0.19}$ \\
        $W_{\rm eq}$  &                            log[$L(0.5-2~{\rm keV})/{\rm erg~s}^{-1}$]   &         30 &      0.256 &     $-0.14_{- 0.20}^{+ 0.21}$ \\
        $W_{\rm eq}$  &          {                 log[$L(2-10~{\rm keV})/{\rm erg~s}^{-1}$]}   &         34 &      0.327 &     $-0.09_{- 0.19}^{+ 0.19}$ \\
        $W_{\rm eq}$  &                           log[$L(0.5-10~{\rm keV})/{\rm erg~s}^{-1}$]   &         30 &      0.220 &     $-0.15_{- 0.19}^{+ 0.20}$ \\
        $W_{\rm eq}$  &                         log[$L(0.5-2~{\rm keV})/L(0.5-10~{\rm keV})$]   &         30 &      0.357 &     $-0.09_{- 0.23}^{+ 0.24}$ \\
        $W_{\rm eq}$  &                                log[$L(0.5-10~{\rm keV})/L_{\rm BOL}$]   &         30 &      0.243 &     $-0.15_{- 0.20}^{+ 0.21}$ \\
         \vavg        &          {                                 log($L_{\rm BOL}/L_\sun$)}   &         26 &      0.168 &     $-0.21_{- 0.21}^{+ 0.22}$ \\
         \vavg        &                              log[$\lambda L_{1125}/{\rm erg~s}^{-1}$]   &         24 &      0.225 &     $-0.17_{- 0.20}^{+ 0.22}$ \\
         \vavg        &                                                          AGN fraction   &         25 &      0.186 &     $ 0.43_{- 0.49}^{+ 0.37}$ \\
         \vavg        &                                             log($L_{\rm AGN}/L_\sun$)   &         25 &      0.142 &     $-0.25_{- 0.20}^{+ 0.23}$ \\
         \vavg        &\underline{                                         \aux             }   &         24 &      0.011 &     $ 0.48_{- 0.19}^{+ 0.15}$ \\
         \vavg        &                                         log($L_{\rm IR}/L_{\rm BOL}$)   &         26 &      0.407 &     $ 0.07_{- 0.28}^{+ 0.29}$ \\
         \vavg        &                                        log($L_{\rm FIR}/L_{\rm BOL}$)   &         26 &      0.280 &     $-0.13_{- 0.21}^{+ 0.21}$ \\
         \vavg        &          {                           log[$N({\rm H})/{\rm cm}^{-2}$]}   &         19 &      0.012 &     $-0.19_{-0.07}^{+0.08}$   \\
         \vavg        &                                                              $\Gamma$   &         22 &      0.402 &     $-0.06_{- 0.24}^{+ 0.24}$ \\
         \vavg        &                            log[$L(0.5-2~{\rm keV})/{\rm erg~s}^{-1}$]   &         21 &      0.126 &     $-0.27_{- 0.20}^{+ 0.23}$ \\
         \vavg        &          {                 log[$L(2-10~{\rm keV})/{\rm erg~s}^{-1}$]}   &         22 &      0.127 &     $-0.26_{- 0.20}^{+ 0.23}$ \\
         \vavg        &                           log[$L(0.5-10~{\rm keV})/{\rm erg~s}^{-1}$]   &         21 &      0.119 &     $-0.27_{- 0.21}^{+ 0.23}$ \\
         \vavg        &                         log[$L(0.5-2~{\rm keV})/L(0.5-10~{\rm keV})$]   &         21 &      0.244 &     $-0.18_{- 0.24}^{+ 0.26}$ \\
         \vavg        &                                log[$L(0.5-10~{\rm keV})/L_{\rm BOL}$]   &         21 &      0.139 &     $-0.26_{- 0.21}^{+ 0.23}$ \\
         \savg        &          {                                 log($L_{\rm BOL}/L_\sun$)}   &         26 &      0.162 &     $ 0.21_{- 0.21}^{+ 0.19}$ \\
         \savg        &                              log[$\lambda L_{1125}/{\rm erg~s}^{-1}$]   &         24 &      0.449 &     $-0.03_{- 0.22}^{+ 0.24}$ \\
         \savg        &                                                          AGN fraction   &         25 &      0.208 &     $-0.33_{- 0.34}^{+ 0.40}$ \\
         \savg        &                                             log($L_{\rm AGN}/L_\sun$)   &         25 &      0.134 &     $ 0.25_{- 0.22}^{+ 0.20}$ \\
         \savg        &\underline{                                         \aux             }   &         24 &      0.005 &     $-0.50_{- 0.16}^{+ 0.18}$ \\
         \savg        &                                         log($L_{\rm IR}/L_{\rm BOL}$)   &         26 &      0.449 &     $ 0.04_{- 0.28}^{+ 0.28}$ \\
         \savg        &                                        log($L_{\rm FIR}/L_{\rm BOL}$)   &         26 &      0.241 &     $ 0.14_{- 0.21}^{+ 0.20}$ \\
         \savg        &          {                           log[$N({\rm H})/{\rm cm}^{-2}$]}   &         19 &      $<$0.001 &     $ 0.17_{-0.07}^{+0.08}$ \\ 
         \savg        &                                                              $\Gamma$   &         22 &      0.437 &     $ 0.04_{- 0.23}^{+ 0.22}$ \\
         \savg        &                            log[$L(0.5-2~{\rm keV})/{\rm erg~s}^{-1}$]   &         21 &      0.082 &     $ 0.32_{- 0.23}^{+ 0.20}$ \\
         \savg        &          {                 log[$L(2-10~{\rm keV})/{\rm erg~s}^{-1}$]}   &         22 &      0.082 &     $ 0.33_{- 0.23}^{+ 0.19}$ \\
         \savg        &                           log[$L(0.5-10~{\rm keV})/{\rm erg~s}^{-1}$]   &         21 &      0.097 &     $ 0.32_{- 0.24}^{+ 0.20}$ \\
         \savg        &                         log[$L(0.5-2~{\rm keV})/L(0.5-10~{\rm keV})$]   &         21 &      0.426 &     $ 0.06_{- 0.27}^{+ 0.27}$ \\
         \savg        &                                log[$L(0.5-10~{\rm keV})/L_{\rm BOL}$]   &         21 &      0.085 &     $ 0.32_{-0.23}^{+ 0.20}$ \\
\enddata
\tablecomments{Column (1): Dependent variable (O~VI/N~V absorption line property). Column (2): Independent variable (quasar/host
  property). Underlined entries under col.\ (2)
  indicate relatively strong correlations with P($r$) $< 0.05$ and $|r| \gtrsim 0.5$. Column (3): Number of data points. Column (4): Probabilities for the correlation coefficients r to be $\leq$0 (for positive correlation) and $\ge$0 (for negative correlation), as defined in the 3rd paragraph in Sec. \ref{61}. Column (5): Correlation
  coefficients $r$ and their 1-$\sigma$ errors.}
\end{deluxetable}

\subsection{Radiation Pressure as the Most Plausible Wind-Driving Mechanism}  \label{63}

Overall, the results from the analyses of the combined ULIRG $+$ quasar sample reinforce the main conclusions of Paper~I: (i) The incidence rate and properties of the O~VI and N~V absorption features (i.e., EWs, outflow velocities and outflow velocity dispersions) are positively correlated with the X-ray weakness of the sources. (ii) The incidence rate of these absorption features is higher in sources with larger X-ray absorbing column densities. The EWs of absorption features may also be higher in such sources.

This dependence of the incidence rate, EWs and kinematic properties of these outflows on the X-ray weakness and/or absorbing columns of the sources can best be explained if these outflows are radiatively driven. As discussed in detail in Section 7.3 of Paper I, the combined radiative force \citep[``force multiplier'';][]{AravLi1994} is greatly suppressed when the gas is over-ionized by the extreme-ultraviolet (EUV)/X-ray photons, becoming too transparent to be radiatively accelerated effectively. The successful launching of a radiatively-driven wind thus depends on whether this ionizing EUV/X-ray radiation is shielded and/or intrinsically weak. 

In the first case, the over-ionized material may serve as a radiative shield to soften the ionizing spectrum enough so that the outflow material downstream can be effectively accelerated \citep[][]{Murray1995,Proga2004,Proga2007,Sim2010}. However, as mentioned in Paper I, the predicted strong near-UV absorption features near systemic velocity produced by the shielding material \citep[e.g.,][]{Hamann2013} are in general not observed in our sample. In the second case, it is proposed that the X-ray emission in weak-lined ``wind-dominated'' quasars are intrinsically faint and unabsorbed \citep{Richards2011,Wu2011,Luo2015,Veilleux2016}.  In our sample, several sources with fast O~VI/N~V outflows show evidence of intrinsically weak X-ray emission, including F07599, PG1001 and PG1004 \citep{Luo2013c,Luo2014}. More sensitive hard X-ray ($>$10 keV) observations of our sample would shed light on the exact origin of this X-ray weakness. 

Apart from the aforementioned trends with X-ray weakness, the new data on the ULIRGs do not add significantly more support to the radiatively driven wind scenario. The lack of positive trends between the maximum velocities of the outflows and the optical, UV, bolometric luminosities or the Eddington ratios is likely due to the limited dynamic range of properties of the combined ULIRG$+$quasar sample, and noise in the predicted correlations associated with projection effects and variance in the launching radius and efficiency of the radiative acceleration associated with the complex microphysics of the photon interaction with the clouds (Paper I).  Furthermore, we find no other case of line-locking among the outflows of ULIRGs (line-locking was observed in the outflows of two quasars in Paper I). Lastly, as in Paper I, no evidence is present that radiation pressure on dust grains is an important contributor to the radiative acceleration in our sample (the outflow properties do not correlate with the mid-, far-, and total (1 -- 1000 \mum) infrared excesses). While the alternative thermal wind and ``blast wave'' models cannot be formally ruled out by our data, these models cannot readily explain the observed connection between the outflow properties and X-ray weakness and absorbing column densities (interested readers are referred to Section 7.3 of Paper I for a more detailed discussion of these models).

\subsection{The Effects of Stochasticity of AGN-Outflow Activity}
\label{64}

Intuitively, the lack of correlations between the properties of the O~VI and/or N~V outflows and those of the AGN is unexpected given that these winds are driven by the AGN. Together with the result that the outflow incidence rate in the ULIRG sample is virtually identical to that in the quasar sample, it may imply that the launching of these quasar-driven outflows is a stochastic phenomenon throughout the late merger stages. This is consistent with the picture that the triggering of AGN activity has a significant chaotic/random component in local gas-rich mergers and AGN \citep[e.g.,][]{Davies2007,Veilleux2009}. Given this stochasticity of AGN activity, \citet{Veilleux2009} warns that a sample size of $\ga$50-100 may be needed to detect any trends with merger phase. Time delays between bursts in AGN activity, the ejection of the material driven this AGN activity, and the detection of the ejected material on pc and kpc scales also likely complicate this picture \citep{Veilleux2017}.

\section{Summary} \label{7}

As part II of an HST/COS FUV spectroscopic study of the QUEST (Quasar/ULIRG Evolutionary Study) sample of local quasars and ULIRGs, we have systematically analyzed a sample of 21 low-redshift (z$<$0.3) ULIRGs, examining both the \lya\ emission line and \ovi\ and \nv\ absorption features. For the \lya\ analysis, the results of the starburst-dominated ULIRGs from \citet{Martin2015} (M15) are combined with ours, when possible. For the analysis of the O~VI and N~V absorption features, the results of the quasar sample from \citet{Veilleux2021} (Paper I) are also combined with ours, when appropriate. The main conclusions of our analyses can be summarized as follows:

\begin{itemize}

\item 
\lya\ line emission is detected in 15 out of the 19 objects where \lya\ lies within the wavelength range of the observations. Blueshifted line centroids and/or wings of \lya\ emission are often seen in our sample, where 12 out of the 14 objects with robustly measured \lya\ profiles show 80-percentile velocities \vba\ $\lesssim$ 0. See Fig. \ref{fig:lyaloopAGN}, Table \ref{tab:lya}, and Section \ref{31}.

\item
The equivalent widths of \lya\ increase with increasing AGN fractions and AGN luminosities. The strength of the \lya\ emission is therefore correlated with that of the AGN. See Fig. \ref{fig:ewlya_AGN}, Fig. \ref{fig:fesc_AGN}, and Section \ref{41}. 

\item
The blueshifted line centroids and/or wings of the \lya\ emission correlate with those of the non-resonant optical emission lines. The 80-percentile velocities \vba\ of \lya\ are positively correlated with those of \oiii\ (or \ha), with the highest statistical significance among all kinematic properties measured from the data. This suggests that the blueshifted wings of \lya\ emission are physically linked to the ionized outflowing gas. There is also a possible connection between the blueshifted \lya\ emission lines and the blueshifted \nad\ absorption lines tracing the cool neutral-atomic gas outflows, although the sample size (3) in this case is very limited. See Fig. \ref{fig:velo3lya}, Fig. \ref{fig:nad}, and Section \ref{42}.

\item
For 6 of the 14 objects with clear \lya\ detections, the \lya\ escape fractions, calculated as the observed \lya\ flux divided by the intrinsic \lya\ flux expected from the extinction-corrected \ha\ flux, are higher than the values expected under Case B recombination adopting \citet{Cardelli1989} reddening law. Weak, positive correlations exist between the \lya\ escape fractions and the AGN strength (e.g. \lagn, \fagn) or outflow velocities (e.g. $-$\vbao). See Fig. \ref{fig:fesc_AGN}, Fig. \ref{fig:v80o3_ewlya}, Section \ref{41}, and Section \ref{42}.

\item
Among the 12 objects with good continuum S/N, at least 6 objects show clear O~VI and/or N~V absorbers. The velocity centroids of these absorbers are all blueshifted and show large ranges of depth-weighted velocities (from $\sim-$12690 to $-$170 \kms) and depth-weighted velocity dispersions (from $\sim$100 to 4600 \kms). They are likely tracing quasar-driven outflows based on their broad and smooth profiles, as well as the evidence for partial covering in several objects. The implied incidence rate of highly ionized gas outflows in our ULIRG sample ($\sim$50\%) is similar to that of the QUEST quasars in Paper I ($\sim$60\%). See Table \ref{tab:abs}, Table \ref{tab:incidence}, Section \ref{33}, and Section \ref{51}.

\item
The locations of these O~VI and N~V outflows are not well constrained, although they are probably located outside of the broad emission line regions since the absorption features are deeper than the underlying continuum level in at least two (and perhaps all six) ULIRGs. The lower limits on the power and momenta of these outflows, based on conservative values of the metal abundances, ionization corrections, and radial distances of the outflowing material, are generally modest compared with the radiative luminosities and momenta of the central energy source (AGN$+$starburst). See Table \ref{tab:energetics} and Section \ref{52}.

\item
 When combining the results on the ULIRGs presented in this paper with those on the QUEST quasar sample from Paper I, we find that the incidence rates of O~VI and/or N~V absorption features are higher in the X-ray weak sources with smaller X-ray-to-UV indices, \aux. Specifically, the incidence rate of either O~VI or N~V or both absorption features is 88\% (1-$\sigma$\ range: 76\%--92\%) in objects with \aux\ $<-$1.3, and 45\% (1-$\sigma$\ range: 35\%--56\%) in objects with \aux\ $\ge-$1.3.  Similarly, the equivalent widths, weighted outflow velocities, and weighted velocity dispersions of these features are higher in the X-ray weak sources. These results reinforce the main conclusions of Paper I and favor radiative acceleration as the dominant driving mechanism. See Fig. \ref{fig:aux}, Table \ref{tab:incidence}, Table \ref{tab:regressions}, and Section \ref{62}.
 
\item
As found in Paper I, the incidence rate of O~VI or N~V or both absorption features for sources with X-ray absorbing column densities \NHxray\ $>$ 10$^{22}$ cm$^{-2}$ is larger (76\%; 1-$\sigma$ range: 65\%--83\%) than the rate among those with lower \NHxray\ (25\%; 1-$\sigma$ range: 17\%--41\%). The equivalent widths of the O~VI and N~V absorption features may also be higher in sources with larger X-ray absorbing column densities. See Fig. \ref{fig:logNH}, Table \ref{tab:incidence}, Table \ref{tab:regressions}, and Section \ref{62}.

\item
Apart from the aforementioned correlations with the X-ray properties of the sources,  the properties of the outflows do not correlate with those of the AGN/host galaxies along the late-stage merger sequence (i.e. from AGN-dominated ULIRGs to quasars). Since the incidence rate of outflows found in our AGN-dominated ULIRGs is also virtually the same as that in the quasars, these results suggest that the launching of these quasar-driven outflows is stochastic throughout the late merger stages. A rigorous exploration of the outflow properties along the merger sequence would require a larger ($\ga$50-100) sample that covers equally well the pre-merger and late-merger stages of ULIRGs. See Section \ref{64}.

\end{itemize}

%\clearpage

\appendix
\section{Notes on Individual Objects} \label{A3}

In this section, we summarize the detections of emission and absorption features in each ULIRG.

F01004$-$2237: Clearly broad, blueshifted wings are seen in both \lya, and N~V emission. An O~VI BAL is present at the edge of the blue side of the spectrum.

Mrk~1014: There are broad \lya, O~VI, and N~V emission. No associated O~VI or N~V absorption lines is visible. 

F04103$-$2838: There is no \lya\ emission at systemic velocity, while a narrow emission line is seen at $\sim -$2000 \kms\ in the rest frame of \lya. It may be a redshifted Si III 1206 emission ($\sim$ 150 \kms), or a narrow \lya\ emission in the foreground. No associated N~V absorption line is visible.

F07599$+$6508: The spectrum is dominated by a prominent N~V BAL with a centroid velocity similar to the blueshifted \nad\ absorption line seen in the optical. There are also multiple blueshifted and redshifted, narrow N~V absorption lines at lower velocities ($\lesssim -$5000 \kms). 

F08572$+$3915\_NW: No signal is detected.

F11119$+$3257: There is broad, blueshifted \lya\ emission and a less blueshifted, narrower \lya\ absorption feature on top of it. Part of the blueshifted N~V emission is also detected. No associated N~V absorption line is visible.

Z11598$-$0112: There is broad, blueshifted \lya\ emission, superimposed with narrower \lya\ absorption line close to the systemic velocity. Associated O~VI and N~V absorption features are also detected.

F12072$-$0444: There are broad, blueshifted \lya\ emission and narrower \lya\ absorption line near the systemic velocity. No associated N~V absorption line is visible.

3C~273: There is broad \lya\ emission, superimposed by multiple narrow foreground absorption features. No associated O~VI or N~V absorption line is visible. 

F13218$+$0552: There is broad O~VI emission, superimposed by narrower O~VI absorption line. \lyb\ absorption line with velocity similar to the O~VI absorption line may exist but can not be confirmed (S/N $\lesssim$2).

F13305$-$1739: The spectrum is dominated by broad \lya\ and N~V emission with blueshifted wings.

Mrk~273: No signal from the source is detected.

F14070$+$0525: No signal from the source is detected. 

F15001$+$1433\_E: No signal from the source is detected.

F15250+3608: The \lya\ line shows a P-Cygni-like profile. N~V emission is also detected. Emission and blueshifted absorption from the N~V 1242 line is visible, whereas the N~V 1239 transition overlaps with the strong geocoronal emission nearby so that no measurements of it can be made. As a result, the overall properties of the N V doublet is highly uncertain. Our estimates for the EW and centroid velocity of the N~V 1242 absorber alone are $\sim$0.3 \AA\ and $-$500 \kms, respectively, which has not taken into account the infilling from the N~V 1238, 1242 emission. There are several blueshifted absorption features from various low ionization species, including \niiuv, \niuv, \siiii, \siiiabc, with \vavg\ $\simeq$ $-$[300, 500] \kms.

F16156$+$0146\_NW: The \lya\ emission is peaked at $\sim +$500 \kms, with a superimposed blueshifted, narrower absorption feature at $\sim -$300 \kms, and a broad emission wing extending to $\lesssim -$2600 \kms. There are also weak, broad N~V emission and a potential detection of \siiii\ absorption line.

F21219$-$1757: There are broad, blueshifted \lya, N~V, and O~VI emission. Highly blueshifted absorption features at $\sim -$4500 \kms are seen for both the N~V doublets and O~VI 1038 line.

F23060$+$0505: There are broad \lya\ and N~V emission lines with blueshifted wings. \lya\ and N~V absorption lines with similar velocities are also seen. 

F23233$+$2817: There is a broad \lya\ emission superimposed by a narrow \lya\ absorption line at the systemic velocity. Broad N~V emission, and possible broad O~VI emission are also visible. The FUV continuum is virtually not detected and no associated O~VI or N~V absorption feature is visible.

\begin{acknowledgements} 

WL acknowledges support for this work provided by NASA through grant numbers HST GO-15662.001-A, GO-15662.001-B and 15915.002-A. SV, and TMT acknowledge partial support for this work provided by NASA through grant numbers HST GO-1256901A and GO-1256901B, GO-13460.001-A and GO-13460.001-B, and GO-15915.004-A from the Space Telescope Science Institute, which is operated by AURA, Inc., under NASA contract NAS 5-26555. Based on observations made with the NASA/ESA Hubble Space Telescope, and obtained from the Hubble Legacy Archive, which is a collaboration between the Space Telescope Science Institute (STScI/NASA), the Space Telescope European Coordinating Facility (STECF/ESA) and the Canadian Astronomy Data Centre (CADC/NRC/CSA). This research has made use of the NASA/IPAC Extragalactic Database (NED),
which is operated by the Jet Propulsion Laboratory, California Institute of Technology, under contract with the National Aeronautics and Space Administration.
\end{acknowledgements}

\software{Astropy \citep{astropy2013, astropy2018}, CALCOS (\url{https://github.com/spacetelescope/calcos}), LINMIX\_ERR \citep{Kelly2007}, LMFIT \citep{lmfit}, NumPy \citep{numpy}, pymccorrelation \citep{Privon2020}, pPXF \citep{ppxf}, SciPy \citep{scipy}.}

\bibliography{cos-ulirg}

\end{document}